\documentclass[11pt,oneside,reqno,xcolor=dvipsnames]{article} 

\usepackage[USenglish]{babel} 
\usepackage[utf8]{inputenc} 

\usepackage[a4paper,top=30mm,bottom=20mm,left=27.5mm,right=27.5mm]{geometry} 
\usepackage{lscape} 

\usepackage{setspace} 
\usepackage{amsmath} 
\usepackage{mathtools} 
\usepackage{array} 
\usepackage{tocloft} 
\usepackage{titlesec} 
    \titleformat*{\section}{\bfseries} 
    \titleformat*{\subsection}{\bfseries} 
    \titleformat*{\subsubsection}{\normalsize\bfseries} 
    \titleformat*{\paragraph}{\normalsize\bfseries} 
    \titleformat{name=\section}{\normalfont\bfseries}{\thesection}{0.7em}{\hspace{0em}}{} 
    \titleformat{name=\subsection}{\normalfont\bfseries}{\thesubsection}{0.7em}{\hspace{0em}}{} 
    \titleformat{name=\subsubsection}{\normalfont\normalsize\bfseries}{\thesubsubsection}{0.7em}{\hspace{0em}}{} 
    \titleformat{name=\paragraph}[runin]{\normalfont\normalsize\bfseries}{\theparagraph}{0.7em}{\hspace{0em}}{} 

\usepackage{titletoc}  

\usepackage[hang,flushmargin,symbol]{footmisc} 
    \renewcommand*{\thefootnote}{\fnsymbol{footnote}} 

\usepackage{lipsum} 

\usepackage{amssymb} 
\usepackage{marvosym} 
\usepackage{ifsym} 
\usepackage{wasysym} 
\usepackage{eurosym} 
\usepackage{textcomp} 
\usepackage{MnSymbol} 

\usepackage{caption} 
\usepackage{subcaption} 
\usepackage[capposition=top]{floatrow} 
\usepackage{float} 
\usepackage{afterpage} 

\usepackage{xcolor} 
\usepackage{soulutf8} 

\usepackage{tabularx} 
    \newcolumntype{L}[1]{>{\raggedright\arraybackslash}p{#1}} 
    \newcolumntype{C}[1]{>{\centering\arraybackslash}p{#1}} 
    \newcolumntype{R}[1]{>{\raggedleft\arraybackslash}p{#1}} 
\usepackage{multirow} 
\usepackage{hhline} 
\usepackage{rotating} 
\usepackage[flushleft]{threeparttable} 
\usepackage{diagbox} 
\usepackage{colortbl} 
\usepackage{arydshln} 
\usepackage{booktabs}

\usepackage{pgfplots} 
    \pgfplotsset{compat=newest}
    \usepgfplotslibrary{statistics} 
    \usepgfplotslibrary{fillbetween} 
    \pgfkeys{/pgf/number format/.cd,1000 sep={}} 
\usepackage{tikz} 
    \usetikzlibrary{decorations.pathreplacing, patterns, intersections} 

\usepackage{enumitem} 

\usepackage{graphicx} 
\usepackage{pdfpages} 

\usepackage{comment} 

\usepackage{bm} 

\usepackage[backend=bibtex,natbib=true,style=authoryear,doi=false,isbn=false,url=false,eprint=false,giveninits=true,uniquename=false,dashed=false,maxnames=3,maxbibnames=30]{biblatex} 

\usepackage{csquotes} 

    \DeclareNameAlias{sortname}{family-given} 
    \DeclareNameAlias{default}{family-given} 

    \DeclareFieldFormat[article,inbook,incollection,inproceedings,patent,thesis,unpublished]{title}{#1\isdot}

    \renewbibmacro{in:}{\ifentrytype{article}{}{\printtext{\bibstring{in}\intitlepunct}}}

    \renewbibmacro*{volume+number+eid}{%
      \printfield{volume}%
      \setunit*{\addnbthinspace}
      \printfield{number}%
      \setunit{\addcomma\space}%
      \printfield{eid}}
    \DeclareFieldFormat[article]{number}{\mkbibparens{#1}}

    \usepackage{xpatch}
    \def\dopatchbibdrivereditorcomma#1{%
      \xpatchbibdriver{#1}
        {\usebibmacro{maintitle+booktitle}%
         \newunit\newblock}
        {\usebibmacro{maintitle+booktitle}%
         \setunit{\addcomma\space}\newblock}
        {}
        {\typeout{failed to patch driver for type #1}}}
    \forcsvlist{\dopatchbibdrivereditorcomma}{inbook,incollection,inproceedings}

    \renewbibmacro*{byeditor+others}{%
      \ifnameundef{editor}
        {}
        {\usebibmacro{byeditor+othersstrg}%
         \setunit{\addspace}%
         \printnames[byeditor]{editor}%
         \clearname{editor}%
         \addcomma}%
      \usebibmacro{byeditorx}%
      \usebibmacro{bytranslator+others}}

    \makeatletter
    \pretocmd{\blx@head@bibintoc}{\phantomsection}{}{\ddt}
    \makeatother

    \AtBeginBibliography{
    }

    \DeclareCiteCommand{\citeyear}
        {}
        {\bibhyperref{\printdate}}
        {\multicitedelim}
        {}
    \DeclareCiteCommand{\citeyearpar}
        {}
        {\mkbibparens{\bibhyperref{\printdate}}}
        {\multicitedelim}
        {}

\usepackage[colorlinks=true,linkcolor=blue,citecolor=blue,urlcolor=black,linkbordercolor=cyan,citebordercolor=cyan,urlbordercolor=cyan,hyperfootnotes=true]{hyperref} 

\pgfplotsset{error bar legend/.style={%
    /pgfplots/legend image code/.prefix code={%
      \pgfkeysgetvalue{/pgfplots/error bars/error mark}{\pgfplotserrorbarsmark}%
      \draw[%
        /pgfplots/every error bar,
        mark=\pgfplotserrorbarsmark,
        /pgfplots/error bars/error mark options,
        sharp plot,
        ##1
      ] plot coordinates {(0.3cm, -0.15cm) (0.3cm, 0.15cm)};%
    }
  }
}

\begin{document}


\begin{titlepage}
\begin{center}


\Large

\textbf{Labor Demand on a Tight Leash}

\normalsize

\vspace*{-2cm}

\normalsize
\begin{minipage}[t][5cm][b]{0.48\textwidth}
\begin{center}
\href{https://www.mariobossler.de/}{\textcolor{black}{\textbf{Mario Bossler}}} \\
\href{https://www.th-nuernberg.de/en/faculties/bw/}{\textcolor{black}{TH Nuremberg}}\textcolor{black}{,} \href{https://iab.de/en/startseite-english/}{\textcolor{black}{IAB}}\textcolor{black}{,} \href{https://www.iza.org/}{\textcolor{black}{IZA}} \textcolor{black}{\&} \href{http://www.laser.uni-erlangen.de/index.php}{\textcolor{black}{LASER}}  \\[-0.2cm]
\end{center}
\end{minipage}
\begin{minipage}[t][5cm][b]{0.48\textwidth}
\begin{center}
\href{https://iab.de/en/employee/?id=12380200}{\textcolor{black}{\textbf{Martin Popp}}} \\
\href{https://iab.de/en/startseite-english/}{\textcolor{black}{IAB}} \textcolor{black}{\&} \href{http://www.laser.uni-erlangen.de/index.php}{\textcolor{black}{LASER}}  \\[-0.2cm]
\end{center}
\end{minipage}

\vspace*{2.5cm}

Outdated Manuscript (click \href{https://doi.org/10.1177/00197939261435961}{\textcolor{blue}{here}} for the published version in \textit{ILR Review})\footnote[1]{Corresponding Address: Martin Popp, Institute for Employment Research (IAB), Regensburger Straße 100, 90478 Nuremberg, Germany, Email: martin.popp@iab.de, Phone: $\plus$49 (0)911/179-3697.
Acknowledgements: Martin Popp is grateful to the Joint Graduate Program of IAB and FAU Erlangen-Nuremberg (GradAB) for financial support of his research.
We particularly thank Lutz Bellmann, Nicole Gürtzgen, Boris Hirsch, Alexander Kubis, Michael Oberfichtner, Andreas Peichl, Thorsten Schank, Claus Schnabel, Philipp vom Berge, and Jürgen Wiemers for helpful discussions and suggestions. Furthermore, we are grateful to Anna Hentschke and Franka Vetter for excellent research assistance. Earlier versions of this paper were presented at the 3rd Workshop of the German Minimum Wage Commission, the 17th CAED Conference (U Coimbra), the 14th Workshop on Microeconomics (U Lueneburg), the 14th Workshop on Labour Economics (IAAEU Trier), the Royal Economic Society Annual Conference 2022 (U Warwick), the 92nd Meeting of the Southern Economic Association (SEA), the International Workshop on Establishment Panel Analyses (IAB) as well as seminars in Nuremberg (IAB, FAU Erlangen-Nuremberg) and Würzburg (U Würzburg).} \\
March 1, 2024
\\[1.5cm]

\begin{abstract} 
\onehalfspacing
\noindent We develop a labor demand model that encompasses pre-match hiring cost arising from tight labor markets. Through the lens of the model, we study the effect of labor market tightness on firms' labor demand by applying novel shift-share instruments to the universe of German firms. In line with theory, we find that a doubling in tightness reduces firms' employment by 5 percent. Taking into account the resulting search externalities, the wage elasticity of firms' labor demand reduces from -0.7 to -0.5 through reallocation effects. In light of our results, pre-match hiring cost amount to 40 percent of annual wage payments. 

\end{abstract}

\vspace*{0.5cm}
\end{center}
\small



\end{titlepage}


\renewcommand*{\thefootnote}{\arabic{footnote}}
\setcounter{footnote}{0}

\newpage

\normalsize

\clearpage
\section{Introduction}
\label{sec:1}

Over the past decade, the German economy experienced a remarkable upswing. Between 2012 and 2019, Germany's real gross domestic product grew on average by 1.5 percent each year. At the same time, the German labor market witnessed the biggest expansion since the 1950s: The number of jobs rose by 3.7 million, reaching a record high of 45.2 million in 2019. As a flip side of this so-called ``German Labor Market Miracle'' \citep{BurdaSeele2020}, labor market tightness -- the ratio of vacancies to job seekers -- doubled. Consequently, German businesses lamented the lack of workers \citep{Handelsblatt2018}. For small and medium-sized companies, the shortage of skilled workers was such a severe problem that some managers were secretly hoping for an economic slowdown to release the strains \citep{FT2019}. In fact, there is no empirical evidence on the extent to which the increasing tightness has prevented firms from retaining or expanding their workforce. By and large, quantifying the employment effects of tightening labor markets would prove insightful as many industrialized economies have been facing labor shortages in recent years. For instance, employers in the U.S.\ find it increasingly hard to fill their vacancies, with labor market tightness reaching its highest level in the last quarter century \citep{AbrahamEtAl2020}.

In tight labor markets, firms compete for a relatively small number of job seekers to fill a relatively large number of vacancies. Such an imbalance in the labor market gives rise to hiring frictions, making it difficult for firms to recruit workers (i.e., their job-filling rate is reduced). To fill their vacancies, firms in tight labor markets must spend more cost on recruitment (e.g., by increasing their search intensity or search duration). As a consequence, firms' reduce their labor demand and, thus, employment falls. Although this channel is key, empirical models of labor demand narrow their analysis to the wage rate but do not consider the role of labor market tightness. The paucity of empirical evidence on the effect of labor market tightness on employment is twofold: On the one hand, detailed information on both vacancies and job seekers per labor market is rarely available. On the other hand, failure to isolate exogenous variation in labor market tightness would lead to spurious estimates.

In this paper, we estimate the causal effects of not only wages but also labor market tightness on labor demand in German firms. For this purpose, we extend the traditional profit-maximization model of firms to include pre-match hiring cost that arise in tight labor markets. These pre-match hiring cost drive a wedge between unit labor cost and the wage rate. The bottom line of our model is that, unlike conventional specifications, higher labor market tightness exerts a negative effect on firms' demand for workers because hiring becomes more costly. We estimate the model on the universe of social security records from the German labor market between 2012 and 2019. For the purpose of our analysis, we enrich these data with official statistics and survey information on vacancies and job seekers to determine firm-specific exposure to labor market tightness in more than 1,200 occupations.

Our data on vacancies and job seekers mirror Germany's favorable economic performance during the last decade. Between 2012 and 2019, the number of job seekers decreased from four to only two per vacancy. Hence, labor market tightness doubled within a span of only seven years. To ascertain whether higher tightness actually impedes recruitment, we inspect the cross-sectional relationship between our measure of labor market tightness and various hiring indicators from a large-scale business survey. In line, labor market tightness is positively correlated with pre-match hiring cost, the number of search channels, and the search duration of firms, while it is negatively correlated with the number of applicants per vacancy.

A naive OLS regression of firms' labor demand on wages and labor market tightness would provide upward-biased elasticities. To rule out reverse causality from uncontrolled shifts in labor demand, we instrument wages and labor market tightness with shift-share instruments, building on the popular instrumental variable (IV) strategy from \citet{Bartik1993}. Typically, Bartik instruments combine national industry shifts with past shares of industries in regions to isolate exogenous variation in variables at the regional level. However, we transfer this shift-share design to the firm level by taking advantage of the fact that a firm employs many occupations, just as a region has many industries. Thus, our novel Bartik-style instruments combine national occupation shifts with past shares of occupations in firms' employment to extract exogenous variation at the firm level. We construct three instruments for average wages, vacancies, and job seekers. By virtue of our shift-share design, exogeneity holds when either the national trends (i.e., the shifts) or firms' predetermined occupational composition (i.e., the shares) are uncorrelated with the differenced error term.

The regression results are in line with our labor demand model that involves positive pre-match hiring cost. As expected, our IV regressions yield more negative elasticities than naive OLS regressions. In our baseline IV regression, we arrive at an own-wage elasticity of labor demand to the single firm of -0.7, which reflects negative substitution and scale effects. Moreover, the elasticity for tightness is -0.05, implying that the observed doubling in labor market tightness lowered firms' employment on average by 5 percent, holding all other things equal. As a favorable feature of our micro-founded model, we can use the wage and tightness elasticities to gauge the relative magnitude of pre-match hiring cost. Through the lens of our model, pre-match hiring cost make up roughly 40 percent of annual wage cost, which lies in the middle of estimates from dynamic labor demand models \citep{Yaman2019}, corroborating the relevance of hiring cost and the plausibility of our identification strategy.

We perform additional empirical checks to validate our identification strategy. First, the underlying first-stage regressions highlight that our three instruments are strong predictors and load correctly on the wage and the tightness variable. Second, we decompose our Bartik estimates along the lines of \citet{Goldsmith-PinkhamEtAl2020} into Rotemberg weights and just-identified IV estimates. The decomposition shows that the wage effect is largely determined by the exogenous introduction of the nation-wide minimum wage in 2015 (i.e., the wage shifts are plausibly exogenous). Third, cross-sectional regressions highlight that the predetermined shares of the most important occupations are hardly correlated with labor demand variables, limiting the scope for reverse causality from uncontrolled shifts in labor demand (i.e., the shares are plausibly exogenous). Fourth, in an alternative specification, we additionally differentiate year fixed effects by industry or region to more rigorously control for common labor demand shocks. Despite the more detailed fixed effects, the results remain in the ballpark of the baseline estimates. Fifth, our results are not affected by different operationalizations of the wage and tightness variable. In particular, the results are robust to using an administrative (rather than a functional) delineation of regions and hold for a broader classification of occupations with 432 (rather than 1,286) occupations. Sixth, we are the first to construct a flow-adjusted measure of labor market tightness that takes into account vacancies and job seekers in neighboring occupations which represent additional outside options in the matching process of workers and firms. When using our flow-adjusted measures of labor market tightness, our baseline elasticities remain unaltered at values of -0.7 (wages) and -0.05 (tightness), respectively.

Our theoretical model and empirical results highlight that higher labor market tightness lowers firms' labor demand. \citet{BeaudryEtAl2018} point out that this negative effect gives rise to search externalities: a reduction in labor demand by one firm lowers labor market tightness and, thus, facilitates recruitment of workers in all other firms, leading to reallocation effects. More broadly speaking, aggregate changes in labor demand feature a self-attenuating feedback mechanism that operates via labor market tightness. When accounting for the resulting reallocation effects, we find that the individual-firm own-wage elasticity of labor demand shrinks from -0.7 to -0.5 at the aggregate level. Put differently, reallocation of workers across firms dampens aggregate changes in labor demand by roughly 30 percent.

In light of our estimates, we perform three further analyses of the German labor market. First, Germany introduced a nation-wide minimum wage in 2015. Based on conventional wage elasticities of labor demand, ex-ante simulations suggested that an hourly wage floor of 8.50 Euro would reduce employment by almost 1 million jobs \citep{KnabeEtAl2014}. However, evidence from ex-post evaluations indicates only modest disemployment effects \citep{BosslerGerner2020, CaliendoEtAl2018, DustmannEtAl2022}. In a simulation exercise, we interact observed wage increases with our aggregate own-wage elasticity of labor demand of -0.5 and find a reduction by 88,000 jobs which represents a similar order of magnitude to the effects of the available ex-post evaluations. Second, we quantify the extent to which the tightening of labor markets in Germany has adversely affected employment. Our simulation implies that the doubling of labor market tightness in Germany between 2012 and 2019 slowed down employment growth by about 1.1 million jobs. Third, we scrutinize additional channels of adjustment and find that firms were willing to make wage and skill concessions only to a limited extent. Hence, the massive increase in tightness did neither result in a substantial wage increase nor a marked downgrading of skill demands.

Our paper contributes to several strands of the literature. First, we join the proliferation of studies that attempt to estimate the own-wage elasticity of labor demand \citep{Hamermesh1986,Nickell1986}. Structural-form models infer elasticities from estimating parameters of cost or profit functions, which reflect the optimization behavior of firms at given factor prices (e.g., \citealp{HijzenEtAl2005,FreierSteiner2010,MuendlerBecker2010,PeichlSiegloch2012}). In contrast, reduced-form models run log-linear regressions of factor demand using wage rates as an explanatory variable (e.g., \citealp{ArellanoBond1991,Slaughter2001,HijzenSwaim2010}). A major concern with both approaches is the endogeneity of wages, namely that unaccounted shifts of the labor demand curve will yield upward-biased elasticities \citep{AngristKrueger2001}. The use of quasi-experimental variation in wages represents a promising method to address problems of endogeneity \citep{AddisonEtAl2014}. Unfortunately, quasi-experimental studies often lack external validity by focusing on rather narrow policy designs (e.g., low-wage workers when studying variation in minimum wages). In this paper, we seek to estimate the causal effect of wages on labor demand while, at the same time, making an externally valid statement about this key relationship. In particular, our novel Bartik-like instruments are designed to isolate exogenous variation at the firm level without requiring us to restrict the analysis to specific groups of workers or submarkets. Since our shift-share design rigorously addresses upward bias, our own-wage elasticity of -0.7 is at the lower end of the values reported in the international and German literature on labor demand \citep{LichterEtAl2015,Popp2023}.

Second, we add to the small but growing literature studying the consequences of the scarcity of labor inputs on firms' labor demand. Several studies exploit commuting policies to examine the impact of positive labor supply shocks on domestic employment in border regions \citep{DustmannEtAl2017,BeerliEtAl2021,Illing2023}. Other work builds on recessions to analyse whether the release of workers affects firms' skill requirements \citep{HershbeinKahn2018,ModestinoEtAl2020}. While studying shifts in labor supply is informative per se, it constitutes only a partial measure for the scarcity of the labor input: the given amount of supply can translate both into slack or tight labor markets depending on whether demand for workers is relatively low or high. To infer scarcity of the labor input, researchers should ideally employ the vacancy-to-job-seeker ratio (i.e., labor market tightness) or, alternatively, use an equilibrium outcome as a proxy thereof.\footnote{A number of studies use firm-level proxies for tightness (such as vacancy rates or self-reported shortages) to study the impact on firms' outcomes, such as productivity \citep{HaskelMartin1993}, labor contracts \citep{Fang2009,HealyEtAl2015}, or investments and capacity utilization \citep{DAcuntoEtAl2020}. Compared to labor market tightness, these firm-level proxies are usually inferior measures because they are plausibly related to firms' conduct and therefore more susceptible to endogeneity.} In a pioneering study on 595 firms, \citet{Stevens2007} shows that employment adjusts slower when industry-wide shortages of skilled labor are reported.

Relying on both supply and demand forces, \citet{BeaudryEtAl2018} perform the most comprehensive analysis to estimate the causal impact of labor market tightness on employment. The authors leverage census data on the U.S.\ economy between 1970 and 2015 and estimate elasticities at the city level using conventional Bartik instruments. The study finds that a 10 percent increase in the employment rate (as a proxy for labor market tightness) reduces employment by about 20 percent. Our paper extends the aforementioned study in several respects: i) Official statistics in Germany allow us to directly measure labor market tightness as the vacancy-to-job-seeker ratio (rather than using a proxy). While most studies are limited to studying tightness at the regional level, ii) our unusually rich data allow us to additionally differentiate tightness between 1,286 occupations at the 5-digit level. What is more, using survey evidence, iii) we verify that our granular measure of tightness positively correlates with several measures of recruitment difficulties. iv) We study the implications of tightness on employers (rather than cities) at the micro level, building on the universe of firms in the German social security register. In this regard, v) we provide shift-share diagnostics to track down the identifying variation behind our results. Moreover, vi) we estimate our model in two-year rather than ten-year differences, focusing on the medium rather than the long run. Finally, in terms of external validity, vii) our results are not limited to urban areas but also reflect the impact on firms in rural areas. Despite conceptual differences and an alternative setting, our findings resonate with \citep{BeaudryEtAl2018}, buttressing that tightness is detrimental to employment.

Third, our paper also speaks to the literature on hiring cost. Whereas business surveys provide direct evidence on hiring cost (e.g., \citealp{Oi1962}), dynamic labor demand models indirectly infer the size and shape of these cost using regression techniques (e.g., \citealp{Nickell1986}). The most detailed evidence from business surveys implies that pre- and post-match hiring cost sum up to roughly 20-30 percent of annual wage payments
\citep{BlatterEtAl2012,MuehlemannPfeifer2016,MuehlemannStruplerLeiser2018}. However, there are two reasons why even highly specialized surveys may underestimate the true magnitude of pre-match hiring cost. On the one hand, it is difficult to inquire quasi-fixed cost of hiring (e.g., the expenditure for renting offices for the human resource department or for participating in job fairs), which are frequently inferred to be large \citep{Hamermesh1989,CaballeroEtAl1997,AbowdKramarz2003,NilsenEtAl2007}. On the other hand, survey information on successful hiring processes is positively selected in the sense that it tends to disregard the more difficult and, hence, more expensive recruitment processes, which were either unsuccessful or not undertaken at all. As a favorable feature of our static profit-maximization model, the ratio of the own-wage to the tightness elasticity of labor demand allows to infer the magnitude of pre-match hiring cost from calibrating only a few model parameters. Our indirect approach is able to capture pre-match hiring cost in their entirety, covering not only successful but also unsuccessful recruitment processes. Based on observed labor demand responses on the universe of German firms, we quantify pre-match hiring cost to amount to roughly 40 percent of annual wage payments. This value falls in the middle of indirect estimates for hiring costs from dynamic labor demand models, which show a wide array of results and range from values near zero \citep{Hall2004,CooperEtAl2007,AsphjellEtAl2014} to more than one year of wage payments \citep{Rota2004,Bloom2009,Yaman2019}.

Fourth, this paper also contributes to the literature on reallocation effects of input factors. Several papers highlight the role of reallocation of workers between occupations \citep{Carrillo-TudelaVisschers2023,KambourovManovskii2009}, industries \citep{GolanEtAl2007}, and firms \citep{DavisHaltiwanger1992,FosterEtAl2016}. In his seminal contribution, \citet{Hamermesh1993} argues that changes in firms' employment overstate aggregate changes in employment to the extent that workers are reallocated between firms within the same aggregate. In recent work, \citet{DustmannEtAl2022} analyze the 2015 minimum wage introduction in Germany and attribute their finding of close-to-zero employment effects to the fact that many laid-off workers were reallocated to other firms. Building on \citet{BeaudryEtAl2018}, our labor demand model with pre-match hiring cost allows to pin down the magnitude of reallocation effects. By means of the tightness channel, we find that search externalities dampen aggregate changes in labor demand by roughly 30 percent through reallocation effects.

The study is organized as follows. In Section \ref{sec:2}, we augment the standard model of labor demand with pre-match hiring cost to highlight the role of labor market tightness. In Section \ref{sec:3}, we set up the empirical specification and develop novel Bartik instruments at the firm level. Section \ref{sec:4} describes the data. In Section \ref{sec:5}, we present descriptive results on labor market tightness in Germany. Section \ref{sec:6} illustrates the regression results, including their implications for the magnitude of pre-match hiring cost and reallocation effects. Section \ref{sec:7} discusses the implications of our results for the 2015 minimum wage introduction in Germany, the doubling of labor market tightness between 2012 and 2019, and the incidence of wage and skill concessions. Finally, we conclude in Section \ref{sec:8}.

\section{Theoretical Model}
\label{sec:2}

We begin with examining the theoretical relationship between wages, labor market tightness, and firms' labor demand to facilitate the later interpretation of our empirical results. Labor demand is a derived demand that originates from firms' ambition to satisfy product demand. The wage rate takes on a key role in the optimization calculus of firms. Under standard assumptions on technology, both negative substitution and negative scale effects imply that the own-wage elasticity of labor demand is less than zero (\citealp{Sakai1974}; \citealp{Hamermesh1986}). In contrast, the theoretical implication of a higher labor market tightness on employment is more subtle. The reason is that traditional models of labor demand assume that employers adjust input factors at no cost \citep{AddisonEtAl2014}. However, given the specific nature of the labor input, the cost of adjusting labor is substantial \citep{Oi1962,HamermeshPfann1996,BlatterEtAl2012,Yaman2019}. Importantly, higher labor market tightness amplifies firms' hiring frictions rendering recruitment more costly \citep{MuehlemannStruplerLeiser2018}. To shed more light on the relationship between labor market tightness and employment, we propose a tractable model that involves positive pre-match hiring cost.

\paragraph{Intertemporal Profit Maximization.} Assume that a representative firm with static expectations seeks to maximize profits. The firm's production function, $Y=F(L,K)$, depends on labor $L$ and capital $K$, each of which exhibits a positive but decreasing marginal product. The firm operates in perfectly competitive product and factor markets. Hence, the firm sells its goods at given price $P$ while employing labor for wage $W$ and purchasing capital at rate $R$. To simplify the model, we abstract from involuntary layoffs and model an exogenous rate $\delta$ at which workers separate from the firm.\footnote{We maintain the assumption of homogeneous labor for simplicity. With labor as a homogeneous input factor, it is not rational for the firm to simultaneously hire and dismiss workers.} The firm may decide to hire new workers but, importantly, must spend unit hiring cost $C$ per hire $H$.\footnote{For ease of presentation, we build on a static framework by assuming that the overall cost of hiring is a linear function of new hires. Under linear adjustment cost, the firm adjusts employment instantaneously to its optimal level $L$ \citep{Hamermesh1993}. In contrast, quadratic adjustment cost slow down the response to shocks that alter $L$ \citep{HoltEtAl1960}. To lower the total cost of adjustment, firms will find it optimal to smoothly adjust labor towards the optimum over several periods \citep{Gould1968}. With quasi-fixed cost of hiring, firms will only move to the new equilibrium level of employment if foregone profits from being out of equilibrium are larger than the respective cost of adjustment \citep{Hamermesh1989}. By and large, implementing these dynamics would add little to the understanding of the effect of labor market tightness on employment.} Let $r$ denote the firm's subjective discount rate for future profits in continuous time. Over time $t$, the firm will choose $H_t$ and $K_t$ so as to maximize the sum of contemporary and discounted future profits
\begin{equation}
\label{eq:1}
\int_{0}^{\infty} \big( P_{t} \cdot Y(L_{t}, K_{t}) \,-\, W_{t} \cdot L_{t} \,-\, R_{t} \cdot K_{t} \,-\, C_{t} \cdot H_{t} \big) \, e^{-rt} \, dt
\end{equation}
subject to the law of motion for employment: $\dot{L}_{t} = H_t \,-\, \delta \cdot L_{t}$. To solve the optimization problem, we set up the following Hamiltonian function $h$ where $L_t$ operates as state variable and $H_t$ as control variable:
\begin{equation}
\label{eq:2}
\max_{L,H,K,\lambda} h \, = \, \big( P \cdot Y(L_{t}, K_{t}) \,-\, W \cdot L_t \,-\, R_t \cdot K_t \,-\, C \cdot H_t \big) \, e^{-rt} \,+\, \lambda_{t} \cdot \big(H_{t} \,-\, \delta \cdot L_{t}\big)
\end{equation}
Using the Euler equation, $\frac{\partial h}{\partial H_{t}}= 0 $, and the maximum principle, $\frac{\partial h}{\partial L_{t}}= -\dot{\lambda}_{t}$, we arrive at the optimality condition for labor:
\begin{equation}
\label{eq:3}
P_{t} \cdot Y_{L}(L_{t},K_{t}) \,\overset{!}{=}\, W_{t} \,+\, ( \delta + r ) \cdot C_{t} \,
\end{equation}
The firm will employ an additional worker as long as the value of the marginal product of labor $Y_{L}$ exceeds the wage rate plus the amortized cost of hiring \citep{Hamermesh1993}. The per-period rate of amortization, $\delta + r$, reflects all factors that lower the future value of a contemporaneous hire. On the one hand, the exogenous separation rate $\delta$ implies that only a fraction of new hires, $(1-\delta)^T$  is still employed in the firm after $T$ periods. On the other hand, the costs of a new hire represent a one-off reduction in today's profit which the company favours with the discount rate $r$ over future profit. Taken together, a higher separation or discount rate shorten the period during which the worker must amortize the cost of hiring.\footnote{In the special case that both the exogenous separation rate and the subjective discount rate is zero, optimal labor demand is no longer a function of hiring cost. By contrast, if the exogenous separation rate is 1, the overall hiring cost must be recovered already in the period of hiring.} Due to the linearity assumption on hiring cost, the optimality condition (\ref{eq:3}) implies that the firm instantaneously adjusts towards a steady state, $\dot{L}_{t} = 0$, where it hires exactly as many workers as it loses exogenously in each period: $H_{t}=\delta \cdot L_{t}$. Thus, for expositional simplicity, we omit time subscripts in the remainder of this section. In the profit-maximizing optimum, unconditional labor demand
\begin{equation}
\label{eq:4}
L \,=\, Y_{L}^{-1}\,\Bigg(\frac{\,W \, + \, ( \delta + r ) \cdot C\,}{P}, K\Bigg)
\end{equation}
depends on unit labor cost, $W^{\text{\textasteriskcentered}} \,\equiv\, W \, + \, ( \delta + r ) \cdot C\,$, the optimal level of the capital stock $K$, and the price level $P$. Under regulatory assumptions on the production function, optimal labor demand is not only decreasing in the wage rate $W$ but also in unit hiring cost $C$.

\paragraph{Pre-Match Hiring Cost.} Following \citet{Yashiv2006}, we decompose unit hiring cost into a pre-match and a post-match component. Pre-match hiring cost $\Phi$ comprises all search costs of filling a vacancy with a suitable candidate. This cost includes expenditures for job advertisement, posting vacancies, screening candidates, interviews, headhunters, signing bonuses, negotiations, or the maintenance of a human resource department. Post-match hiring cost $\Psi$ involves all costs after the contract was signed, namely costs of onboarding new workers.\footnote{Specifically, post-match hiring cost comprise adaption and disruption costs. Adaption costs arise because new hires have an initially lower productivity and require formal training. Disruption costs result from informal instruction of new hires by incumbent workers which hampers the production process.}

In the following, we take into account that filling a vacancy becomes more difficult in tighter labor markets where the number of open vacancies $V$ is relatively high compared to the number of unemployed persons $U$ \citep{MortensenPissarides1999, Pissarides2000}. To attract workers in tight labor markets, the firm must search longer or increase its search effort (e.g., by using more costly search channels or interviewing more candidates). Accordingly, we assume that labor market tightness, $\theta=\frac{V}{U}$, is positively associated with pre-match hiring cost \citep{Pissarides2009}. Furthermore, we assume that pre-match hiring costs are a function of the wage rate \citep{Pissarides1994,BeaudryEtAl2018}, but do no specify a sign restriction since it is unclear whether higher wages ultimately require more screening (positive sign) or ease recruitment (negative sign). For ease of presentation, we assume fixed post-match hiring costs which are plausibly unrelated to labor market tightness.\footnote{We are aware of only one study that examines the impact of labor market tightness on hiring cost. Using survey data on Switzerland, \citet{MuehlemannStruplerLeiser2018} find a positive relationship between labor market tightness and pre-match hiring cost, but no such association for post-match hiring cost.}
Overall, the functional form for unit hiring cost is
\begin{equation}
\label{eq:5}
 C \,\,=\,\, \Phi \,\,+\,\, \Psi \,\,=\,\, \,c \cdot W^{\phi_1} \cdot \theta^{\phi_2} \,\,+\,\, \Psi
\end{equation}
where $c \geq 0$, $\phi_1 \gtreqless 0$, $\phi_2 \geq 0$, and $\Psi \geq 0$. To support our formulation of hiring cost, we will later provide empirical validation for our postulated relationships between wages, labor market tightness, and pre-match hiring cost using cross-sectional information on successful hiring processes (see Section \ref{sec:5}).

\paragraph{Wages, Labor Market Tightness, and Labor Demand.} The fundamental law of labor demand describes the key determinants of firms' profit-maximizing labor demand. The law decomposes the own-wage elasticity of labor demand into substitution and scale effects \citep{Hamermesh1993}:
\begin{equation}
\label{eq:6}
\eta^{L}_{W} \,\equiv\, \frac{\partial\, ln\,L}{\partial\, ln\,W}  \,=\, -\,(1-s_{L}) \cdot \sigma \,-\, s_{L} \cdot \eta^{Y}_{P}
\end{equation}
Absent hiring cost, the law states that labor demand is more elastic (i.e., the unconditional own-wage elasticity of labor demand becomes more negative), ...
\begin{itemize}[leftmargin=2cm]
\item[1.] ... the higher the elasticity of substitution between labor and capital, $\sigma$, and
\item[2.] ... the higher the price elasticity of product demand, $\eta^{Y}_{P}=- \frac{Y_{P} \, P}{Y}$, and
\item[3.] ... the higher the share of labor in total cost, $s_{L}$.\footnote{These relationships describe three of Marshall's \citeyearpar{Marshall1890} ``Four Laws of Derived Demand''. Regarding the third law, Hicks \citeyearpar{Hicks1932} notes that a higher labor share renders labor demand more elastic only when, in absolute terms, the price elasticity of product demand exceeds the elasticity of substitution.}
\end{itemize}
In Appendix \ref{sec:A}, we combine the optimality condition for labor (\ref{eq:3}), the optimality condition for the capital stock, a product market equilibrium condition, and our formulation of unit hiring cost (\ref{eq:5}) to derive a hiring-cost adjusted version of the fundamental law of labor demand:
\begin{equation}
\label{eq:7}
\eta^{L}_{W} \,\equiv\, \frac{\partial\, ln\,L}{\partial\, ln\,W}  \,=\,\, \Bigg(\,\frac{W}{\,W^{\text{\textasteriskcentered}}\,} \,+\, \phi_{1} \cdot \frac{ (\delta + r) \cdot c \cdot W^{\phi_{1}} \cdot \theta^{\phi_{2}}}{W^{\text{\textasteriskcentered}}} \Bigg) \,\cdot\, \vphantom{\Bigg(}\Big( \,-(1-s_{L}) \cdot \sigma \,-\, s_{L} \cdot \eta^{Y}_{P}  \, \Big)\vphantom{\Bigg)}
\end{equation}
In this formulation, the prevalence of positive unit hiring cost drives a wedge between unit labor cost and the wage rate. Thus, the first term in (\ref{eq:7}) describes the elasticity of unit labor cost with respect to the wage rate whereas the second term refers to the elasticity of labor demand to unit labor cost. Specifically, the elasticity of unit labor cost with respect to the wage rate is the weighted wage elasticity of the three components of unit labor cost: namely the wage rate (entering with elasticity 1), pre-match hiring costs (entering with elasticity $\phi_{1}$), and post-match hiring costs (entering with elasticity 0). The weights refer to the share of the wage rate, the share of amortized pre-match hiring costs, and the share of amortized post-match hiring costs in unit labor cost. Thus, in addition to (\ref{eq:6}), labor demand becomes more elastic, ...
\begin{itemize}[leftmargin=2cm]
\item[4.] ... the higher the share of the wage rate in unit labor cost, $\frac{W}{\,W^{\text{\textasteriskcentered}}}$, and
\item[5.] ... the higher the elasticity of pre-match hiring cost to the wage rate, $\phi_{1}$, and
\item[6.] ... the higher the share of amortized pre-match hiring costs in unit labor cost, $\frac{ (\delta + r) \, c \, W^{\phi_{1}} \,\theta^{\phi_{2}}}{W^{\text{\textasteriskcentered}}}$, provided that $\phi_{1}>0$.
\end{itemize}

Since we model tightness as a main determinant of hiring cost, we further derive the elasticity of labor demand with respect to labor market tightness (see Appendix \ref{sec:A}):
\begin{equation}
\label{eq:8}
\eta^{L}_{\theta} \,\equiv\, \frac{\partial\, ln\,L}{\partial\, ln\,\theta}  \,=\, \phi_{2} \cdot \frac{ (\delta + r) \cdot c \cdot W^{\phi_{1}} \cdot \theta^{\phi_{2}}}{W^{\text{\textasteriskcentered}}} \,\cdot\, \Big( \,-(1-s_{L}) \cdot \sigma \,-\, s_{L} \cdot \eta^{Y}_{P}  \, \Big)
\end{equation}
Analogously to (\ref{eq:7}), the elasticity equals the elasticity of labor demand with respect to unit labor cost multiplied by the elasticity of unit labor cost to labor market tightness. The latter elasticity is the elasticity of labor market tightness on pre-match hiring cost, $\phi_{2}$, weighted by the share of amortized pre-match hiring costs in unit labor cost. Overall, unconditional labor demand reacts more elastically to changes in labor market tightness, ...
\begin{itemize}[leftmargin=2cm]
\item[1.] ... the higher the elasticity of pre-match hiring cost with respect to labor market tightness, $\phi_{2}$, and
\item[2.] ... the higher the share of amortized pre-match hiring costs in unit labor cost, $\frac{ (\delta + r) \, c \,  W^{\phi_{1}} \, \theta^{\phi_{2}}}{W^{\text{\textasteriskcentered}}}$, and
\item[3.] ... the higher the elasticity of labor demand with respect to unit labor cost.
\end{itemize}

So far, our theoretical model postulates that labor market tightness purely raises hiring cost. However, when labor markets tighten, firms may also raise wages to attract workers and, thus, ease congestion in the hiring process \citep{BassierEtAl2023}. In search-and-matching, bargaining, or efficiency-wage models, higher labor market tightness will also exert a positive effect on wages. In Appendix \ref{sec:A}, we derive a version of Equation (\ref{eq:8}) that also incorporates a direct effect of tightness on the wage rate. Since, in our baseline regression model, we jointly estimate the effect of wages and tightness on labor demand, we are conditioning on the wage level when inferring the elasticity of labor demand with respect to tightness. Thereby, we do not capture effects of wage changes that originate from a change in tightness. To substantiate this approach, we will later show empirically that a tightness-induced rise in wages plays only a minor role in our setting: First, we point out in Section \ref{sec:6} that the omission of the wage variable in the regression leaves the effect of higher labor market tightness on labor demand essentially unchanged. Second, in Section \ref{sec:7}, we directly estimate the effect of labor market tightness on wages which turns out to be rather small.

\begin{figure}[!ht]
\centering
\caption{Feedback Cycle via Labor Market Tightness}
\label{fig:1}
\begin{center}
\scalebox{0.8}{
            \begin{tikzpicture}[ dot/.style={circle,,minimum size=4pt,inner sep=0pt,outer sep=-1pt}]

                        \draw[->,xshift=0cm, rounded corners] plot coordinates  {(0,0) (5,0)} ;
                        \draw[->,xshift=0cm, rounded corners] plot coordinates {(6.4,-0.2) (6.4,-2) (1.5,-2)} ;
                        \draw[->,xshift=0cm, rounded corners] plot coordinates {(0,-1.7) (0,-1) (5.9,-1) (5.9,-0.2)} ;

                        \node[left] at (0,0) {Wage Rate};
                        \node[right] at (5,0) {Labor Demand};
                        \node[left] at (1.5,-2) {Labor Market Tightness};
                        \node[above] at (2.5,0) {--};
                        \node[right] at (6.4,-1.1) {+};
                        \node[above] at (2.825,-1) {--};
\end{tikzpicture}
}
\end{center} \vspace*{-0.7cm}
\floatfoot{\footnotesize\textsc{Note. ---} The figure visualizes how the labor demand response to a changing wage rate triggers a self-attenuating feedback cycle, as implied by our theoretical model. Source: Own illustration.}
\end{figure}
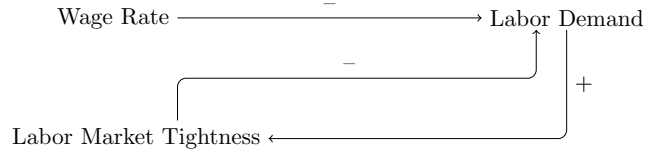

\paragraph{Search Externalities and Feedback Cycle.} Our model highlights that not only higher wages but also higher labor market tightness reduce firms' labor demand. Unlike traditional models of labor demand, this framework implicitly incorporates search (or congestion) externalities from firms' labor demand decisions, namely that an increased employment in one firm complicates recruitment in other firms by intensifying hiring frictions (and vice versa). Due to these externalities, aggregate changes in labor demand feature a self-attenuating feedback mechanism via labor market tightness, as illustrated in Figure \ref{fig:1}.

As a matter of fact, the impact of a single firm's change in employment on labor market tightness is certainly negligible when the firm is small in relation to the overall size of the labor market. However, even when the labor market is atomistic, the feedback mechanism becomes relevant when many firms alter their labor demand simultaneously (e.g., from responding to an increase in a nation-wide minimum wage). When firms act in concert, an aggregate decline in labor demand will reduce labor market tightness which in turn will stimulate labor demand of individual firms. Ultimately, this self-dampening feedback cycle implies that the aggregate reduction in labor demand becomes less negative than the sum of firms' individual first-round responses due to reallocation of workers across firms.

To derive the feedback mechanism formally, suppose that there is a constant-returns-to-scale Cobb-Douglas matching function: $M(U,V) \,=\, \kappa \, \cdot \, U^{\mu} \, \cdot \, V^{1-\mu}$ with $\kappa > 0$ and $0 < \mu < 1$.\footnote{\citet{PetrongoloPissarides2001} provide a review on studies that seek to estimate matching functions. The authors conclude that the majority of studies find support for the constant-returns-to-scale assumption.} We assume that the economy is in the steady state: $\dot{L} \,=\, H \, - \, \delta \cdot L \, = \, 0$. By substituting hires $H$ by the number of matches $M(U,V)$, we reformulate the steady-state condition as:
\begin{equation}
\label{eq:9}
\theta \,\equiv\, \frac{V}{U} \, = \, \bigg({\frac{\delta \cdot L}{\,\, \kappa \cdot U\,\,} \bigg)^{\frac{1}{1 -\mu}}}
\end{equation}
Thus, labor market tightness $\theta$ is increasing in aggregate employment. Specifically, a first-round reduction in aggregate labor demand by 1 percent lowers labor market tightness by
\begin{equation}
\label{eq:10}
\nu \,\equiv \, \frac{\partial \, ln \,\theta}{\partial \, ln \, L} \,=\,\frac{1}{1-\mu} \cdot \Big( 1 - \frac{\partial \,ln \,U}{\partial \,ln \,L} \Big)
\end{equation}
percent.
This reduction in labor market tightness relieves some congestion in the hiring process and, by virtue of Equation (\ref{eq:8}), stimulates the demand for labor in all other firms by $\omega \,=\, \nu \,\cdot\, \eta^{L}_{\theta}$ percent.\footnote{Conventional models of labor demand rule out such a built-in feedback cycle by assuming a priori that labor market tightness has no effect on labor demand: $\eta^{L}_{\theta}=0$.} This second-round increase in labor demand will in turn raise labor market tightness by $\omega^2$ percent, which again translates into a third-round reduction in labor demand by $\omega^3$ (and so forth). Overall, aggregate changes in labor demand (regardless of their trigger) generate a self-dampening feedback cycle to the extent that aggregate labor demand creates congestion in the hiring process via increased tightness. Provided that $|\omega|<1$, the resulting reallocation effects across firms reduce the first-round response in aggregate labor demand by factor $(1 -\frac{1}{1-\omega})$.\footnote{Using a geometric series, we can derive that only a fraction of the first-round response in labor demand survives the self-attenuating feedback cycle: $ 1 +  \omega +  \omega^{2} + \dots  = \sum_{t=0}^{\infty} \omega^{t} = \lim_{T \to \infty} \sum_{t=0}^{T} \omega^{t} = \lim_{T \to \infty} \frac{1-\omega^{T+1}}{1-\omega} = \frac{1}{\,\,1-\omega\,\,} $. Note that the geometric series converges only when $|\omega| \, = \, |\nu \cdot \eta^{L}_{\theta}|<1$.}

In terms of the wage effect, knowledge about the relative strength of the feedback mechanism (i.e., $\omega$), allows us to derive the aggregate own-wage elasticity of labor demand
\begin{equation}
\label{eq:11}
\tilde{\eta}^{L}_{W} \,=\, \frac{\eta^{L}_{W}}{\,\,\,1 \,-\, \omega \,\,\,} \,=\, \frac{\eta^{L}_{W}}{\,\,\,1 \,-\, ( \nu \cdot \eta^{L}_{\theta} ) \,\,\,}
\end{equation}
which, by accounting for search externalities, captures the ultimate response of aggregate labor demand to aggregate wage changes \citep{BeaudryEtAl2018}. When, according to theory, $\omega$ is below zero, the aggregate own-wage elasticity of labor demand $\tilde{\eta}^{L}_{W}$ becomes less negative than the own-wage elasticity of labor demand $\eta^{L}_{W}$ of an individual firm due to reallocation effects.

\section{Empirical Design}
\label{sec:3}

\paragraph{Empirical Model.} According to theory, employers reduce their labor demand not only when wages rise but also when labor markets tighten. To test these propositions, we set up the following empirical model in first differences
\begin{equation}
\label{eq:12}
\Delta\,ln\,L_{it} \, = \, \eta_{0} \,\, + \,\,  \eta^{L}_{W} \cdot \Delta \, ln \, W_{it} \,\, + \,\, \eta^{L}_{\theta} \cdot \Delta \,ln \,\theta_{it} \,\, + \,\, \zeta_{t} \,\, + \,\, \Delta \, \varepsilon_{it}
\end{equation}
to estimate the causal effect of both wages and labor market tightness on firms' labor demand. Specifically, we regress the log difference of labor demand of firm $i$ in year $t$, $\Delta \,ln\, L_{it}$, on the respective log differences in wages,  $\Delta \,ln\, W_{it}$, and labor market tightness, $\Delta \,ln\, \theta_{it}$. Moreover, the model includes an intercept $\eta_{0}$, a set of year fixed effects $\zeta_{t}$, and a differenced idiosyncratic error term $\Delta \,\varepsilon_{it}$.

Labor market tightness is typically measured at the level of regional labor markets. However, regional labor market tightness may not be a precise measure for firms' demand if a firm only recruits from specific occupational labor markets within a region.\footnote{Occupations play a particularly important role in Germany. Under the dual vocational system, apprenticeship training determines the occupation of workers when they enter the labor market with only little re-training and occupational mobility later on in a career \citep{RheinEtAl2013}.} To take into account the occupational demands of individual firms, we develop a measure of firm-specific labor market tightness
\begin{equation}
\label{eq:13}
\theta_{it} = \sum_{o=1}^{O} \frac{L_{oit}}{L_{it}}  \cdot  \frac{V_{ort}}{U_{ort}}
\end{equation}
where $\frac{V_{ort}}{U_{ort}}$ is the ratio between vacancies in an occupation $o$, in region $r$, at year $t$ and the number of job seekers in the same occupation-by-region-by-year cell. We weight these occupational measures of labor market tightness in a firm's region by the respective shares of these occupations in each firm's overall employment, $\frac{L_{oit}}{L_{it}}$. As such, our measure simply postulates that firms occupational structure of employment mirrors their structure of vacancies. In Section \ref{sec:5}, we substantiate empirically that firms' occupational composition of vacancy shares resembles their composition of employment shares by showing that our employment-based measure of labor market tightness is highly correlated with an analogous vacancy-based measure from survey information (see Figure \ref{fig:D7}).

In the baseline specification, we estimate our empirical model (\ref{eq:12}) in two-year differences of dependent and independent variables. These biennial changes allow adjustments to take two years to materialize, identifying firms' responses in the longer run. By contrast, in the short run (e.g., for one-year differences), open vacancies are not necessarily filled yet and the stock of capital remains fixed. Hence, the full response may not be observed after one year, implying that adjustments would be underestimated. In line, \citet{Jung2014} reports that German firms need about 1.8 years to complete half of the desired adjustment towards their optimal labor demand. Therefore, we will later present a robustness check related to the choice of the lag difference (see Section \ref{sec:6}).

\paragraph{Threats to the Identification.} Our empirical model delivers reduced-form effects in the sense that it captures effects of wages without conditioning on product prices, levels of production or capital input. Hence, we are estimating an unconditional own-wage elasticity of labor demand that operates through substitution effects and scale effects. This framework yields a comprehensive effect of wages on labor demand, which is desirable provided that the variation in wages and labor market tightness is exogenously identified. While differencing eliminates unobserved time-invariant heterogeneity capturing all permanent differences between firms (including the industry, the location, or firms' permanent growth potential), we still require exogeneity of the differenced independent variables.\footnote{From Equation (\ref{eq:4}), we know that changes in total factor productivity, capital, and the product price enters the differenced error term. Hence, we must ensure that our variation in wages or tightness does not stem from changes in the omitted variables at the respective firm.}

In our differenced model (\ref{eq:12}), the threats of identification are twofold. First, a major source of endogeneity arises from the interplay between labor demand and labor supply. When estimating own-wage elasticities of labor demand, we seek to determine the inverse slope of the labor demand curve. Hence, variation in wages should represent movements along the labor demand curve rather than shifts of the curve itself. Given the positive relationship between wages and labor supply, uncontrolled shifts of the labor demand curve will result in an upward bias \citep{Wright1928,AngristKrueger2001}. 
For instance, a positive firm-specific productivity shock will stimulate labor demand of the firm (i.e., shift the labor demand curve of the firm and the market rightwards), leading to a simultaneous increase in the market wage. Traditional models of labor demand attempt to mitigate this problem by relying on micro-level data \citep{Hamermesh1993,LichterEtAl2015}. Under the assumption of perfect competition, labor demand of an individual firm has a negligible effect on market wages (i.e., the price-taking firm faces an exogenously given wage rate). In reality, however, firms are not necessarily small in relation to the market, and, hence, the assumption of competitive labor markets may not hold.

The second threat of identification stems from reverse causality between labor demand and labor market tightness (see Figure \ref{fig:1}). Specifically, we are interested in the effects of labor market tightness on firms' employment. At the same time, however, Equation (\ref{eq:9}) implies that any change in employment directly impacts labor market tightness. As an increase in labor demand will raise tightness, the feedback mechanism leads to an upward bias. 
While the feedback effect on tightness is certainly stronger when entire regions (rather than only the single firm) adjust their labor demand, there is no guarantee that single firms always have a negligible effect on labor market tightness.

\paragraph{Identification Strategy.} Since a naive OLS estimation of Equation (\ref{eq:12}) is likely to provide biased results, we estimate our model based on variation from instrumental variables using two-stage least squares (2SLS) estimation (see Appendix \ref{sec:B} for further details). We propose three new shift-share instruments in the tradition of \citet{Bartik1993}. Bartik instruments exploit the inner product structure of endogenous variables to deliver plausibly exogenous variation at the regional level \citep{Goldsmith-PinkhamEtAl2020}. These instruments became popular through a wide range of applications, such as identifying the labor market effects of regional demand shocks \citep{BlanchardKatz1992}, the effects of local migration shocks \citep{Card2001}, the effects of region-specific import competition \citep{AutorEtAl2013,DauthEtAl2014}, or the analysis of regional labor demand \citep{BeaudryEtAl2012,BeaudryEtAl2018}. However, for the purpose of analyzing reallocation effects of workers across employers, we do not seek to identify employment effects at the level of regions but for individual firms. For this purpose, we develop novel Bartik instruments that provide variation at the firm level. In particular, we take advantage of the fact that firms differ in the occupational composition of their workforce and, thus, are differently exposed to common shocks.

We begin with developing a Bartik instrument for exogenous wage changes at the firm level. A firm's change in wages can be described by the following accounting identity:
\begin{equation}
\label{eq:14}
\Delta \, ln\,W_{it} \,=\, \sum_{o=1}^{O} \, s_{iot} \cdot g_{W_{iot}}
\end{equation}
A firm's growth in average wages $\Delta\,ln\, W_{it}$ equals the growth in average wages in each occupational group of the firm $g_{W_{iot}}$ weighted by the respective occupational employment shares in the firm $s_{iot}$. The firm-specific growth rate of wages in an occupation can be decomposed in a nation-wide occupation-level growth rate and an additive idiosyncratic firm- and occupation-specific growth rate
\begin{equation}
\label{eq:15}
g_{W_{iot}} \,=\, g_{W_{ot}} \,+\, \tilde{g}_{W_{iot}}
\end{equation}
where $\tilde{g}_{W_{iot}}$ is designed to capture firm $i$'s divergence from the national growth rate of occupation-specific wages $g_{W_{ot}}$. This divergence in wage growth may, for instance, capture firm-specific trends in variables that co-determine firms' labor demand - namely factor productivity, capital stock, or product prices. Thus, the idiosyncratic component of firms' wage growth may correlate with uncontrolled factors that shift firms' labor demand curve and, thus, may result in a spurious estimate. Building on Equations (\ref{eq:14}) and (\ref{eq:15}), our firm-level Bartik instrument which is meant to exogenously predict wage changes looks as follows:
\begin{equation}
\label{eq:16}
Z_{W_{it}} \,=\, \sum_{o=1}^{O} s_{io\tau} \cdot g_{W_{ot}} \,=\, \sum_{o=1}^{O} \frac{L_{io\tau}}{L_{i\tau}}  \cdot \Delta \, ln \, W_{ot}
\end{equation}
Specifically, our shift-share instrument is the inner product of past employment shares of occupations within firms, $\frac{L_{io\tau}}{L_{i\tau}}$, and occupation-specific growth rates at the national level, $\Delta \,ln\, W_{ot}$. To ensure exogeneity, the instrument (\ref{eq:16}) departs from (\ref{eq:14}) in two dimensions: On the one hand, to avoid endogeneity from firm-specific labor demand shocks, we rely only on nation-wide occupation-specific wage growth, which is - under certain conditions stated below - immune against reverse causality (i.e., the national wage growth cannot be reasonably altered by a single firm's labor demand). On the other hand, we fix firm-specific occupation shares at an early period $\tau$, implying that the shares are predetermined, and thus, unlikely to correlate with differences in the contemporaneous error term.

To isolate Bartik-style variation for firm-specific labor market tightness $\theta_{it}$, we rewrite Equation (\ref{eq:13}) as follows:
\begin{equation}
\label{eq:17}
\Delta\,ln\,\theta_{it}\,=\, \sum_{o=1}^{O} \, s_{iot} \cdot g_{\theta_{ort}}
\end{equation}
Our firm-specific measure of labor market tightness relies on tightness at the market level and becomes firm-specific by weighting with occupational shares in firms' employment. Hence, the growth rate $g_{\theta_{ort}}$ features subscripts for both occupations and regions, which define the labor market.
Rewriting the growth rate of labor market tightness as the difference between the growth rate of vacancies and the growth rate of job seekers yields a difference of two shift-share expressions:
\begin{equation}
\label{eq:18}
\Delta\,\theta_{it}\,=\, \sum_{o=1}^{O} \, s_{iot} \cdot (g_{V_{ort}} \,-\, g_{U_{ort}})\,=\, \sum_{o=1}^{O} \, s_{iot} \cdot g_{V_{ort}} - \sum_{o=1}^{O} \, s_{iot} \cdot g_{U_{ort}}
\end{equation}
Building on this identity, we analogously define two separate Bartik instruments for vacancies and job seekers, $Z_V$ and $Z_U$, which are meant to generate exogenous variation in firm-specific tightness:
\begin{equation}
\label{eq:19}
 Z_{V_{it}} \,=\, \sum_{o=1}^{O} s_{io\tau} \cdot g_{V_{ot}} \,=\, \sum_{o=1}^{O} \frac{L_{io\tau}}{L_{i\tau}}  \cdot \Delta \, ln \, V_{ot}
\end{equation}
\begin{equation}
\label{eq:20}
 Z_{U_{it}} \,=\, \sum_{o=1}^{O} s_{io\tau} \cdot g_{U_{ot}} \,=\, \sum_{o=1}^{O} \frac{L_{io\tau}}{L_{i\tau}}  \cdot \Delta \, ln \, U_{ot}
\end{equation}
Note that we replace the occupational growth rate of vacancies and unemployment in the regional labor market by the national growth rates and, again, we harness predetermined occupation shares $\frac{L_{io\tau}}{L_{i\tau}}$ from the base year $\tau$.

Bartik instruments can be straightforwardly constructed from accounting identities but do not necessarily provide a valid identification strategy. The exclusion restriction of our three Bartik-style instruments $Z_{W}$, $Z_{V}$, and $Z_{U}$ is fulfilled when either the national growth rates (i.e., the shifts) or the predetermined firm-specific occupational composition (i.e., the shares) are uncorrelated with the error term \citep{Goldsmith-PinkhamEtAl2020, BorusyakEtAl2022}.

Given the design of our firm-level Bartik instruments, it is plausible to assume that at least one of the two conditions holds. On the one hand, national growth patterns might stem from exogenous sources (e.g., wage growth due to a higher minimum wage or an increase in job seekers from a sudden influx of migrants) or labor supply shocks (e.g., higher female labor force participation). In both cases, the explanatory variable would not be correlated with the error term. Moreover, when exposure to labor demand shocks is not correlated between units, a single firm's labor demand decision cannot reasonably shape national growth patterns, that is, the use of the Bartik instrument protects against reverse causality. On the other hand, even if labor demand shocks correlate across units (e.g., a common technology shock), exogeneity is maintained when the predetermined employment shares (i.e., differential exposure to common shocks) are uncorrelated with the error term. By the choice of a base year $\tau$ that lies far in the past, we minimize the possibility that the shares exert an effect on firms' contemporary changes in labor demand other than through the channel of the explanatory variables (i.e., the level of past shares is uncorrelated with changes in uncontrolled determinants of firms' contemporary labor demand). By and large, the identifying idea behind our Bartik instrument is that firms face different exogenous exposure to national growth in the variable of interest based on their assigned occupational composition from the past.

\citet{Goldsmith-PinkhamEtAl2020} show that the Bartik estimator can be decomposed into a weighted sum of just-identified IV estimators that use each share as an instrument. The so-called Rotemberg weights for the separately identified IV estimates (i.e., for each occupation-by-year combination in our setting) depend on the product of the national growth rate and the covariance between share and endogenous regressor. The Rotemberg weights sum up to 1 while, in certain circumstances, some weights can take negative values which complicate a LATE interpretation of the Bartik estimates. In the empirical part of the paper, we will decompose our Bartik estimators for wages and labor market tightness into weights and just-identified IV estimates to assess the plausibility of our identification strategy (see Section \ref{sec:6}). In particular, the Rotemberg weights will highlight the subset of occupations to which the final Bartik estimator is most sensitive. For these occupations, we will i) inspect the sources of identifying variation and ii) examine whether past shares are uncorrelated with contemporaneous shifts in labor demand.

Finally, we must clarify that, besides exogeneity, the identification also requires that the three instruments need to be relevant. In other words, the instrumental variables need to be strong predictors for the endogenous explanatory variables. We illustrate and test this final but crucial assumption empirically.

\section{Data}
\label{sec:4}

\paragraph{Integrated Employment Biographies.} To bring our empirical model to the data, we assemble information from three independent data sources on the German labor market: the Integrated Employment Biographies, the Official Statistics from the German Federal Employment Agency, and the IAB Job Vacancy Survey (see Appendix \ref{sec:C} for further details). The Integrated Employment Biographies (IEB) compiles manifold sources of administrative labor market records on Germany \citep{MuellerWolter2020}. From the IEB, we use the universe of employment records of all workers subject to social security contributions, which are collected from employers in Germany as part of the mandatory reporting requirement. In particular, the data cover the entirety of regular full-time, regular part-time, and marginal part-time workers.\footnote{The data exclude only self-employed persons, civil servants, and family workers because these groups are not obliged to pay social security contributions.} The IEB provides day-to-day information on workers' employment histories, such as workers' establishment, daily gross wages (which we impute above the censoring limit), type of contract, place of work as well as an indicator whether workers have a full- or part-time contract. For the years 2010-2014, the IEB additionally includes information on the number of working hours. Importantly, the IEB data also offer exceptionally rich information on workers' 5-digit occupation, distinguishing between a total of $O=1,286$ occupational categories.\footnote{In particular, we utilize information on the German Classification of Occupations (KldB) from the year 2010. The four leading digits describe the type of occupation whereas the fifth digit designates the level of skill requirement (helper, professional, specialist, or expert).}

For June 30 of the years 2012-2019, we construct a panel dataset by calculating the number of workers and average daily wages for each establishment.\footnote{Although data availability would permit to construct our detailed measures of labor market tightness from 2010 onward, we chose 2012 as the starting year for our period of analysis since there was a structural break in 2011/12 in the occupation variable in the IEB data.} The term ``establishment'' comprises all plants of a company that share the same economic activity within a municipality.\footnote{Throughout this study, we use the terms ``establishment'' and ``firm'' interchangeably.} For lack of systematic information on individual working hours, we follow standard practice and restrict our baseline analysis to full-time workers in regular employment (who are supposed to work a similar number of hours). In a further check, we also include regular part-time and marginal part-time workers by approximating average hourly wages from the available information on hours (between 2012 and 2014). Throughout the study, we exclude apprentices and people in partial retirement schemes.

\paragraph{Labor Market Tightness.} We define labor markets as combinations of 5-digit occupations and commuting zones. In terms of occupations, we employ the 5-digit classification for two reasons: First, the leading four digits differentiate between 700 types of occupations in the highest available level of detail. Second, the fifth digit further delivers valuable information on the level of skill requirement, namely whether workers are helpers, professionals, specialists, or experts. It is highly important to distinguish between requirement levels since tasks with different levels of complexity plausibly define segregated labor markets even if the underlying 4-digit occupation is identical.\footnote{In this respect, \citet{DemingKahn2018} show that that skill requirements are key predictors of wage patterns. In addition, \citet{Ziegler2021} finds that job postings with higher skill requirements offer higher remuneration but involve a longer vacancy duration.} In terms of regions, we employ the graph-theoretical method from \citet{KroppSchwengler2016} and merge 401 administrative districts (3-digit NUTS regions) to more appropriate functional labor market regions that reflect commuting patterns (see Figure \ref{fig:C1}). Taken together, our baseline labor markets constitute combinations of $O=1,286$ occupations and $R=51$ commuting zones. To ascertain that our results are not driven by this specific delineation, we show in the empirical part of the paper that the later empirical results also hold for more broadly or narrowly defined labor markets.

We gather process data on posted vacancies and job seekers from the Federal Employment Agency (FEA) to construct our measure of firm-specific labor market tightness. For each June 30 between 2012 and 2019, we draw official statistics on the stock of registered vacancies \citep{FEA2019}, including the targeted 5-digit occupation and commuting zone (in terms of workplace). In Germany, there is no obligation for firms to register vacancies with the Federal Employment Agency. To quantify the overall stock of registered plus unregistered vacancies for each labor market and year, we divide the number of registered vacancies by the yearly share of registered vacancies from the IAB Job Vacancy Survey \citep{BosslerEtAl2020}. The IAB Job Vacancy Survey (IAB-JVS) is a representative establishment survey with a focus on recruitment behavior and, in particular, asks firms about their number of registered and unregistered vacancies. When dividing by the yearly shares of registered vacancies, we differentiate between three levels of skill requirement: occupations for helpers, for professionals, and for specialists along with experts (see Table \ref{tab:C1}).\footnote{From the IAB Job Vacancy Survey, we calculate the following notification shares for vacancies, averaged over 2012-2019: helpers (46.1 percent), professionals (45.6 percent), specialists and experts (31.1 percent).}

In contrast to vacancies, it is mandatory to register as unemployed with the Federal Employment Agency to be eligible for benefits from unemployment insurance or social assistance. For the same labor market and years, we extract official information on the number of job seekers \citep{FEA2018}, namely registered unemployed plus employed workers searching for a job via the Federal Employment Agency.\footnote{\citet{AbrahamEtAl2020} find that the number of effective job searchers features a higher explanatory power in the matching function than the mere stock of unemployed persons.} Upon registration, job-seeking individuals must submit their targeted 5-digit occupation with the Federal Employment Agency. For each labor market and year, we divide the overall stock of registered plus unregistered vacancies by the stock of job seekers. Next, we apply (\ref{eq:13}) and weight these ratios with contemporaneous shares of 5-digit occupations in firms' overall employment from the IEB to arrive at our measure of firm-specific labor market tightness.

\paragraph{Shift-Share Instruments.} In a final step, we build our firm-level shift-share instruments from Equations (\ref{eq:16}), (\ref{eq:19}) and (\ref{eq:20}). To this end, we interact biennial national changes in average wages, stock of vacancies, and stock of job seekers per occupation with IEB information on firms' shares of occupations in their employment from the past. When choosing the base year of these employment shares, we face a trade-off between maximizing the time lag to the estimation period (i.e., 2012-2019) and minimizing structural breaks in the data over time. Due to a major redesign of the IEB data, we calculate the predetermined employment shares only from 1999 onward.\footnote{In principle, IEB information is available from 1975 (West Germany) and 1993 (East Germany) onward. However, we refrain from analyzing information from before 1999 because information on marginal employment has not yet been available. For the years 1999-2011, we rely on available crosswalks to translate the information on the 1988's occupational classification (KldB-1988) into time-consistent information on the 2010's occupational classification (KldB-2010), which is available in the IEB from 2012 onward.} Hence, in most cases, the base year refers to 1999 (35.6 percent) or, alternatively, the year of birth for firms that entered the labor market at a later stage (0.9-5.0 percent per year from 2000 onward). 

\paragraph{Structure of the Final Dataset.} Our final dataset (including employment, average wages, labor market tightness, and instrumental variables) refers to the near-universe of establishments in Germany and contains a total of 21,689,291 establishment-year observations. The panel covers 4,205,183 establishments, which we monitor, on average, 5.2 times between 2012 and 2019. Tracked establishments employ a total of 278,633,024 workers, which equals 32.9-36.7 million workers per year or 78-82 percent of overall employment in Germany.

\section{Results: Labor Market Tightness}
\label{sec:5}

Before coming to our regression results, we inspect our measure of (firm-specific) labor market tightness and assess its performance as a main determinant for pre-match hiring cost (see Appendix \ref{sec:D} for further details).

\paragraph{The German Economy.} After having risen steadily for several decades, Germany's unemployment rate reached a peak of 13 percent in the mid-2000s. During that time, the muted economic environment also deterred many firms from posting vacancies. Thus, labor market tightness in Germany reached an all-time low in 2004 (see Figure \ref{fig:D1}). Since then, the German labor market has undergone a remarkable transformation, accompanied by significant employment growth. \citet{DustmannEtAl2014} attribute this reversal to the flexibility and decentralization of the wage-setting process resulting in higher labor demand from lower real wages. In addition, a comprehensive reform of German labor market institutions in the years 2003-2005 (the so-called Hartz laws) contributed to the labor market upswing \citep{KrauseUhlig2012,HochmuthEtAl2021}. Among others, these laws re-structured the Federal Employment Agency and reduced the generosity of unemployment benefits to increase workers' incentive to accept jobs, thus further weakening workers' bargaining position. A number of studies demonstrate that the Hartz reforms also came along with an increased matching efficiency \citep{FahrSunde2009,KlingerRothe2012,LaunovWaelde2016}. As a side effect of the favorable employment growth, labor market tightness started to increase \citep{BurdaSeele2020}. The economic prosperity continued in the following decade. Simultaneously, demographic change led to a decline in the number of unemployed, especially in East Germany \citep{SchneiderRinne2019}. As a result, the increase in labor market tightness accelerated during the 2010s.

\paragraph{Beveridge Curve.} Figure \ref{fig:2} displays the Beveridge curve for Germany for our period of analysis. Between 2012 and 2013, labor market tightness decreased slightly during the sovereign debt crisis in the Euro area. From 2013 to 2019, we observe a sharp increase in the number of vacancies by 800,000 while the number of job seekers declined in a similar order of magnitude. In this period, the economy-wide ratio of vacancies to job seekers rose steadily from 0.23 to 0.47: while we report four job seekers per vacancy in 2013, there were only two job seekers per vacancy in 2019, implying a doubling in labor market tightness.

\pgfkeys{/pgf/number format/.cd,1000 sep={,}}
\begin{figure}[!ht]
\centering
\caption{Beveridge Curve}
\label{fig:2}
\scalebox{0.80}{
\begin{tikzpicture}
\begin{axis}[ytick={1000,1200,1400,1600,1800,2000}, height=14cm, width=14cm, grid=major, grid style = dotted, xtick={4200,4400,4600,4800,5000}, xlabel={Job Seekers (in Thousands)}, ylabel={Vacancies (in Thousands)}, ymin=900, ymax=2100, xmin=4100, xmax=5100, legend pos = north east, y tick label style={/pgf/number format/.cd,fixed,fixed zerofill, precision=0,/tikz/.cd}]
\addplot[mark=*, solid, color=black, mark options={fill=white, scale=1}, line width=0.25mm] coordinates {  (4933.403,1206.575) (5029.943,1123.425) (4978.978,1225.132) (4838.533,1326.414) (4777.521,1390.879) (4698.161,1656.227) (4363.391,1860.812) (4226.751,1999.79) };
\node[mark=none, left] at (4933.403,1206.575) {2012};
\node[mark=none, below] at (5029.943,1123.425) {2013};
\node[mark=none, above right] at (4978.978,1225.132) {2014};
\node[mark=none, below left] at (4838.533,1326.414) {2015};
\node[mark=none, below left] at (4777.521,1390.879) {2016};
\node[mark=none, above right] at (4698.161,1656.227) {2017};
\node[mark=none, below left] at (4363.391,1860.812) {2018};
\node[mark=none, above] at (4226.751,1999.79) {2019};
\end{axis}
\end{tikzpicture}
}
\floatfoot{\footnotesize\textsc{Note. ---} The figure shows the Beveridge curve for Germany between 2012 and 2019. The numbers of registered vacancies and job seekers stem from notifications to the German Federal Employment Agency. We divide the stock of registered vacancies by the yearly share of registered vacancies per requirement level from the IAB Job Vacancy Survey to account for unregistered vacancies. Sources: Official Statistics of the German Federal Employment Agency $\plus$ IAB Job Vacancy Survey, 2012-2019.}
\end{figure}
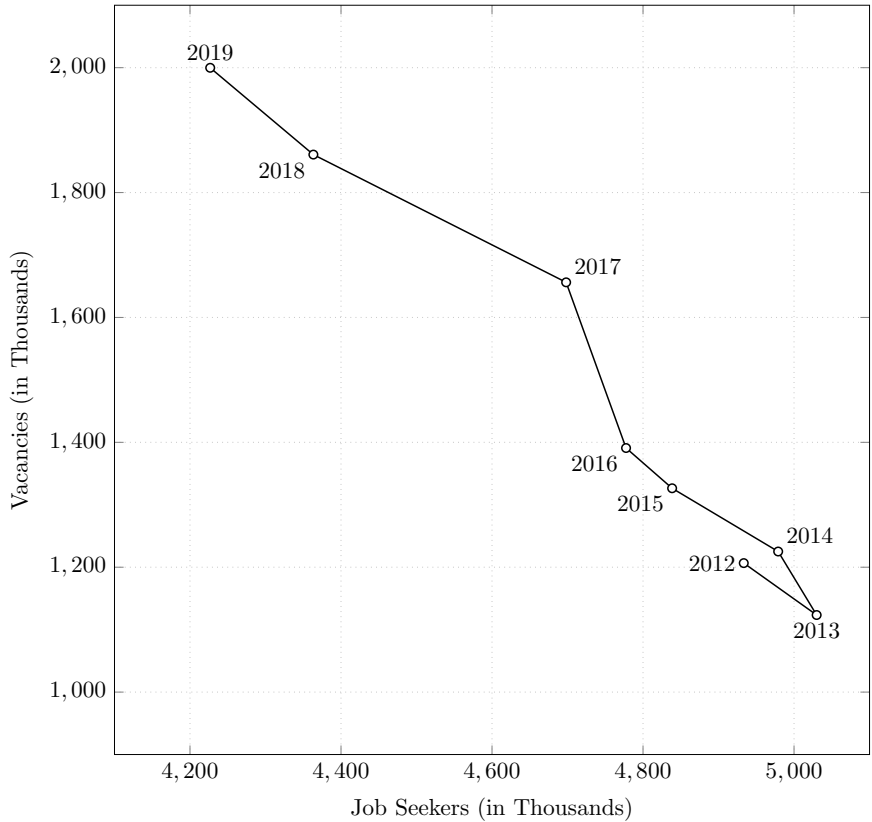

\pgfkeys{/pgf/number format/.cd,1000 sep={}}

Importantly, the increase in labor market tightness is not just driven by a certain subset of occupations or regions. Despite some idiosyncrasies, all occupational areas (see Figure \ref{fig:D2}) and commuting zones (see Figure \ref{fig:D3}) moved towards a higher tightness during the period of analysis. As explained above, the increase in tightness coincides with a significant and long-lasting phase of prosperity of the German economy, which came along with a rapid employment expansion, rising from 42.0 million employees in 2012 to 45.2 million employees in 2019. Despite the rise in employment, the increase in labor market tightness was accompanied by a markedly higher share of firms that face labor shortages (see Figure \ref{fig:D4}), implying that firms' employment could have grown by even more if the tightness stayed constant.

\paragraph{Labor Market Tightness, Wages, and Recruitment Indicators.}

As outlined in Section \ref{sec:2}, labor market tightness is hypothesized to exert an effect on the demand for labor through increased hiring cost. When labor market tightness increases, it becomes more costly to hire additional workers. As a consequence of these hiring frictions, firms' employment grows by less or falls, because vacancies either remain unfilled (i.e., the firm has not spent enough hiring cost) or are not posted at all (i.e., hiring cost are prohibitively high).

The IAB Job Vacancy Survey allows us to shed some light on the mediating channels underlying the relationship between labor market tightness and labor demand. In repeated cross-sections, the survey includes questions on the respective firm's most recent successful process of hiring.\footnote{Crucially, this survey information is selective since it does not include information on the recruitment process when a vacancy remains unfilled or is not posted at all due to prohibitively high hiring cost. We suspect that correlates of tightness and hiring indicators would be even larger if unfilled positions were included because the available information on successful hires is a positive selection of all hiring processes.} The survey includes information on the following recruitment indicators: direct pre-match hiring cost (in Euro), search effort (in working hours), the number of applicants, the number of search channels, and search duration (in days).\footnote{Unfortunately, the IAB Job Vacancy Survey does not inquire information on post-match hiring cost.} In addition, we combine the survey information on search effort (in working hours) with IEB information on the respective firm's average wage rate to measure of indirect pre-match hiring cost (in Euro).\footnote{We calculate indirect pre-match hiring cost (in Euro) by multiplying the search effort (in working hours) by the firm's average hourly wage rate (in the IEB) of i) workers in human resource management (KldB-2010 Code: 715) or, if not available, ii) managers or, if not available, iii) all workers.}

\afterpage{\begin{figure}[H]
\caption{Labor Market Tightness and Pre-Match Hiring Costs}
\label{fig:3}
\begin{subfigure}{1\textwidth}
\caption{Direct Pre-Match Hiring Costs}
\centering
\scalebox{0.725}{
\begin{tikzpicture}
\begin{axis}[xlabel=Log Labor Market Tightness, ylabel=Log Hiring Costs, xmin=-3.5, xmax=2, ymin=5.25, ymax=7.75, height=8cm, width=16cm, grid=major, legend pos = south east, xtick={-3,-1.5,0,1.5}, ytick={5.5,6.5,7.5}, grid style = dotted, y tick label style={/pgf/number format/.cd,fixed,fixed zerofill, precision=2,/tikz/.cd}]
\addplot[only marks, mark=o, mark options={solid,scale=1}, color=black] coordinates {  (-2.8439567,6.7720466) (-2.6897449,6.1657219) (-2.5329204,5.6672106) (-2.3661563,5.9811201) (-2.2607296,5.5286999) (-2.1800339,6.5807476) (-2.1062729,6.1004372) (-2.0487239,5.1458197) (-1.9779736,6.3059731) (-1.8876545,6.0406351) (-1.8497854,6.1914196) (-1.7840927,6.8072104) (-1.7171773,6.8048449) (-1.6361536,6.4235139) (-1.5689162,5.9622474) (-1.5248942,6.7371445) (-1.4795882,6.4382067) (-1.4438325,5.7410412) (-1.3833739,6.1230693) (-1.3367449,6.308784) (-1.2956235,6.6872244) (-1.2423723,6.2731419) (-1.1892105,6.218843) (-1.1521565,5.9778194) (-1.1355114,6.1276207) (-1.1170281,6.144278) (-1.0860486,6.2881384) (-1.0518066,5.6599622) (-1.0219617,6.593224) (-.99496615,6.1063066) (-.9560796,6.4297385) (-.92095321,6.4734564) (-.89544821,6.8840923) (-.86211187,6.6112003) (-.8327648,6.0994449) (-.81363308,6.1966076) (-.78284484,6.7790785) (-.74244756,6.9841471) (-.71714228,6.3644972) (-.69210559,6.6178317) (-.66140592,6.6556587) (-.63114065,6.5634089) (-.60940748,6.3620043) (-.57618093,6.2067952) (-.5405857,6.6793475) (-.51249468,6.3786058) (-.4850944,6.2596545) (-.45851117,6.7606273) (-.43561104,7.1040702) (-.4047676,5.9757233) (-.38514224,6.5228505) (-.36026663,6.9962268) (-.33400455,6.7631793) (-.30708879,6.2495251) (-.28422472,6.279006) (-.25026038,4.7415848) (-.22995971,6.3375483) (-.21303284,6.7079601) (-.1828147,6.0098867) (-.15957972,6.5711932) (-.14139999,6.564043) (-.10870646,6.3958664) (-.07890847,6.2831311) (-.0691351,7.0681629) (-.04793186,6.1439476) (-.01212671,6.6348653) (.01152034,6.2773137) (.02905312,6.4508381) (.06025831,6.4832296) (.09121963,6.1694918) (.11113751,7.1095219) (.13113309,6.1992688) (.15471861,6.026907) (.18461996,7.1757665) (.21358325,6.6539092) (.24108398,6.5046411) (.2703377,6.0205202) (.29338241,6.4733443) (.32143784,6.3013039) (.35493359,6.8999052) (.38107571,6.1573148) (.41167533,6.5347328) (.448643,6.1518683) (.48572436,6.3312931) (.53185195,6.3850737) (.56267923,6.5105653) (.59692258,6.2494254) (.63775229,7.0187654) (.68106645,6.242013) (.72284108,6.7600174) (.77235997,6.5088229) (.82023537,6.9039893) (.85199994,6.4874411) (.89065534,6.9274354) (.93265468,6.286777) (.99805886,6.5765781) (1.0696896,6.252944) (1.1376294,6.8158641) (1.1839761,6.5076647) (1.2407242,6.7687945) };
\addplot[domain=-3.25:1.75, color=black, solid, line width=0.75mm] {6.447765+0.1216047*x};
\legend{~Observations,~Linear Fit}
\end{axis}
\end{tikzpicture}
}
\end{subfigure}
\vskip \baselineskip
\begin{subfigure}{1\textwidth}
\caption{Indirect Pre-Match Hiring Costs}
\centering
\scalebox{0.725}{
\begin{tikzpicture}
\begin{axis}[xlabel=Log Labor Market Tightness, ylabel=Log Hiring Costs, xmin=-3.5, xmax=2, ymin=4.00, ymax=6.5, height=8cm, width=16cm, grid=major, legend pos = south east, xtick={-3,-1.5,0,1.5}, ytick={4.25,5.25,6.25}, grid style = dotted, y tick label style={/pgf/number format/.cd,fixed,fixed zerofill, precision=2,/tikz/.cd}]
\addplot[only marks, mark=o, mark options={solid,scale=1}, color=black] coordinates {  (-2.8382397,5.0619826) (-2.6977744,5.0337291) (-2.5563872,5.0364523) (-2.4179361,4.6805401) (-2.3125789,5.1260996) (-2.2468143,4.8019004) (-2.1836779,5.5045481) (-2.1186323,4.9926615) (-2.0581505,5.1234975) (-1.9987442,4.9930558) (-1.9202688,5.0447288) (-1.8764051,5.4021211) (-1.8438745,5.2329659) (-1.7799048,5.2920442) (-1.727143,5.5646996) (-1.6657057,5.4234805) (-1.6108757,5.1283255) (-1.5582762,5.0009151) (-1.5124028,5.2190933) (-1.4657832,5.2888227) (-1.4369979,4.9606209) (-1.3964905,5.1609464) (-1.3528367,5.3108006) (-1.3160785,4.6672587) (-1.2783226,5.2116842) (-1.2303957,5.428134) (-1.1874353,5.4483118) (-1.1533523,5.3419194) (-1.1303509,5.3313217) (-1.1055951,5.3528042) (-1.0749007,5.652504) (-1.0389194,5.4307628) (-1.0053627,5.4270687) (-.97237295,5.5466866) (-.93164039,5.2114196) (-.90340894,5.2786245) (-.87474459,5.4084878) (-.84131706,5.2883034) (-.81709528,5.2899418) (-.79062414,5.5808878) (-.75118023,5.2679477) (-.73030651,5.5356536) (-.71073502,5.1588612) (-.6821546,5.41716) (-.65310186,5.3712392) (-.62508243,4.9571652) (-.60548532,5.3365326) (-.57474893,4.5512328) (-.55080783,5.1649165) (-.52532452,5.3308773) (-.49902412,5.1541758) (-.46323499,5.3876038) (-.43868503,5.495038) (-.40879789,5.2689013) (-.38526452,5.3554797) (-.36076692,5.4700241) (-.33333191,5.56989) (-.30528972,4.9326339) (-.28480822,5.3483505) (-.25550666,5.0375857) (-.22796589,5.2792039) (-.20405155,5.3542719) (-.17328474,5.4258008) (-.14656252,5.6395264) (-.12013846,5.2894192) (-.08430254,5.3615751) (-.06648713,5.2259574) (-.0431586,5.3133702) (-.00910963,5.5576081) (.01475797,5.1700945) (.03934921,5.324604) (.06979547,5.4850173) (.09968755,5.4666924) (.12378148,5.4213839) (.15081976,5.2184415) (.17680357,5.3045807) (.20494294,5.5417833) (.23956151,5.2368245) (.26716125,5.2206683) (.29724386,5.0910368) (.33091608,5.2686143) (.36495274,5.5688949) (.39974302,5.3709002) (.43186,5.4253769) (.46712533,5.3594856) (.50610006,5.0544224) (.54878265,5.4587345) (.58655524,5.2669439) (.62477893,5.4245858) (.67709684,5.0460386) (.72863293,5.4928274) (.77838111,5.4983978) (.83046305,5.2988667) (.86929989,5.5556688) (.91112745,5.335804) (.97321886,5.2295666) (1.0353245,5.2380819) (1.109601,5.2633801) (1.1740952,5.2199092) (1.2371784,5.3658991) };
\addplot[domain=-3.25:1.75, color=black, solid, line width=0.75mm] {5.319275+0.069096*x};
\legend{~Observations,~Linear Fit}
\end{axis}
\end{tikzpicture}
}
\end{subfigure}
\vskip \baselineskip
\begin{subfigure}{1\textwidth}
\centering
\caption{Overall Pre-Match Hiring Costs}
\scalebox{0.725}{
\begin{tikzpicture}
\begin{axis}[xlabel=Log Labor Market Tightness, ylabel= Log Hiring Costs, xmin=-3.5, xmax=2, ymin=4.75, ymax=7.25, height=8cm, width=16cm, grid=major, legend pos = south east, xtick={-3,-1.5,0,1.5}, ytick={5,6,7}, grid style = dotted, y tick label style={/pgf/number format/.cd,fixed,fixed zerofill, precision=2,/tikz/.cd}]
\addplot[only marks, mark=o, mark options={solid,scale=1}, color=black] coordinates {  (-2.8424644,5.7774391) (-2.7144654,5.4321823) (-2.573441,5.8159404) (-2.4280119,5.0387087) (-2.318562,5.5863624) (-2.2440021,5.5973196) (-2.1775162,6.200809) (-2.1187725,5.6216588) (-2.0580857,5.5681534) (-1.9986315,5.6830959) (-1.9256192,5.3506165) (-1.8787163,5.7469201) (-1.8462605,5.7839689) (-1.7858982,6.1255331) (-1.737396,6.1034913) (-1.6806248,5.8434424) (-1.6280144,5.5831375) (-1.5700999,5.4271593) (-1.5259469,5.9953642) (-1.4792275,6.1969967) (-1.447107,5.0985637) (-1.414688,5.6740084) (-1.3642813,5.8645573) (-1.3244512,4.9313178) (-1.2951392,5.9524961) (-1.2458988,5.9936123) (-1.1999756,5.8606682) (-1.1690181,5.9266233) (-1.1413056,6.2476525) (-1.1166871,6.0924997) (-1.0900578,5.5541368) (-1.0566919,6.077466) (-1.0217652,6.2196283) (-.99040675,5.9800401) (-.95257747,6.0756578) (-.91860038,5.7687006) (-.89313769,6.0511084) (-.85984415,6.2017789) (-.83326113,6.1303988) (-.81308532,5.8197637) (-.78098452,6.0415354) (-.74682099,5.751966) (-.72825718,6.2041121) (-.70465904,6.086071) (-.67711616,6.1511173) (-.6498903,6.1576085) (-.62361336,5.2828302) (-.6049968,5.9229407) (-.57370692,6.210176) (-.54241288,5.6419716) (-.51786435,6.0789437) (-.48893172,5.9365215) (-.45545229,6.3359132) (-.43250972,6.2015328) (-.40243152,5.8865066) (-.38037673,6.0877452) (-.35386798,6.5289645) (-.32303232,6.1064205) (-.29356006,5.7854447) (-.2752834,6.0399871) (-.24742585,5.2216239) (-.22216973,6.2169523) (-.19360597,5.9449153) (-.16551811,6.3083253) (-.1424599,6.3424401) (-.11124116,5.8809795) (-.07926229,6.4407067) (-.0614009,6.032567) (-.03503328,6.0967669) (-.00407578,6.2848482) (.01921485,6.0049429) (.0469714,5.8328905) (.08066473,6.3098049) (.10741307,6.3656497) (.1314263,5.8575177) (.15676507,5.6346641) (.18148047,6.3604345) (.2123156,6.3396444) (.2463645,5.8200121) (.27823442,6.2321596) (.31196177,6.0820289) (.34906527,6.3282285) (.38245887,6.1591887) (.41297334,6.0036283) (.44769025,6.0313401) (.48583323,5.7867684) (.52952898,6.0621052) (.56783658,6.2534399) (.60666531,6.1903324) (.66100264,5.9343157) (.71401429,6.4876757) (.77032405,6.1933913) (.82246059,6.3633256) (.86215413,6.3721609) (.90850401,6.2409825) (.97303593,6.1385322) (1.0387533,5.7230382) (1.1108729,6.0534797) (1.1695299,6.2882738) (1.2339407,6.4095707) };
\addplot[domain=-3.25:1.75, color=black, solid, line width=0.75mm] {6.070835+0.1719*x};
\legend{~Observations,~Linear Fit}
\end{axis}
\end{tikzpicture}
}
\end{subfigure}
\linespread{1} \floatfoot{\footnotesize\textsc{Note. ---} The figures show binned scatterplots with 100 markers to depict cross-sectional correlations between log labor market tightness and the log of direct, indirect, and overall pre-match hiring costs. Whereas direct pre-match hiring costs (in Euro) are asked separately in the IAB Job Vacancy Survey, we calculate indirect pre-match hiring costs (in Euro) by multiplying the search effort (in working hours) by the firm's average hourly wage rate of i) workers in human resource management (KldB-2010 Code: 715) or, if not available, ii) managers or, if not available, iii) all workers. Pre-match hiring costs were deflated with base year 2015. Labor markets are combinations of 5-digit KldB occupations and commuting zones. We trim labor market tightness at the 5th and 95th percentile. The numbers of observed successful hires are: 12,348 for direct, 29,641 for indirect, and 24,884 for overall pre-match hiring costs. Sources: Integrated Employment Biographies $\plus$ Official Statistics of Federal Employment Agency $\plus$ IAB Job Vacancy Survey, 2014-2015, 2017-2019.} \linespread{1.5}
\end{figure}
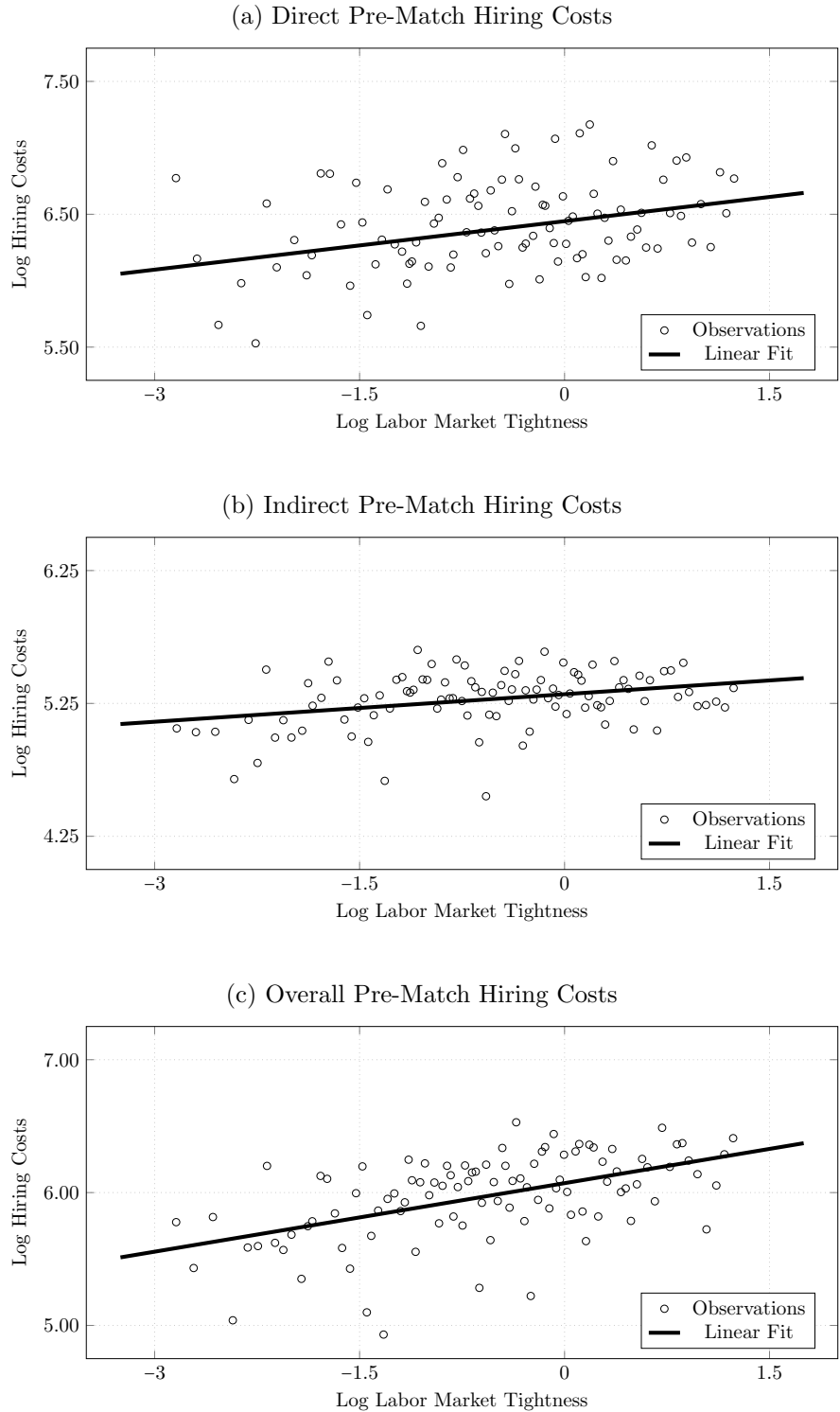}

Based on the firm's location, its targeted 5-digit occupation, and the year, we enrich our survey-based recruitment indicators with the respective labor market tightness. Figure \ref{fig:3} illustrates the cross-sectional correlations between labor market tightness and pre-match hiring cost.
\footnote{For lack of longitudinal information in the IAB Job Vacancy Survey, we cannot control for unobserved time-invariant heterogeneity and, thus, refrain from interpreting the magnitude of the underlying coefficients.} Panel a) shows that higher labor market tightness is associated with higher direct pre-match hiring cost, such as expenditures for posting vacancies, job advertisement, or headhunters. In addition, Panel b) implies that increased tightness also involves higher indirect pre-match hiring cost, namely for screening candidates, job interviews, and signing contracts. Taken together, Panel c) suggest that an increase in labor market tightness raises overall pre-match hiring cost of successful hiring processes. In a similar fashion, higher wages also come along with higher pre-match hiring costs (see Figure \ref{fig:D5}), suggesting that the cost of intensified screening dominate the higher ease of recruitment when paying higher wages (i.e., $\phi_{1}>0$). Overall, the cross-sectional relationships of pre-match hiring cost with labor market tightness and wages provide an empirical foundation for our formulation of hiring cost in Equation (\ref{eq:5}).

In addition, our measure of labor market tightness performs well with regard to the three remaining recruitment indicators that drive pre-match hiring cost (see Figure \ref{fig:D6}): the number of applicants, the number of search channels, and the search duration. Labor market tightness and the number of applicants are negatively correlated, mirroring the definition of labor market tightness, which includes the number of job seekers in the denominator. Higher labor market tightness also makes firms use more search channels (such as using private placement services), which further raises pre-match hiring cost.\footnote{Between 2012 and 2019, the IAB Job Vacancy includes time-consistent information on the following ten search channels for successful hiring processes: advertisement in newspapers or magazines, advertisement on own website, advertisement on online job boards, contact to the federal employment agency (FEA), internet services of FEA, pool of applicants to other positions, private placement services, internal job advertisement, personal contacts of employees, and selection among apprentices, leased workers, or interns.} Higher labor market tightness also entails a longer search duration for hires, which is the temporal dimension of pre-match hiring cost.

Building on Equation (\ref{eq:13}), we weight our measures of labor market tightness by firms' occupational employment shares to construct a firm-specific measure of labor market tightness. In order for this measure to capture firms' difficulty to fill their vacancies, firms' occupational composition of vacancies must resemble their composition of employment. To address this concern, we gather information on the occupational structure of firms' open vacancies from the IAB Job Vacancy Survey and construct an analogous measure of firm-specific labor market tightness that is based on the vacancy shares of occupations. Using record linkage, we contrast this vacancy-based measure with our employment-based measure of labor market tightness (Figure \ref{fig:D7}). Favourably, we find a strongly positive and linear relationship between both measures. We view this pattern as supporting evidence in favor of our employment-based firm-specific labor market tightness that we are able to leverage for the universe of firms in the IEB (see Section \ref{sec:6}).

\section{Results: Labor Demand Effects}
\label{sec:6}

In the following section, we quantify the extent to which the tightening of labor markets reduces firms' labor demand (and ultimately employment), while simultaneously determining the own-wage elasticity of labor demand.

\paragraph{Baseline Results.} Table \ref{tab:1} displays the baseline estimates, including potentially endogenous OLS estimates as well as instrumental variable estimates from 2SLS. The first column presents results from a naive OLS estimation of Equation (\ref{eq:12}). While the own-wage elasticity of labor demand is negative, albeit small, the elasticity with respect to tightness turns out to be positive, unlike suggested by theory. However, as pointed out in Section \ref{sec:3}, the OLS estimates may feature an upward bias.

To address the bias in either case, we use our Bartik-style instruments (\ref{eq:16}), (\ref{eq:19}), and (\ref{eq:20}) to insulate plausibly exogenous variation in wages and labor market tightness in a second, third and fourth specification. To begin with, we solely estimate the effect of wages on labor demand, using the wage instrument (\ref{eq:16}). As expected in comparison to OLS, the wage elasticity of labor demand turns out to be more negative, implying that firms lower employment by 0.73 percent when wages increase by 1 percent. Building on the instruments for vacancies (\ref{eq:19}) and job seekers (\ref{eq:20}), Column (3) displays the IV effect of tightness on employment without conditioning on wages. In contrast to OLS, the estimated elasticity turns negative, indicating that an increase in labor market tightness by 1 percent reduces employment by 0.05 percent on average.

\begin{table}[!ht]
\centering
\scalebox{0.90}{
\begin{threeparttable}
\caption{Effects of Wages and Labor Market Tightness on Employment}
\label{tab:1}
\begin{tabular}{L{4cm}C{2,7cm}C{2,7cm}C{2,7cm}C{2,7cm}} \hline
\multirow{3.4}{*}{} & \multirow{3.4}{*}{\shortstack{(1) \\ $\Delta$ Log L$^{\text{FT}}$ }} & \multirow{3.4}{*}{\shortstack{(2) \\ $\Delta$ Log  L$^{\text{FT}}$ }} & \multirow{3.4}{*}{\shortstack{(3) \\ $\Delta$ Log  L$^{\text{FT}}$ }} & \multirow{3.4}{*}{\shortstack{(4) \\ $\Delta$ Log  L$^{\text{FT}}$ }} \\
&&&& \\
&&&& \\[0.2cm] \hline
&&&& \\[-0.3cm]
\multirow{2.4}{*}{$\Delta$ Log W$^{\text{FT}}$} &  \multirow{2.4}{*}{\shortstack{\hphantom{***}-0.136***\hphantom{-} \\ (0.002)}}  & \multirow{2.4}{*}{\shortstack{\hphantom{***}-0.733***\hphantom{-} \\ (0.022)}} &   &  \multirow{2.4}{*}{\shortstack{\hphantom{***}-0.730***\hphantom{-} \\ (0.022)}} \\
&&&& \\
 \multirow{2.4}{*}{$\Delta$ Log V/U} &  \multirow{2.4}{*}{\shortstack{\hphantom{***}0.047***\\ (0.000)}}  &  &  \multirow{2.4}{*}{\shortstack{\hphantom{***}-0.054***\hphantom{-}\\ (0.002)}}  &   \multirow{2.4}{*}{\shortstack{\hphantom{***}-0.051***\hphantom{-}\\ (0.002)}} \\
&&&& \\[0.2cm] 
 Fixed Effects  & Year &  Year  & Year & Year \\[0.2cm] \hline
&&&& \\[-0.2cm]
 Instruments  & None & Z\(_{\text{W}^{\text{FT}}}\)  & Z\(_{\text{V}}\), Z\(_{\text{U}}\) & Z\(_{\text{W}^{\text{FT}}}\), Z\(_{\text{V}}\), Z\(_{\text{U}}\) \\[0.2cm]
 Observations &  7,993,993          & 7,993,993        &  7,993,993          & 7,993,993             \\[0.2cm]
 Clusters &  1,801,671          & 1,801,671        &  1,801,671          & 1,801,671             \\[0.2cm]
 F: $\Delta$ Log W$^{\text{FT}}$ &  & 9,952   &  & 3,322  \\[0.2cm]
 F: $\Delta$ Log V/U &  &  & 45,522 & 30,380 \\[0.2cm] \hline
\end{tabular}
\begin{tablenotes}[para]
\footnotesize\textsc{Note. ---} The table displays OLS and IV regressions of differences in log employment (of regular full-time workers) per establishment on differences in the log of average daily wages and the log of labor market tightness. The instrumental variables refer to shift-share instruments of biennial national changes in occupations weighted by past occupational employment in the respective establishment. The lag difference is two years. Labor markets are combinations of 5-digit KldB occupations and commuting zones. Standard errors (in parentheses) are clustered at the establishment level. F = F Statistics of Excluded Instruments. FT = Full-Time. KldB = German Classification of Occupations. L = Employment. U = Job Seekers. V = Vacancies. W = Average Daily Wages. Z = Shift-Share Instrument. * = p$<$0.10. ** = p$<$0.05. *** = p$<$0.01. Sources: Integrated Employment Biographies $\plus$ Official Statistics of the German Federal Employment Agency $\plus$ IAB Job Vacancy Survey, 1999-2019.
\end{tablenotes}
\end{threeparttable}
}
\end{table}

Finally, Column (4) displays effects of wages and tightness from a joint IV model, in which both variables exert an additive impact on firms' labor demand. We refer to this model as our baseline specification, which is described by Equation (\ref{eq:12}). In our baseline estimation, the own-wage elasticity of labor demand is -0.73. Since our shift-share design rigorously addresses upward bias, this elasticity is at the lower end of the values found in the international and German literature \citep{LichterEtAl2015,Popp2023}. The elasticity of labor demand with respect to tightness is -0.05, implying that the observed doubling in tightness between 2012 and 2019 (i.e., an increase by 100 percent) reduced firms' employment ceteris paribus by 5 percent on average. The effects are statistically significant at 1 percent levels. Interestingly, both elasticities remain largely unchanged compared to the separate regressions in Columns (2) and (3). On the one hand, this finding highlights that the instruments for wages and tightness do not interact with each other. On the other hand, by comparing Column (4) with Column (3), we can rule out that labor market tightness substantially affects labor demand through changes in wages because controlling for the wage channel does not alter the tightness effect.\footnote{The similarity of the tightness effect between Column (3) and Column (4) implies that, in Equation (\ref{eq:A34}) in Appendix \ref{sec:A}, the relative effect of labor market tightness on wages, $\gamma$, is close to zero. In this case, Equation (\ref{eq:A34}) collapses to our baseline version of the elasticity of labor demand with respect to tightness (\ref{eq:8}).}

\paragraph{Magnitude of Pre-Match Hiring Cost.} Favourably, our underlying theoretical model in Section \ref{sec:2} allows us to validate the plausibility of our baseline results by quantifying the magnitude of pre-match unit hiring cost. Given Equations (\ref{eq:7}) and (\ref{eq:8}), it is easy to show that the ratio of the tightness elasticity of labor demand to the own-wage elasticity of labor demand simply reflects the ratio of the tightness elasticity of unit labor cost to the own-wage elasticity of unit labor cost (i.e., the elasticity of labor demand to unit labor cost cancels out):
\begin{equation}
\label{eq:21}
\frac{\eta^{L}_{\theta}}{\eta^{L}_{W}} = \frac{\phi_{2} \cdot (\delta + r) \cdot \Phi}{  W + \phi_{1} \cdot (\delta + r) \cdot \Phi  }
\end{equation}
Thus, the ratio of baseline estimates in Column (4) of Table \ref{tab:1} implies that the relative effect of tightness on unit labor cost is about 1/14 of the relative wage effect on unit labor cost.\footnote{We calculate this fraction as: $\frac{\eta^{L}_{\theta}}{\eta^{L}_{W}}=\frac{-0.051}{-0.730}\approx\frac{1}{14}.$} Given this proportion, we can approximate the magnitude of pre-match hiring cost (as a fraction of annual wage payments) by collecting additional information on only few model parameters:
\begin{equation}
\label{eq:22}
\frac{\Phi}{W}\,=\,  \Bigg( (\delta + r) \cdot \bigg( \phi_2 \cdot \frac{\eta^{L}_{W}}{\eta^{L}_{\theta}}  \,-\, \phi_1 \bigg) \Bigg)^{-1}
\end{equation}
Table \ref{tab:2} displays the calibration of Equation (\ref{eq:22}). Using IEB data for our period of study, we calculate a yearly separation rate $\delta$ of 33.1 percent.\footnote{To minimize bias from temporal aggregation \citep{Nordmeier2014}, we estimate a day-to-day separation rate of 0.10999 percent, which translates into a yearly separation rate of 33.1 percent: $1-(1-0.0010999)^{365.25}=0.331$.} Building on evidence from business surveys \citep{JagannathanEtAl2016,Graham2022}, we assume that firms' yearly subjective discount rate $r$ equals 15 percent. Furthermore, we rely on the evidence from \citet{MuehlemannStruplerLeiser2018} who, by leveraging panel data, provide the most convincing estimates on the effect of wages ($\phi_{1}=1.852$) and labor market tightness ($\phi_{2}=0.468$) on pre-match hiring cost.\footnote{To arrive at $\phi_{2}$, we multiply their reported semi-elasticity for labor market tightness of 0.323 with their reported average tightness of 1.45, which yields: $0.323\cdot1.45=0.468$.} Given this calibration, our baseline estimates imply that pre-match unit hiring cost amount to 42.9 percent of annual wage payments for the average full-time worker. This value lies in the middle of the wide array of indirect estimates derived from dynamic labor demand models \citep{Yaman2019} which range from values close to zero to more than one year of wage payments. At the same time, our value is somewhat higher than evidence on successful hiring processes from business surveys which plausibly form a lower bound due to non-consideration of unsuccessful recruitment processes and the difficulty of inquiring quasi-fixed cost of hiring.

\begin{table}[!ht]
\centering
\scalebox{0.90}{
\begin{threeparttable}
\caption{Magnitude of Pre-Match Hiring Cost}
\label{tab:2}
\begin{tabular}{L{7.5cm}C{0.7cm}C{1.15cm}L{6cm}} \hline
\multirow{2.4}{*}{Parameter} &  & \multirow{2.4}{*}{Value} & \multirow{2.4}{*}{Source} \\
& & &   \\[0.2cm] \hline
& & &   \\[-0.3cm]
Yearly Separation Rate &    $\delta$ &    ~0.331 &  IEB, Own Calculations \\[0.1cm]
\multirow{2.4}{*}{Yearly Discount Rate} &  \multirow{2.4}{*}{$r$} &    \multirow{2.4}{*}{~0.150} &  \multirow{2.4}{*}{\shortstack[l]{ \citet{JagannathanEtAl2016}: Table 3 \\ \citet{Graham2022}: Table 2}} \\[0.1cm]
& & &   \\[0.2cm]
Wage Elasticity of Labor Demand &  $\eta^{L}_{W}$ &  -0.730 &  Table \ref{tab:1}, Column 4 \\[0.4cm]
Tightness Elasticity of Labor Demand &  $\eta^{L}_{\theta}$ &  -0.051 &  Table \ref{tab:1}, Column 4  \\[0.1cm]
\multirow{2.4}{*}{Wage Elasticity of Pre-Match Hiring Cost} &  \multirow{2.4}{*}{$\phi_{1}$} &  \multirow{2.4}{*}{~1.852} &  \multirow{2.4}{*}{\shortstack[l]{Muehlemann/Strupler Leiser \citeyearpar{MuehlemannStruplerLeiser2018}:\\ Table 8}}  \\[0.1cm]
& & &   \\[-0.6cm]
\multirow{2.4}{*}{Tightness Elasticity of Pre-Match Hiring Cost} &  \multirow{2.4}{*}{$\phi_{2}$} &  \multirow{2.4}{*}{~0.468} &  \multirow{2.4}{*}{\shortstack[l]{Muehlemann/Strupler Leiser \citeyearpar{MuehlemannStruplerLeiser2018}:\\ Page 125 and Table 8 }} \\
& & &  \\[0.1cm] \hline
& & &  \\[-0.4cm]
\multirow{2.4}{*}{\shortstack[l]{Pre-Match Hiring Cost \\ (as Fraction of Annual Wage Payments)}} &  \multirow{2.4}{*}{$\frac{\Phi}{W}$} &  \multirow{2.4}{*}{~0.429} &  \multirow{2.4}{*}{Equation (\ref{eq:22})}  \\[0.1cm]
& & &   \\ \hline
\end{tabular}
\begin{tablenotes}[para]
\footnotesize\textsc{Note. ---} The table displays the calibration of Equation (\ref{eq:22}) to approximate the magnitude of pre-match hiring cost.
\end{tablenotes}
\end{threeparttable}
}
\end{table}

\paragraph{Empirical Checks on Identification Strategy.} In a next step, we decompose our Bartik estimator to shed more light on the identifying variation underlying our estimates, as proposed by \citet{Goldsmith-PinkhamEtAl2020}. We fully present these shift-share diagnostics in Appendix \ref{sec:E}. Specifically, we deconstruct our Bartik estimates into Rotemberg weights (see Table \ref{tab:E1}) and just-identified IV estimates (see Figure \ref{fig:E1}), separately for the wage instrument, the vacancy instrument, and the job seeker instrument.\footnote{For ease of computation, we carry out the decomposition using a random 50 percent sample of firms for the second and third specification (of Table \ref{tab:1}).} The key messages from these diagnostics are as follows: First, the quantitative magnitude of negative Rotemberg weights is small in all cases, thus allowing for a LATE interpretation of our Bartik estimates.\footnote{As the vacancy and the job seeker instrument exert an opposite impact on labor market tightness (i.e., the vacancy-to-job-seekers ratio), the joint normalization of their Rotemberg weights to 1 implies that the weights for the job seeker instrument feature opposite signs (i.e., negative weights for the job seeker instrument must be interpreted as positive ones and vice versa).} Second, the largest positive weights for the wage effects relate to the intervals 2013-2015 and 2014-2016, which both cover the first-time introduction of national minimum wage in 2015. By contrast, the identifying variation is more evenly distributed across years for the vacancy and job seeker instrument, which is in line with a steadily tightening labor market during our period of analysis. Third, the distribution of Rotemberg weights across occupations is highly skewed. The top five occupations with the largest weight account for 44.5 (wage instrument), 33.6 (vacancy instrument), and 27.1 percent (job seeker instrument) of the sum of absolute Rotemberg weights. For the wage effect, the top five occupations comprise gastronomy workers, medical assistants, hairdressers, cooks, and farmers. In line with the distribution of weights across years, earnings in these low-wage occupations were highly affected by the 2015 minimum wage, corroborating that this policy intervention drives a significant fraction of our identifying variation. For the tightness effect, sales workers receive by far the largest weight, seemingly because employment of these workers strongly follows the business cycle. Fourth, the vacancy instrument determines about three quarters of the labor market tightness effect whereas the remaining quarter stems from the job seeker instrument. Fifth, the just-identified IV estimates show substantial heterogeneity across occupations.

Given the logic of the Bartik estimator, exogeneity of either the national growth rates or the predetermined employment shares would suffice to establish unbiasedness. Favorably, our decomposition highlighted that a large part of the variation underlying our estimated wage effect stems from an exogenous event, namely the first-time introduction of a statutory nation-wide minimum wage in Germany. On top, we follow \citet{Goldsmith-PinkhamEtAl2020} and perform a further empirical check to examine whether predetermined occupational shares are uncorrelated with the error term (i.e., with uncontrolled determinants of changes in labor demand). Using survey information from the IAB Establishment Panel, we regress firms' predetermined occupational employment shares on a set of labor demand and labor supply variables in the very same year. If the cross-sectional variation between shares and the level of labor demand variables turns out to be low, then the correlation with changes in labor demand variables (in the far-off future) should be even smaller. We narrow our analysis to the top five occupations with the largest Rotemberg weights.

The supply-determining variables such as the female employment share and the share of foreign citizens are strongly correlated with shares in the top five occupations from the wage effect (see Panel a) of Table \ref{tab:E2}). By contrast, demand shifters like capital investment or business expectations are hardly correlated with the occupational shares. An exception is firms' labor productivity which is negatively correlated with most relevant occupations. However, this negative cross-sectional correlation simply reflects that firms with high shares in these low-wage occupations are less productive and, accordingly, are more strongly affected by the exogenous wage increase from the minimum wage introduction in 2015. Finally, for the vacancy and job seeker instrument (see Panel b) and c) of Table \ref{tab:E2}), the predetermined shares of top five occupations are more strongly correlated with supply rather than demand variables, supporting that we observe an effect that is identified from exogenous changes in tightness rather than demand-driven reverse causality.

Table \ref{tab:3} presents the underlying first-stage regressions for the second specification in Column (1), the third specification in Column (2), and the fourth and baseline specification in Columns (3) and (4). The first-stage estimates show that our wage instrument is a good predictor for wage changes. A national wage increase by 10 percent (weighted by firms' past occupational employment shares) raises firms' wages by 6.4 percent, which is large and positive. Similarly, the instruments for vacancies and unemployed predict well shifts in labor market tightness, with signs of the coefficients featuring the expected directions: a national increase in vacancies (job seekers) by 10 percent raises (lowers) firm-specific labor market tightness by 4.52 (4.47) percent. The F statistics for the joint exclusion of the instruments are sufficiently large in all cases. Hence, the irrelevance of the instruments is clearly rejected, demonstrating that our shift-share design delivers strong instruments.\footnote{In Appendix \ref{sec:F}, we provide a visual inspection of the first stages, which illustrates favorable correlation patterns between the instruments and their respective endogenous variable (see Figure \ref{fig:F1}). In addition, we further display reduced-form regressions of the outcome variable on the instruments (see Table \ref{tab:F1}). Each instrument shows the expected signs implied by the baseline IV estimates. The reduced-form effects turn out lower than second-stage elasticities, which is in line with the intuition that national shifts should exert an effect on firms' employment smaller than effects of direct changes at the firm level.}

\begin{table}[!ht]
\centering
\scalebox{0.90}{
\begin{threeparttable}
\caption{First-Stage Regressions}
\label{tab:3}
\begin{tabular}{L{4cm}C{2.7cm}C{2.7cm}C{2.7cm}C{2.7cm}} \hline
\multirow{3.4}{*}{} & \multirow{3.4}{*}{\shortstack{(1) \\ $\Delta$ Log W$^{\text{FT}}$ \vphantom{/}  }} & \multirow{3.4}{*}{\shortstack{(2) \\ $\Delta$ Log V/U \vphantom{/}  }} & \multirow{3.4}{*}{\shortstack{(3) \\ $\Delta$ Log W$^{\text{FT}}$ \vphantom{/}  }} & \multirow{3.4}{*}{\shortstack{(4) \\ $\Delta$ Log V/U\vphantom{/}  }} \\
&&&& \\
&&&& \\[0.2cm] \hline
&&&& \\[-0.3cm]
\multirow{2.4}{*}{ Z\(_{\text{W}^{\text{FT}}}\)  } &  \multirow{2.4}{*}{\shortstack{\hphantom{***}0.636*** \\ (0.006)}} & &  \multirow{2.4}{*}{\shortstack{\hphantom{***}0.635*** \\ (0.006)}} &   \multirow{2.4}{*}{\shortstack{\hphantom{***}-0.260***\hphantom{-} \\ (0.019)}} \\
&&&&  \\
\multirow{2.4}{*}{ Z\(_{\text{V}}\)  } & & \multirow{2.4}{*}{\shortstack{\hphantom{***}0.452*** \\ (0.002)}}  &  \multirow{2.4}{*}{\shortstack{\hphantom{***}0.001*** \\ (0.000)}} & \multirow{2.4}{*}{\shortstack{\hphantom{***}0.452*** \\ (0.002)}}    \\
&&&&  \\
\multirow{2.4}{*}{ Z\(_{\text{U}}\)  } & & \multirow{2.4}{*}{\shortstack{\hphantom{***}-0.447***\hphantom{-} \\ (0.003)}}    &  \multirow{2.4}{*}{\shortstack{0.001 \\ (0.001)}} & \multirow{2.4}{*}{\shortstack{\hphantom{***}-0.447***\hphantom{-} \\ (0.003)}} \\
&&&&  \\[0.2cm]
Fixed Effects  & Year &  Year & Year &  Year   \\[0.2cm] \hline
&&&& \\[-0.2cm]
Observations &  7,993,993          & 7,993,993        &  7,993,993          & 7,993,993             \\[0.2cm]
Clusters &  1,801,671          & 1,801,671        &  1,801,671          & 1,801,671             \\
\multirow{2.4}{*}{\shortstack[l]{F Statistics of \\ Excluded Instruments}} & \multirow{2.4}{*}{9,952 } & \multirow{2.4}{*}{45,522}  & \multirow{2.4}{*}{3,322 } & \multirow{2.4}{*}{30,380}  \\
&&&& \\[0.2cm] \hline
\end{tabular}
\begin{tablenotes}[para]
\footnotesize\textsc{Note. ---} The table displays the underlying first-stage regressions of the IV estimations in Column (2), (3), and (4) from Table \ref{tab:2}. The instrumental variables refer to shift-share instruments of biennial national changes in occupations weighted by past occupational employment in the respective establishment. The lag difference is two years. Labor markets are combinations of 5-digit KldB occupations and commuting zones. Standard errors (in parentheses) are clustered at the establishment level. FT = Full-Time. KldB = German Classification of Occupations. L = Employment. U = Job Seekers. V = Vacancies. W = Average Daily Wages. Z = Shift-Share Instrument. * = p$<$0.10. ** = p$<$0.05. *** = p$<$0.01. Sources: Integrated Employment Biographies $\plus$ Official Statistics of the German Federal Employment Agency $\plus$ IAB Job Vacancy Survey, 1999-2019.
\end{tablenotes}
\end{threeparttable}
}
\end{table}

\paragraph{Sensitivity and Heterogeneity.} We examine the sensitivity and underlying heterogeneity of our baseline estimates in various respects. Figure \ref{fig:4} visualizes the corresponding results whereas the detailed regression results are assembled in Appendix \ref{sec:F}.

First, we address the choice of the lag difference. As elaborated in Section \ref{sec:3}, we specify the empirical model in two-year differences to estimate long-run elasticities. The results show that the wage effect in the specification with one-year differences is slightly lower, indicating that overall labor demand responses do not fully materialize in the short run due to adjustment cost \citep{Nickell1986}. However, the elasticity barely increases for three-year differences. 
We observe a similar pattern for the tightness effect, which turns out to be smaller for one-year differences as well. Again, the smaller coefficient for one-year differences likely reflects sluggish responses in labor demand.

\afterpage{
\begin{landscape}
\begin{figure}[!ht]
\centering
\caption{Sensitivity and Heterogeneity of Labor Demand Effects}
\label{fig:4}
\vspace*{0.5cm}
\begin{subfigure}{0.5125\textwidth}
\captionsetup{oneside,margin={4cm,0cm}}
\caption{Wage Rate}
\centering
\scalebox{0.80}{
\begin{tikzpicture}
\begin{axis}[xlabel=Wage Elasticity of Labor Demand, ylabel=Specification, height=13.5cm, width=13cm, grid=major, grid style=dotted, xmin=-1.7, xmax=0.5, ymin=-1, ymax=22.5, xtick={-1.6,-1.2,-0.8,-0.4,0,0.4}, extra x tick style={grid=none}, extra x ticks={0}, x tick label style={/pgf/number format/.cd,fixed,fixed zerofill, precision=1,/tikz/.cd},y tick label style = {font=\small}, ytick={1,2,3.5,4.5,6,7,8,9,10.5,11.5,12.5,13.5,14.5,15.5,16.5,18,19,20.5}, yticklabels={{High Firm-AKM Effects\vphantom{/}},{Low Firm-AKM Effects\vphantom{/}},{East Germany\vphantom{/}},{West Germany\vphantom{/}},{Large Firms\vphantom{/}},{Medium-Sized Firms\vphantom{/}},{Small Firms\vphantom{/}},{Weighted by Firm Size\vphantom{/}},{Districts\vphantom{/}},{Adj.\ KldB-5 Occupations\vphantom{/}},{KldB-3/5 Occupations\vphantom{/}},{Only Registered Vacancies\vphantom{/}},{Median Wage Rate\vphantom{/}},{With Year-by-Zone FE\vphantom{/}},{With Year-by-Industry FE\vphantom{/}},{3-Year Lag Difference\vphantom{/}},{1-Year Lag Difference\vphantom{/}},{\textbf{Baseline}\vphantom{/}},}]
\addplot[only marks,mark=square*,mark options={fill=black},color=black, error bar legend, error bars/.cd, x dir=both, x explicit] coordinates {   (-.35129306,1)+-(.07974237,0) (-.6412245,2)+-(.06603992,0) (-.84191066,3.5)+-(.06465137,0) (-.65688986,4.5)+-(.05296612,0) (-.76051015,6)+-(.38424081,0) (-.90574944,7)+-(.07771176,0) (-.51756936,8)+-(.05256897,0) (-.83035278,9)+-(.22442973,0) (-.75960356,10.5)+-(.04655671,0) (-.75143027,11.5)+-(.0432905,0) (-.74273723,12.5)+-(.04377896,0) (-.75893199,13.5)+-(.04366881,0) (-.71666741,14.5)+-(.04235357,0) (-.7188856,15.5)+-(.04267502,0) (-.78292727,16.5)+-(.07468998,0) (-.77204239,18)+-(.04436547,0) (-.61657321,19)+-(.04330373,0) (-.73040599,20.5)+-(.04334217,0) };
\addplot[loosely dashed] coordinates {(0,-1.5) (0,23.5)};
\end{axis}
\end{tikzpicture}
}
\end{subfigure}
\begin{subfigure}{0.4125\textwidth}
\captionsetup{oneside,margin={0cm,0cm}}
\caption{Labor Market Tightness}
\centering
\scalebox{0.80}{
\begin{tikzpicture}
\begin{axis}[xlabel=Labor Market Tightness Elasticity of Labor Demand, height=13.5cm, width=13cm, grid=major, grid style=dotted, xmin=-0.17, xmax=0.05, ymin=-1, ymax=22.5, xtick={-0.16,-0.12,-0.08,-0.04,0,0.04}, extra x tick style={grid=none}, extra x ticks={0}, x tick label style={/pgf/number format/.cd,fixed,fixed zerofill, precision=2,/tikz/.cd},y tick label style = {font=\small}, ytick={1,2,3.5,4.5,6,7,8,9,10.5,11.5,12.5,13.5,14.5,15.5,16.5,18,19,20.5}, yticklabels={}]
\addplot[only marks,mark=square*,mark options={fill=black},color=black, error bar legend, error bars/.cd, x dir=both, x explicit] coordinates {   (-.07012868,1)+-(.00449394,0) (-.02839668,2)+-(.00526478,0) (-.05768122,3.5)+-(.00845801,0) (-.04928338,4.5)+-(.00373343,0) (-.05925097,6)+-(.0211648,0) (-.08205613,7)+-(.00647161,0) (-.04497844,8)+-(.00402651,0) (-.05393401,9)+-(.02073069,0) (-.05914257,10.5)+-(.00402196,0) (-.04955246,11.5)+-(.00404051,0) (-.05312726,12.5)+-(.00371074,0) (-.04561866,13.5)+-(.00362173,0) (-.05169965,14.5)+-(.00340627,0) (-.05043299,15.5)+-(.00347815,0) (-.06721862,16.5)+-(.00457751,0) (-.05802297,18)+-(.00389846,0) (-.02258315,19)+-(.00306529,0) (-.05126658,20.5)+-(.00343224,0) };
\addplot[loosely dashed] coordinates {(0,-1.5) (0,23.5)};
\end{axis}
\end{tikzpicture}
}
\end{subfigure}
\floatfoot{\footnotesize\textsc{Note. ---} The figure illustrates estimated labor demand elasticities for a variety of specifications. The baseline IV estimation regresses differences in log employment (of regular full-time workers) per establishment on differences in the log of average daily wages and the log of labor market tightness. The instrumental variables for average wages, vacancies, and job seekers refer to shift-share instruments of biennial national changes in occupations weighted by past occupational employment in the respective establishment. In the baseline regression, the lag difference is two years. Baseline labor markets are combinations of 5-digit KldB occupations and commuting zones. Each point estimate features a 95 percent confidence interval. Adj. = Flow-Adjusted. KldB-3/5 = 3-Digit KldB Occupation by Level of Skill Requirement. Sources: Integrated Employment Biographies $\plus$ Official Statistics of the German Federal Employment Agency $\plus$ IAB Job Vacancy Survey, 1999-2019.}
\end{figure}
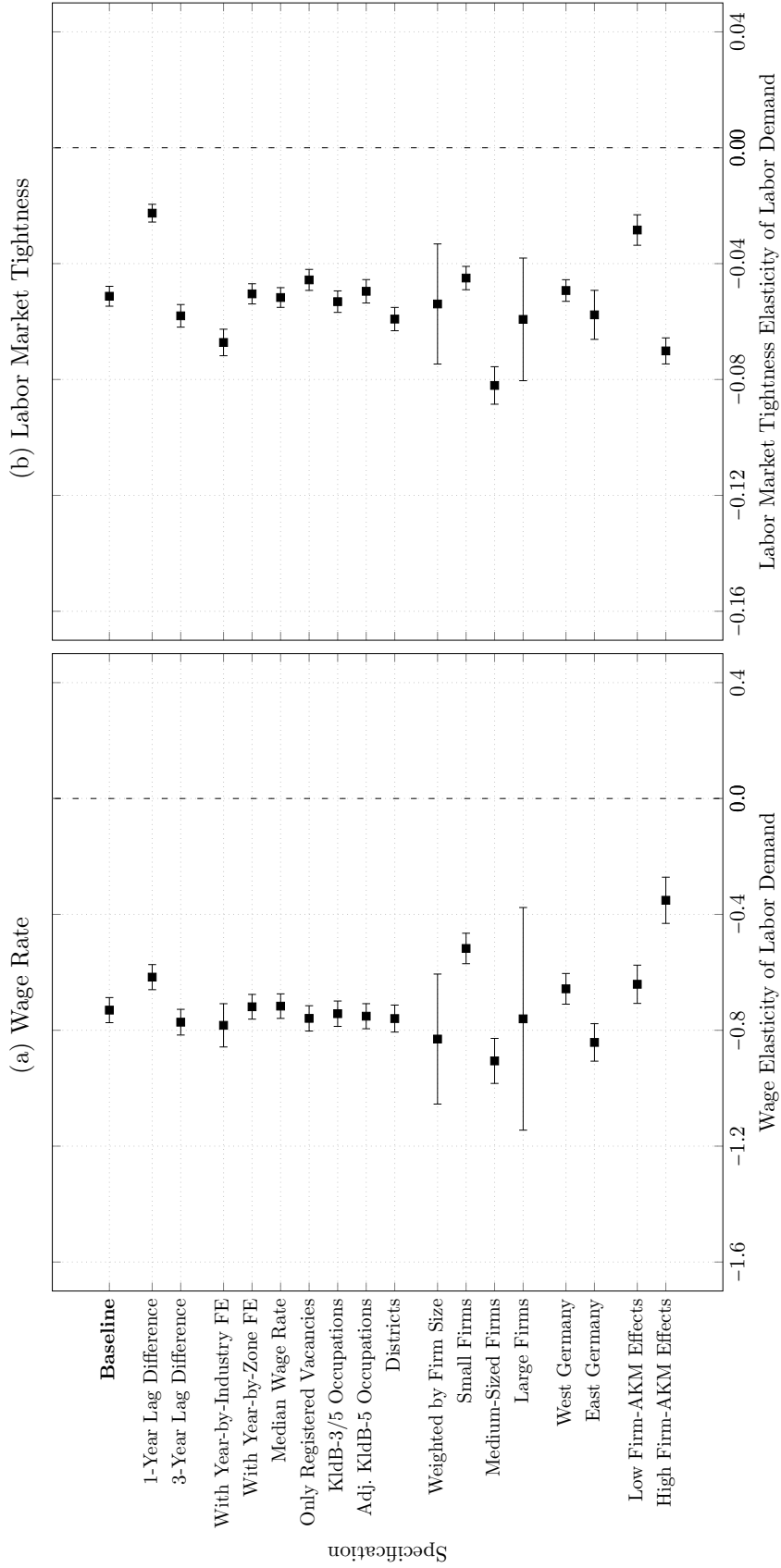

\end{landscape}
}

Next, we carry out robustness checks relating to the regression specification and the measurement of the model variables. Specifically, we additionally differentiate the year fixed effects by 38 industries or 51 commuting zones to more rigorously control for common labor demand shocks (which, if the predetermined shares were not exogenous, the Bartik instrument may not protect against), such as technological change. The results do not change substantially when including the more disaggregated fixed effects.\footnote{Including industry-by-year fixed effects is idiosyncratic in the sense that the underlying variation stems only from firms whose occupational structure is different from the industry average. Nevertheless, the elasticities remain in the ballpark of our baseline estimates.} This robustness is well in line with our correlation analysis that attributed only a minor role to labor demand shocks across firms (see Table \ref{tab:E2}), lending further credence to a causal interpretation of our elasticity estimates. Regarding the measurement of the model variables, we replace the log average wage by the log median wage which is robust against outliers and the top-coding of wages at the social security limit. Furthermore, we use registered vacancies instead of total vacancies for the measurement of labor market tightness. In both cases, the elasticities retain a negative sign and feature a similar order of magnitude, corroborating that our results are not driven by a certain operationalization of the model variables.

We also test the sensitivity of the results with respect to our baseline definition of a labor market, which are combinations of 1,286 5-digit occupations and 51 commuting zones. In terms of the occupational delineation, we begin with differentiating 3-digit occupations merely by the level of skill requirement (i.e., the 5th digit of the occupational classification), which reduces the number of occupations from 1,286 to only 432. While this robustness check provides a broader definition of occupational labor markets, it is still possible that firms (or workers) substitute away to neighboring occupations to fill (find) their vacancies (job). To address the problem of adequately delineating occupational labor markets more rigorously, we construct a novel flow-based measure of firm-specific labor market tightness (see Appendix \ref{sec:G} for a detailed description). Using weights that build on observed transitions probabilities between 5-digit occupations, our flow-based measure of labor market tightness additionally takes into account vacancies and job seekers from occupations other than the focal occupation.\footnote{If an occupation is classified overly narrow, the weights of vacancies and job seekers in similar but differently classified occupations turn out correspondingly higher, thus minimizing potential measurement error.} In terms of the regional delineation, we also replace our functional delineation of 51 commuting zones by an administrative delineation of 401 districts (i.e, 3-digit NUTS regions). In each of the three robustness checks, the results closely mirror our baseline effects, buttressing that our findings are not driven by a certain labor market definition.

Next, we scrutinize whether wage and tightness effects differ by firm size. Up to now, the coefficients have expressed average effects across firms. First, we weight observations by employment to assign larger firms more importance. The results remain fairly robust. Second, we differentiate between three establishment size categories: small (1-9 workers), medium-sized (10-99), and large establishments (more than 100 workers). Small establishments feature a less negative wage elasticity (-0.5) than the average but the tightness effect is only slightly smaller than the effect in the baseline estimation. Medium-sized establishments exhibit above-average effects (-0.9 and -0.08). The elasticities of large establishments resemble those from the overall sample. In all three size classes, the ratio of the tightness to the wage effect is close to 1/14, as implied by our baseline estimates.

We additionally differentiate between firms from West and East Germany as well as those with low and high productivity. Labor demand in East Germany reacts more sensitively to wage shifts, mirroring that East Germany lags behind West Germany in terms of productivity \citep{Mueller2013}. But, in line with the literature \citep{Schnabel2016}, the difference between these elasticities is not substantial. The tightness effects in West and East Germany are not significantly different. We approximate productivity by firm fixed effects from log-linear wage regressions in the spirit of Abowd, Kramarz, and Margolis (\citeyear{AbowdEtAl1999}, hereafter `AKM') for the years 2012-2019. These AKM effects reflect a relative wage premium paid to all regular full-time workers within the firm, conditional on individual and year fixed effects. In line with rent sharing, the own-wage elasticity of labor demand for firms with high productivity (i.e., above the median AKM effect) turns out to be less negative than for low productive firms. At the same time, the negative effect of labor market tightness is nearly three times smaller for low-productivity firms, reflecting that labor shortage poses a more severe problem to highly productive firms. We observe descriptively that these highly productive firms expand in terms of employment during 2012-2019 whereas the group of low-productive firms tends to remain stable on average. Hence, the presented AKM heterogeneity suggests that the rise in tightness restricts additional employment growth rather than forcing firms to shrink.

For lack of adequate data, conventional labor demand studies on administrative data from Germany usually report own-wage elasticities of labor demand only for full-time workers. However, we make use of the available IEB information on individual working hours from the years 2012 to 2014 to approximate hourly wage rates (see Appendix \ref{sec:C} for futher details). This approximation enables us to analyze labor demand for (regular or marginal) part-time workers. Table \ref{tab:4} shows own-wage elasticities and tightness effects by labor outcome. In Column (1), we first present results for our baseline of full-time workers but based on our constructed measure of hourly (instead of daily) wage rates. Reassuringly, we arrive at quantitatively similar elasticities for full-time workers, namely -0.713 for wages and -0.048 for tightness. Interestingly, in Column (2), the own-wage elasticity for part-time workers turns out to be only slightly negative and insignificant. While this small elasticity could be driven by measurement error in our hourly wage rate, the result is in line with \citet{FreierSteiner2010} who find that the demand for part-time employees is quite unresponsive to wage changes for male workers in West Germany. In contrast, we observe a significantly negative tightness effect on part-time workers which is similar in size to that of full-time workers.

\begin{table}[!ht]
\centering
\scalebox{0.90}{
\begin{threeparttable}
\caption{Labor Demand Effects by Labor Outcome}
\label{tab:4}
\begin{tabular}{L{4cm}C{5.8cm}C{5.8cm}} \hline
\multirow{3.4}{*}{} & \multirow{3.4}{*}{\shortstack{(1) \\ $\Delta$ Log L$^{\text{FT}}$ }} & \multirow{3.4}{*}{\shortstack{(2) \\ $\Delta$ Log  L$^{\text{PT}}$ }}  \\
&& \\
&& \\[0.2cm] \hline
&& \\[-0.3cm]
\multirow{2.4}{*}{$\Delta$ Log W$^{\text{FT}}$} &  \multirow{2.4}{*}{\shortstack{\hphantom{***}-0.713***\hphantom{-} \\ (0.021)}}  &   \\
&& \\
\multirow{2.4}{*}{$\Delta$ Log W$^{\text{PT}}$} &   & \multirow{2.4}{*}{\shortstack{-0.067\hphantom{-} \\ (0.060)}}     \\
&& \\
\multirow{2.4}{*}{$\Delta$ Log V/U} &  \multirow{2.4}{*}{\shortstack{\hphantom{***}-0.048***\hphantom{-}\\ (0.002)}}  & \multirow{2.4}{*}{\shortstack{\hphantom{***}-0.043***\hphantom{-}\\ (0.002)}}  \\
&& \\[0.2cm]
Fixed Effects  & Year &  Year   \\[0.2cm]  \hline
&& \\[-0.2cm]
Instruments  & Z\(_{\text{W}^{\text{FT}}}\), Z\(_{\text{V}}\), Z\(_{\text{U}}\) & Z\(_{\text{W}^{\text{PT}}}\), Z\(_{\text{V}}\), Z\(_{\text{U}}\)  \\[0.2cm]

Observations &  7,993,993          & 11,448,610                 \\[0.2cm]
Clusters &  1,801,671          & 2,693,588              \\[0.2cm]
F: $\Delta$ Log W$^{\text{FT}}$ & 2,952  &    \\[0.2cm]
F: $\Delta$ Log W$^{\text{PT}}$ &  & 69       \\[0.2cm]
F: $\Delta$ Log V/U & 30,360 & 31,085  \\[0.2cm] \hline
\end{tabular}
\begin{tablenotes}[para]
\footnotesize\textsc{Note. ---} The table displays IV regressions of differences in log employment per establishment on differences in the log of average hourly wages and the log of labor market tightness. The instrumental variables refer to shift-share instruments of biennial national changes in occupations weighted by past occupational employment in the respective establishment. The lag difference is two years. Labor markets are combinations of 5-digit KldB occupations and commuting zones. Full-time employment includes regular full-time workers whereas part-time employment encompasses regular part-time and marginal part-time workers. Standard errors (in parentheses) are clustered at the establishment level. F = F Statistics of Excluded Instruments. FT = Full-Time. KldB = German Classification of Occupations. L = Employment. PT = Part-Time. U = Job Seekers. V = Vacancies. W = Average Hourly Wages. Z = Shift-Share Instrument. * = p$<$0.10. ** = p$<$0.05. *** = p$<$0.01. Sources: Integrated Employment Biographies $\plus$ Official Statistics of the German Federal Employment Agency $\plus$ IAB Job Vacancy Survey, 1999-2019.
\end{tablenotes}
\end{threeparttable}
}
\end{table}

\paragraph{Reallocation Effects.} \citet{Hamermesh1993} emphasizes that firm-level responses overstate aggregate changes in employment to the extent that workers transition between firms within the aggregate.\footnote{ In line, meta-regressions indicate that own-wage elasticities of labor demand at the industry level are smaller than estimates based on wage variation within single firms \citep{LichterEtAl2015}.} In the presence of hiring frictions (i.e., $\eta^{L}_{\theta}<0$), an employer's labor demand decision affects the hiring decision in other firms through search externalities (see Section \ref{sec:2}). In such a setting, the feedback effect on labor market tightness will partly offset any first-round response in labor demand due to reallocation effects. As a consequence, the aggregate own-wage elasticity of labor demand (i.e., including these so-called search or congestion externalities) will be less negative than the own-wage elasticity of labor demand of the single firm to the extent that higher (lower) aggregate employment amplifies (reduces) hiring frictions \citep{BeaudryEtAl2018}.

To gauge the magnitude of the feedback effect, $\omega\,=\, \nu \cdot \eta^{L}_{\theta}$, we estimate the following auxiliary regression, which is a log-linearized version of Equation (\ref{eq:9}):
\begin{equation}
\label{eq:23}
\Delta\,ln\,\theta_{rt} \, = \, \zeta \,\, + \,\, \nu \cdot \Delta \, ln \, L_{rt} \,\,  + \,\, \Delta \, \varepsilon_{rt}
\end{equation}
Specifically, we estimate the impact of changes in aggregate employment, $L_{rt}$, on regional labor market tightness, defined as $\theta_{rt}=\frac{V_{rt}}{U_{rt}}$. To rule out bias from reverse causality, we construct a conventional Bartik instrument for employment at the regional level: $Z_{L_{rt}}  \,=\, \sum_{o=1}^{O} \frac{L_{ro\tau}}{L_{r\tau}}  \cdot \Delta \, ln \, L_{ot}$.\footnote{As the auxiliary regressions do not refer to the firm but the more aggregated regional level, we construct traditional Bartik instruments at the level of commuting zones.} The regressions refer to the years 2012-2019 and we set the base period $\tau$ at year 1999 to ensure predetermined employment shares. We specify regions in terms of our 51 commuting zones and estimate Equation (\ref{eq:23}) in one-year differences.

Columns (1) and (2) in Table \ref{tab:5} display the IV results of the feedback effect. In line with theory, the feedback effect of aggregate employment on labor market tightness, $\nu$, turns out to be significantly positive both for regular full-time and for part-time workers. Specifically, we find that a 1 percent increase in regional employment of regular full-time workers raises labor market tightness by 9.3 percent. With a value of 10.4, we arrive at a similar order of magnitude for part-time workers. Such a positive impact of aggregate employment on labor market tightness gives rise to a self-dampening feedback cycle when aggregate employment shifts. Both effects are in line with the descriptive observation that employment increased by roughly 10 percent during our period of analysis while labor market tightness increased by about 100 percent.

\begin{table}[!ht]
\centering
\scalebox{0.90}{
\begin{threeparttable}
\caption{Feedback Regression of Tightness on Aggregate (Un-)Employment}
\label{tab:5}
\begin{tabular}{L{4cm}C{2.7cm}C{2.7cm}C{2.7cm}C{2.7cm}} \hline
\multirow{3.4}{*}{} & \multirow{3.4}{*}{\shortstack{(1) \\ $\Delta$ Log V/U }} & \multirow{3.4}{*}{\shortstack{(2) \\ $\Delta$ Log V/U}} & \multirow{3.4}{*}{\shortstack{(3) \\ $\Delta$ Log U}}  & \multirow{3.4}{*}{\shortstack{(4) \\ $\Delta$ Log U}}   \\
&&&& \\
&&&& \\[0.2cm] \hline
&&&& \\[-0.3cm]
\multirow{2.4}{*}{$\Delta$ Log L$^{\text{FT}}$} &   \multirow{2.4}{*}{\shortstack{\hphantom{***}9.285*** \\ (0.969)}} &   & \multirow{2.4}{*}{\shortstack{\hphantom{***}-4.039***\hphantom{-} \\ (0.386)}}  & \\
&&&& \\
\multirow{2.4}{*}{$\Delta$ Log L$^{\text{PT}}$} &   &  \multirow{2.4}{*}{\shortstack{\hphantom{***}10.36*** \\ (1.697)}} &   &  \multirow{2.4}{*}{\shortstack{\hphantom{***}-4.264***\hphantom{-} \\ (0.667)}}    \\
&&&& \\[0.2cm] \hline
&&&& \\[-0.2cm]
Instruments  & Z\(_{\text{L}^{\text{FT}}}\) & Z\(_{\text{L}^{\text{PT}}}\) & Z\(_{\text{L}^{\text{FT}}}\) & Z\(_{\text{L}^{\text{PT}}}\)  \\[0.2cm]
Observations &   357               &      357            &   357               &      357                 \\[0.2cm]
Clusters &     51              &         51            &     51              &         51             \\[0.2cm]
F: $\Delta$ Log L$^{\text{FT}}$ & 155    &  & 155    &  \\[0.2cm]
F: $\Delta$ Log L$^{\text{PT}}$ &  &  65     &  &  65     \\[0.2cm] \hline
\end{tabular}
\begin{tablenotes}[para]
\footnotesize\textsc{Note. ---} The table displays IV regressions of differences in log labor market tightness per commuting zone on differences in the log of aggregate full-time/part-time (un-)employment in the respective commuting zone. The instrumental variables refer to shift-share instruments of yearly national changes in occupations weighted by occupational employment in the respective commuting zone as of 1999. The lag difference is one year. Labor markets are combinations of 5-digit KldB occupations and commuting zones. Full-time employment includes regular full-time workers whereas part-time employment encompasses regular part-time and marginal part-time workers. Standard errors (in parentheses) are clustered at the commuting-zone level. F = F Statistics of Excluded Instruments. FT = Full-Time. KldB = German Classification of Occupations. L = Employment. LM = Labor Market. PT = Part-Time. U = Job Seekers. V = Vacancies. Z = Shift-Share Instrument. * = p$<$0.10. ** = p$<$0.05. *** = p$<$0.01. Sources: Integrated Employment Biographies $\plus$ Official Statistics of the German Federal Employment Agency $\plus$ IAB Job Vacancy Survey, 1999-2019.
\end{tablenotes}
\end{threeparttable}
}
\end{table}

To check the plausibility of these values, we make use of Equation (\ref{eq:10}) and decompose $\nu$ into the elasticity of matching with respect to the stock of vacancies, $1-\mu$, and the elasticity of regional number of job seekers with respect to regional employment, $\frac{\partial \,ln \,U}{\partial \,ln \,L}$.  We quantify the latter effect by estimating the impact of regional employment on the number of job seekers per region in an analogous specification to (\ref{eq:23}). As before, we make use of the traditional Bartik instrument for regional employment. Columns (3) and (4) in Table \ref{tab:5} show that higher employment significantly reduces the number of job seekers in a region. Quantitatively, an increase in full-time (part-time) employment by 1 percent is associated with a reduction in job seekers by 4.0 (4.3) percent. Given the estimates from Table \ref{tab:5}, we solve Equation (\ref{eq:10}) for $1-\mu$. We arrive at a matching elasticity with respect to the stock of vacancies of 0.54 for full-time and 0.51 for part-time workers.\footnote{$1-\mu = \frac{1-\frac{\partial ln U}{\partial ln L}}{\nu}=\frac{1-(-4.039)}{9.285}=0.54$ for full-time and $\frac{1-(-4.264)}{10.36}=0.51$ for part-time workers.} Favorably, these implied matching elasticities are well in line with the empirical literature on the matching function. Specifically, Fahr and Sunde (\citeyear{FahrSunde2006a,FahrSunde2006b}) estimate matching functions from regional panel data of Germany and find matching elasticities in the range between 0.4 and 0.5 for the stock of vacancies.

Finally, we insert our estimates $\hat{\eta}^{L}_{W}$ and $\hat{\eta}^{L}_{\theta}$ from Table \ref{tab:4} as well as $\hat{\nu}$ from Table \ref{tab:5} into Equation (\ref{eq:11}) to calculate the aggregate own-wage elasticity of labor demand. Thus, by virtue of the self-correcting feedback mechanism, the individual-firm own-wage elasticity of labor demand for full-time workers shrinks from -0.71 to -0.49 when accounting for search externalities at the aggregate level.\footnote{$\tilde{\eta}^{L}_{W} = \frac{\eta^{L}_{W}}{\,\,\,1 \,-\, ( \nu \cdot \eta^{L}_{\theta} ) \,\,\,} =\vphantom{\Big(}\frac{-0.713}{1-(-0.048 \,\cdot\, 9.285)}=-0.49$ for full-time and $\frac{-0.067}{1-(-0.043 \,\cdot\, 10.36)}=-0.05$ for part-time workers.} In a similar fashion, the aggregate own-wage elasticity of labor demand for part-time workers shrinks from -0.07 to -0.05. In both cases, by factoring in the feedback cycle, the wage elasticities shrink by 30.8 percent. Overall, the feedback cycle follows an infinite geometric series, but it effectively dies off after two cycles (i.e., the converging value is only about 10 percent off the limit value after two periods).

\section{Discussion}
\label{sec:7}

In this section, we discuss the implications of our findings in three further analyses. For more detailed information on the analyses, we refer the reader to Appendix \ref{sec:H}.

\paragraph{Minimum Wage Introduction in 2015.} In Germany, a national minimum wage was introduced on January 1, 2015. We use our estimated own-wage elasticities of labor demand for full- and part-time workers to simulate the employment effects of this policy. Using a worker-level difference-in-difference-in-difference specification, we estimate that the hourly minimum wage of 8.50 Euro raised the aggregate wage level by 0.7 percent for full-time and 3.3 percent for part-time workers, respectively (see Table \ref{tab:H1}). We multiply these effects with our estimated own-wage elasticities of labor demand to arrive at the aggregate minimum wage effect on employment. Our baseline simulation absent reallocation effects (i.e., when using the individual-firm own-wage elasticities for full- and part-time workers from Table \ref{tab:4}) yields a negative effect on employment of $-126,299$ workers (see Table \ref{tab:H1}). However, this effect disregards that an aggregate reduction in labor demand also lowers labor market tightness via search externalities. When incorporating this feedback cycle (i.e., by using the aggregate own-wage elasticities of labor demand), the disemployment effect reduces to $-87,844$ workers due to reallocation of workers across firms. Overall, this effect mirrors evidence from ex-post evaluations of the 2015 minimum wage which unanimously find that employment effects were small \citep{BosslerGerner2020,CaliendoEtAl2019}.

The literature offers several explanations to rationalize the absence of large disemployment effects of minimum wages \citep{Schmitt2015}, namely changes along the hours margin, product price adjustments, productivity increases, non-compliance, and monopsony power. Our analysis provides an additional explanation for this puzzle: after an aggregate reduction in employment, search externalities lower labor market tightness which, in turn, facilitates recruitment for firms. This mechanism closely mirrors findings from \citet{DustmannEtAl2022}, who show that most of the firm-level employment reduction is offset by reallocation of workers to competing employers. Our estimates suggest that reallocation of workers reduces the disemployment effect of minimum wages by around 30 percent.

\paragraph{Employment Trends and Labor Market Tightness.} In Section \ref{sec:5}, we have shown that the German labor market has considerably tightened between 2012 and 2019. Specifically, labor market tightness doubled within only seven years. Our baseline results in Table \ref{sec:3} imply that a 100 percent increase in tightness lowers firm-level employment of full- and part-time workers by around 5 percent. Using these elasticities, we quantify the impact of the doubling in labor market tightness on aggregate employment in Germany for the period of our analysis. Specifically, in a counterfactual analysis, we compare the observed aggregate employment growth with a hypothetical scenario in which labor market tightness had not changed (i.e., we fix the ratio of vacancies to job seekers at its 2012 level). In this hypothetical scenario, we multiply the observed relative change in tightness by our estimated elasticities of labor demand with respect to tightness (separately for full- and part-time workers), and subtract the respective sum from the factual stock of employment.

While observed total employment rose from 32.9 million employees in 2012 to 36.7 million jobs in 2019, our simulation implies that it could have risen to 37.8 million jobs if labor market tightness had not changed. Thus, in the absence of increasing labor market tightness, employment could have grown by an additional 1.1 million jobs until 2019. Overall, our results imply that the increase in labor market tightness considerably dampened the positive employment trend, underlining the importance of hiring frictions in tight labor markets.

\paragraph{Wage and Skill Concessions.} Finally, firms facing higher labor market tightness do not necessarily have to settle for lower employment levels. Instead, these firms could still manage to retain or expand their workforce by making concessions, for instance, by raising wages or by recruiting workers with lower skills. We empirically address the conjecture that firms need to make concessions to maintain their employment (growth) in tight labor markets. Building on the same instrumental variable approach as in our analysis of labor demand, we regress the average wage level of firms and the fraction of unskilled workers in a firm on our measure of labor market tightness (see Table \ref{tab:H3}).

On average, the doubling in tightness raises average wages of full-time workers in a firm by almost 1 percent. While the wage response is significantly positive, the magnitude of this effect is fairly small, namely just about a fifth of the negative employment response.\footnote{Note that our baseline estimates in Column (4) of Table \ref{tab:1} do not capture wage adjustments in the course of an increased tightness because we are conditioning on wages while estimating the effect of tightness on labor demand. However, even when discarding the wage level in Column (3) of Table \ref{tab:1}, the tightness effect on employment shows a similar order of magnitude, reflecting the relatively small extent of wage concessions.} However, firms' positive but small wage response is in line with the empirical literature on the wage curve for Germany, which relates wages to the unemployment rate \citep{BaltagiEtAl2009,BellmannBlien2001}. Regarding skill demand, we also observe only a limited extent of concessions. Starting from an average share of 6.4 percent, our results imply that the doubling in tightness raised the share of low-skill workers in firms' employment by only 0.3 percentage points. Overall, the estimates suggest that the extent of firms' wage and skill concessions was fairly small in practice, providing an explanation for the markedly negative effect of labor market tightness on employment.

\section{Conclusion}
\label{sec:8}

We develop a labor demand model in which employers' labor demand depends not only on wages but also on pre-match hiring cost arising from tight labor markets. In light of our model, we determine the effect of wages and labor market tightness on firms' demand for labor by leveraging the universe of administrative employment records in Germany along with official statistics and survey data on vacancies and job seekers. To address issues of endogeneity, we construct novel Bartik instruments at the firm level. We take advantage of the fact that, due to their occupational composition, firms are differently exposed to shocks at the national level. We report an own-wage elasticity of demand for full-time workers of -0.7. Further, we find that the observed doubling in labor market tightness between 2012 and 2019 reduced firms' employment by 5 percent.

Our finding that the vacancy-to-job-seeker ratio amplifies hiring frictions highlights the relevance of search externalities: A reduction in labor demand by one firm improves the recruitment opportunities for all other firms in the same market. As a consequence, aggregate changes in labor demand (e.g. due to a change in wages) alter labor market tightness which, in turn, gives rise to a self-weakening feedback cycle. When incorporating the negative feedback effect via search externalities, the aggregate own-wage elasticity of labor demand reduces to -0.5, which is a reduction by 30 percent from reallocation effects. This mechanism helps to reconcile evidence on the wage elasticity of labor demand with ex-post evaluations of minimum wages: Despite evidence that firms reduce their labor demand when facing higher wages, the reallocation of workers to other firms facilitates explaining why minimum wages are frequently found to have only limited disemployment effects \citep{DustmannEtAl2022}.

Our effects allow us to shed more light on the importance and magnitude of pre-match hiring cost. So far, the literature has determined hiring cost from either survey questions or dynamic labor demand models which harness the sluggishness of labor demand responses. By modeling labor market tightness as the main determinant of pre-match hiring cost, our static profit-maximization model offers a new alternative to quantify pre-match hiring cost without the necessity of modelling dynamics. By calibrating only few model parameters, we find that pre-match hiring cost amount to roughly 40 percent of annual wage payments.

\clearpage

\printbibliography[heading=bibintoc] 
%


\clearpage
\begin{appendix}
\begin{refsection} 

\renewcommand\thetable{\thesection\arabic{table}} 
\renewcommand\thefigure{\thesection\arabic{figure}} 

\pdfbookmark[0]{Appendix}{appendix} 

\begin{center}

\vspace*{1cm}

\Large
Appendix \\
\textbf{Labor Demand on a Tight Leash}

\normalsize

\vspace*{-2cm}

\normalsize
\begin{minipage}[t][5cm][b]{0.48\textwidth}
\begin{center}
\href{https://www.mariobossler.de/}{\textcolor{black}{\textbf{Mario Bossler}}} \\
\href{https://www.th-nuernberg.de/en/faculties/bw/}{\textcolor{black}{TH Nuremberg}}\textcolor{black}{,} \href{https://iab.de/en/startseite-english/}{\textcolor{black}{IAB}}\textcolor{black}{,} \href{https://www.iza.org/}{\textcolor{black}{IZA}} \textcolor{black}{\&} \href{http://www.laser.uni-erlangen.de/index.php}{\textcolor{black}{LASER}}  \\[-0.2cm]
mario.bossler@th-nuernberg.de
\end{center}
\end{minipage}
\begin{minipage}[t][5cm][b]{0.48\textwidth}
\begin{center}
\href{https://iab.de/en/employee/?id=12380200}{\textcolor{black}{\textbf{Martin Popp}}} \\
\href{https://iab.de/en/startseite-english/}{\textcolor{black}{IAB}} \textcolor{black}{\&} \href{http://www.laser.uni-erlangen.de/index.php}{\textcolor{black}{LASER}}  \\[-0.2cm]
martin.popp@iab.de
\end{center}
\end{minipage}

\vspace*{2cm}

\Large
\textbf{Content}
\normalsize

\startcontents[sections] 
\printcontents[sections]{l}{1}{\setcounter{tocdepth}{2}} 

\end{center}

\clearpage

\counterwithin{equation}{section}
\newcommand\ddfrac[2]{\frac{\displaystyle #1}{\displaystyle #2}}
\allowdisplaybreaks

\section{Hiring-Cost Adjusted Version of Fundamental Law of Labor Demand}
\label{sec:A}

In a first step, we derive a hiring-cost adjusted version of the fundamental law of derived demand, that is, we identify the determinants of the own-wage elasticity of labor demand in a model with positive hiring cost.
In a second step, we derive the elasticity of labor demand with respect to labor market tightness, which we model as a determinant of hiring cost.

\paragraph{The Wage Effect on Labor Demand.} We begin with combining the optimality condition for labor (\ref{eq:3}) with our proposed formulation for unit hiring cost (\ref{eq:5}):
\begin{equation}
\label{eq:A1}
\underbrace{\,P \cdot Y_{L}(L,K)\,}_{\substack{\text{marginal value \vphantom{/}} \\ \text{product of labor\vphantom{/}}}} \,\,\,=\,\,  \underbrace{\vphantom{(}\,\,\,\,\,\,\,W\,\,\,\,\,\,\,\vphantom{)}}_{\substack{\text{wage\vphantom{/}} \\ \text{rate}}} \,\,+\,\, \underbrace{(\delta + r) \cdot  c \cdot W^{\phi_1} \cdot \theta^{\phi_2} }_{\substack{\text{amortized\vphantom{/}} \\ \text{pre-match\vphantom{/}} \\ \text{hiring costs\vphantom{/}}}} \,\,+\,\, \underbrace{(\delta + r) \cdot  \Psi \,}_{\substack{\text{amortized\vphantom{/}} \\ \text{post-match\vphantom{/}} \\ \text{hiring costs\vphantom{/}}}} \,\,\equiv\,\, \underbrace{\vphantom{(}\,\,\,\,\,\,W^{\text{\textasteriskcentered}\,\,\,\,\,\,\,\vphantom{)}}}_{\substack{\text{unit\vphantom{/}} \\ \text{ labor cost\vphantom{/}} }}
\end{equation}
Thus, the marginal value product of labor must equal the unit cost of labor $W^{\text{\textasteriskcentered}}$ which is the sum of the wage rate, the amortized pre-match hiring costs, and amortized post-match hiring costs. In the long run, the firm can also optimize the level of the capital stock. Thus, the marginal value product of capital must equal the capital rate:
\begin{equation}
\label{eq:A2}
P \cdot Y_{K}(L,K) \,=\,  R
\end{equation}
Moreover, we assume that the product market is cleared:
\begin{equation}
\label{eq:A3}
\underbrace{Y(L,K)}_{\substack{\text{product} \\ \text{supply}}} \,=\, \underbrace{Y^{d}(P)}_{\substack{\text{product}\\ \text{demand}}}
\end{equation}
We totally differentiate Equations (\ref{eq:A1}), (\ref{eq:A2}), and (\ref{eq:A3}) with respect to the wage rate $W$:
\vspace*{-0.25cm}
\begin{align}
\label{eq:A4}
Y_{L} \cdot P_{W} + P \cdot (Y_{LL} \cdot L_W + Y_{LK} \cdot K_W) \,&=\, 1 \,+\, \phi_1 \cdot (\delta + r) \cdot c \cdot W^{\phi_{1}-1} \cdot \theta^{\phi_{2}} \\
\label{eq:A5}
Y_{K} \cdot P_{W} + P \cdot (Y_{KL} \cdot L_W + Y_{KK} \cdot K_W) \,&=\, 0
 \\
\label{eq:A6}
  Y_{L} \cdot L_{W} + Y_{K} \cdot K_{W} + \eta^{Y}_{P} \cdot P_{W} \cdot  \frac{Y}{P}  \,&=\, 0
\end{align}
Given the definition of the price elasticity of product demand,  $\eta^{Y}_{P} = - Y_{P} \cdot \frac{P}{Y}$, the derivative of the right-hand side of Equation (\ref{eq:A3}), $Y_{P} \cdot P_{W}$, can be rearranged to $-\eta^{Y}_{P} \cdot \frac{Y}{P} \cdot P_{W}$, culminating in Equation (\ref{eq:A6}).

In the following, we assume that the production technology is characterized by constant returns to scale. As a consequence, the following two properties apply:
\begin{align}
\label{eq:A7}
Y_{LL} \,&=\, - \frac{K}{L} \cdot Y_{LK} \\
\label{eq:A8}
Y_{KK} \,&=\, - \frac{L}{K} \cdot Y_{LK}
\end{align}
Under the additional assumption of perfect competition, Euler's product exhaustion theorem holds:
\begin{equation}
\label{eq:A9}
Y = L \cdot Y_{L} + K \cdot Y_{K}
\end{equation}
The elasticity of substitution describes the percentage change in the relative use of labor and capital, when the marginal productivity ratio between both factors increases by 1 percent. Intuitively, the elasticity describes the ease of substituting labor by capital without changing output. Given Equations (\ref{eq:A7}), (\ref{eq:A8}), and (\ref{eq:A9}), the elasticity of substitution becomes:
\begin{equation}
\label{eq:A10}
\sigma \,\equiv\, \frac{ln \, \frac{K}{L}}{\,ln \, \frac{Y_L}{Y_K}\,} \,=\, \frac{Y_{L} \cdot Y_{K}}{Y \cdot Y_{LK}} \,\geq\, 0
\end{equation}

Inserting Equations (\ref{eq:A1}), (\ref{eq:A2}), (\ref{eq:A7}), (\ref{eq:A8}), and (\ref{eq:A10}) into the system of total derivatives (\ref{eq:A4}), (\ref{eq:A5}), and (\ref{eq:A6}) yields after rearrangement:
\begin{align}
\label{eq:A11}
Y \cdot \sigma \cdot P_{W} - \frac{K}{L} \cdot R \cdot L_{W} +  R \cdot K_{W} \,&=\, Y \cdot P \cdot \sigma \cdot \frac{1 \,+\, \phi_1 \cdot (\delta + r) \cdot c \cdot W^{\phi_{1}-1} \cdot \theta^{\phi_{2}}}{W^{\text{\textasteriskcentered}}} \\
\label{eq:A12}
Y \cdot \sigma \cdot P_{W} +  W^{\text{\textasteriskcentered}} \cdot L_{W} - \frac{L}{K} \cdot W^{\text{\textasteriskcentered}} \cdot K_{W} \,&=\, 0 \\
\label{eq:A13}
Y \cdot \eta^{Y}_{P} \cdot P_{W} + W^{\text{\textasteriskcentered}} \cdot L_{W} +  R \cdot K_{W} \,&=\, 0
\end{align}
In matrix notation, the system looks as follows:
\begin{equation}
\label{eq:A14}
\underbrace{
\begin{pmatrix}
Y \cdot \sigma      & -\frac{K}{L} \cdot R   & R \\
Y \cdot \sigma      & W^{\text{\textasteriskcentered}}                      & -\frac{L}{K} \cdot W^{\text{\textasteriskcentered}} \\
\,Y \cdot \eta^{Y}_{P} & W^{\text{\textasteriskcentered}}                      & R
 \end{pmatrix}}_{\textbf{A}}
\cdot
\underbrace{
\begin{pmatrix}
P_{W} \\
L_{W} \\
K_{W}
 \end{pmatrix}}_{\textbf{x}}
=
\underbrace{
\begin{pmatrix}
\,\,Y \cdot P \cdot \sigma \cdot \frac{1 \,+\, \phi_1 \, (\delta + r) \, c \, W^{\phi_{1}-1} \, \theta^{\phi_{2}}}{W^{\text{\textasteriskcentered}}} \,\, \\
0 \\
0
 \end{pmatrix}}_{\textbf{a}}
\end{equation}
Given this equation system, we calculate the derivative of unconditional labor demand with respect to the wage rate using Cramer's rule: $L_{W} = \frac{|\textbf{A}_{2}|}{|\textbf{A}|}$. To arrive at the numerator matrix \(\textbf{A}_{2}\), we replace the second column in the denominator matrix \(\textbf{A}\) by \(\textbf{a}\):

\begin{equation}
\label{eq:A15}
 L_{W} = \frac{|\textbf{A}_{2}|}{|\textbf{A}|}=
\frac{
\begin{vmatrix}
Y \cdot \sigma      & Y \cdot P \cdot \sigma \cdot \frac{1 \,+\, \phi_1 \, (\delta + r) \, c \, W^{\phi_{1}-1} \, \theta^{\phi_{2}}}{W^{\text{\textasteriskcentered}}}   & R \\
Y \cdot \sigma      & 0                     & -\frac{L}{K} \cdot W^{\text{\textasteriskcentered}} \\
\,\,\,Y \cdot \eta^{Y}_{P} & 0                      & R
\end{vmatrix}
}{
\begin{vmatrix}
Y \cdot \sigma      & -\frac{K}{L} \cdot R   & R \\
Y \cdot \sigma      & W^{\text{\textasteriskcentered}}                      & -\frac{L}{K} \cdot W^{\text{\textasteriskcentered}} \\
\,Y \cdot \eta^{Y}_{P} & W^{\text{\textasteriskcentered}}                      & R
\end{vmatrix}}
\end{equation}
Given the rule of Sarrus, the ratio of determinants becomes:
\begin{equation}
\label{eq:A16}
L_{W} \,=\, \frac{-Y^{2} \cdot P \cdot \sigma \cdot \frac{1 \,+\, \phi_1 \, (\delta + r) \, c \, W^{\phi_{1}-1} \, \theta^{\phi_{2}}}{W^{\text{\textasteriskcentered}}} \cdot W^{\text{\textasteriskcentered}} \cdot \frac{L}{K} \cdot \eta^{Y}_{P} - Y^{2} \cdot \sigma^{2} \cdot P \cdot R \cdot \frac{1 \,+\, \phi_1 \, (\delta + r) \, c \, W^{\phi_{1}-1} \, \theta^{\phi_{2}}}{W^{\text{\textasteriskcentered}}}  }{2 \cdot R \cdot Y \cdot \sigma \cdot W^{\text{\textasteriskcentered}} + Y \cdot \eta^{Y}_{P} \cdot R \cdot W^{\text{\textasteriskcentered}} - W^{\text{\textasteriskcentered}} \cdot R \cdot Y \cdot \eta^{Y}_{P} + \frac{L}{K} \cdot Y \cdot \sigma \cdot (W^{\text{\textasteriskcentered}})^2  + Y \cdot \sigma \cdot \frac{K}{L} \cdot R^{2} }
\end{equation}
Building on Equations (\ref{eq:A1}), (\ref{eq:A2}), and (\ref{eq:A9}), the derivative simplifies to:
\begin{align}
\label{eq:A17}
L_{W} \,&=\, \ddfrac{\frac{-Y^{2}  \cdot P \cdot \sigma \cdot ( R \cdot K \cdot \sigma + W^{\text{\textasteriskcentered}} \cdot L \cdot \eta^{Y}_{P})}{ K \cdot \frac{W^{\text{\textasteriskcentered}}}{1 \,+\, \phi_1 \, (\delta + r) \, c \, W^{\phi_{1}-1} \, \theta^{\phi_{2}}}  }  }{  \frac{P^{2} \cdot \sigma \cdot Y^{3} }{L \cdot K}  } \\[0.5cm]
\label{eq:A18}
\,&=\, - L \cdot \frac{1 \,+\, \phi_1 \cdot (\delta + r) \cdot c \cdot W^{\phi_{1}-1} \cdot \theta^{\phi_{2}}}{W^{\text{\textasteriskcentered}}} \cdot \Bigg( \,\underbrace{\frac{R \cdot K}{P \cdot Y}}_{1-s_{L}} \cdot \, \sigma \,+\,  \underbrace{\frac{W^{\text{\textasteriskcentered}} \cdot L}{P \cdot Y}}_{s_{L}} \cdot \, \eta^{Y}_{P} \, \Bigg)  \\[0.5cm]
\label{eq:A19}
\,&=\, L \cdot \frac{1 \,+\, \phi_1 \cdot (\delta + r) \cdot c \cdot W^{\phi_{1}-1} \cdot \theta^{\phi_{2}}}{W^{\text{\textasteriskcentered}}} \cdot \Big( \,-(1-s_{L}) \cdot \sigma \,-\,  s_{L} \cdot \eta^{Y}_{P} \, \Big)
\end{align}
Finally, multiplying Equation (\ref{eq:A19}) by $\frac{W}{L}$ yields the unconditional own-wage elasticity of labor demand. In doing so, we arrive at a hiring-cost adjusted version of the fundamental law of labor demand:
\begin{equation}
\label{eq:A20}
\eta^{L}_{W} \,=\,\, \underbrace{\Bigg(\,\frac{W}{\,W^{\text{\textasteriskcentered}}\,} \,+\, \phi_{1} \cdot \frac{ (\delta + r) \cdot c \cdot W^{\phi_{1}} \cdot \theta^{\phi_{2}}}{W^{\text{\textasteriskcentered}}} \Bigg)}_{\substack{\text{elasticity of unit labor cost\vphantom{/}} \\ \text{with respect to\vphantom{/}} \\ \text{wage rate\vphantom{/}}}} \,\cdot\, \underbrace{\vphantom{\Bigg(}\Big( \,-(1-s_{L}) \cdot \sigma \,-\, s_{L} \cdot \eta^{Y}_{P}  \, \Big)\vphantom{\Bigg)}}_{\substack{\text{elasticity of labor demand\vphantom{/}} \\ \text{with respect\vphantom{/}} \\ \text{to unit labor cost\vphantom{/}}}}
\end{equation}
The standard version (\ref{eq:6}) of the fundamental law of labor demand \citep{Hamermesh1993} represents the elasticity of labor demand with respect to unit labor cost. In our model, however, the prevalence of positive hiring cost drives a wedge between unit labor cost $W^{\text{\textasteriskcentered}}$ (i.e., the cost of an additional unit labor) and the wage rate $W$. Thus, in the hiring-cost adjusted formulation (\ref{eq:A20}), the standard version of the fundamental law of labor demand is scaled by the elasticity of unit labor cost with respect to the wage rate. This elasticity is a weighted sum of the relative wage effects on the three components of unit labor cost: namely the wage rate (entering with elasticity 1), pre-match hiring costs (entering with elasticity $\phi_{1}$), and post-match hiring costs (entering with elasticity 0). The weights refer to the share of the wage rate, the share of amortized pre-match hiring costs, and the share of amortized post-match hiring costs in unit labor cost. By and large, the hiring-cost adjusted version of the fundamental law of labor demand implies that labor demand reacts more elastically to wage changes, ...
\begin{itemize}[leftmargin=2cm]
\item[1.] ... the higher the share of the wage rate in unit labor cost, $\frac{W}{\,W^{\text{\textasteriskcentered}}}$, and
\item[2.] ... the higher the elasticity of pre-match hiring cost with respect to the wage rate, $\phi_{1}$, and
\item[3.] ... the higher the share of amortized pre-match hiring costs in unit labor cost, $\frac{ (\delta + r) \, c \, W^{\phi_{1}} \,\theta^{\phi_{2}}}{W^{\text{\textasteriskcentered}}}$, provided that $\phi_{1}>0$, and
\item[4.] ... the higher the elasticity of labor demand with respect to unit labor cost.
\end{itemize}
In the absence of both pre-match hiring costs, $c=0$, and post-match hiring costs, $\Omega=0$, unit labor cost equal the wage rate, $W^{\text{\textasteriskcentered}}=W$, and equation (\ref{eq:A20}) would collapse to (\ref{eq:6}).

\paragraph{The Tightness Effect on Labor Demand.} In addition to the wage rate, our model also allows us to derive the elasticity of labor demand with respect to labor market tightness. For this purpose, we totally differentiate Equations (\ref{eq:A1}), (\ref{eq:A2}), and (\ref{eq:A3}) with respect to $\theta$:
\begin{align}
\label{eq:A21}
Y_{L} \cdot P_{\theta} + P \cdot (Y_{LL} \cdot L_{\theta} + Y_{LK} \cdot K_{\theta}) \,&=\,  \phi_2 \cdot (\delta + r) \cdot c \cdot W^{\phi_{1}} \cdot \theta^{\phi_{2}-1} \\
\label{eq:A22}
Y_{K} \cdot P_{\theta} + P \cdot (Y_{KL} \cdot L_{\theta} + Y_{KK} \cdot K_{\theta}) \,&=\, 0
 \\
\label{eq:A23}
  Y_{L} \cdot L_{\theta} + Y_{K} \cdot K_{\theta} + \eta^{Y}_{P} \cdot P_{\theta} \cdot  \frac{Y}{P}  \,&=\, 0
\end{align}
Using (\ref{eq:A1}), (\ref{eq:A2}), (\ref{eq:A7}), (\ref{eq:A8}), and (\ref{eq:A10}), we rearrange the equation system as follows:
\begin{align}
\label{eq:A24}
Y \cdot \sigma \cdot P_{\theta} - \frac{K}{L} \cdot R \cdot L_{\theta} +  R \cdot K_{\theta} \,&=\, Y \cdot P \cdot \sigma \cdot \frac{ \phi_2 \cdot (\delta + r) \cdot c \cdot W^{\phi_{1}} \cdot \theta^{\phi_{2}-1}}{W^{\text{\textasteriskcentered}}} \\
\label{eq:A25}
Y \cdot \sigma \cdot P_{\theta} +  W^{\text{\textasteriskcentered}} \cdot L_{\theta} - \frac{L}{K} \cdot W^{\text{\textasteriskcentered}} \cdot K_{\theta} \,&=\, 0 \\
\label{eq:A26}
Y \cdot \eta^{Y}_{P} \cdot P_{\theta} + W^{\text{\textasteriskcentered}} \cdot L_{\theta} +  R \cdot K_{\theta} \,&=\, 0
\end{align}
In matrix notation, the system of derivatives is:
\begin{equation}
\label{eq:A27}
\underbrace{
\begin{pmatrix}
Y \cdot \sigma      & -\frac{K}{L} \cdot R   & R \\
Y \cdot \sigma      & W^{\text{\textasteriskcentered}}                      & -\frac{L}{K} \cdot W^{\text{\textasteriskcentered}} \\
\,Y \cdot \eta^{Y}_{P} & W^{\text{\textasteriskcentered}}                      & R
 \end{pmatrix}}_{\textbf{B}}
\cdot
\underbrace{
\begin{pmatrix}
P_{\theta} \\
L_{\theta} \\
K_{\theta}
 \end{pmatrix}}_{\textbf{x}}
=
\underbrace{
\begin{pmatrix}
\,\,Y \cdot P \cdot \sigma \cdot \frac{\phi_2 \, (\delta + r) \, c \, W^{\phi_{2}} \, \theta^{\phi_{2}-1}}{W^{\text{\textasteriskcentered}}} \,\, \\
0 \\
0
 \end{pmatrix}}_{\textbf{b}}
\end{equation}
As before, we calculate the derivative of unconditional labor demand using Cramer's rule, the rule of Sarrus, and Equations (\ref{eq:A1}), (\ref{eq:A2}), and (\ref{eq:A9}):
\begin{align}
\label{eq:A28}
 L_{\theta} &= \frac{|\textbf{B}_{2}|}{|\textbf{B}|}=
\frac{
\begin{vmatrix}
Y \cdot \sigma      & Y \cdot P \cdot \sigma \cdot \frac{ \phi_2 \, (\delta + r) \, c \, W^{\phi_{1}} \, \theta^{\phi_{2}-1}}{W^{\text{\textasteriskcentered}}}   & R \\
Y \cdot \sigma      & 0                     & -\frac{L}{K} \cdot W^{\text{\textasteriskcentered}} \\
\,\,\,Y \cdot \eta^{Y}_{P} & 0                      & R
\end{vmatrix}
}{
\begin{vmatrix}
Y \cdot \sigma      & -\frac{K}{L} \cdot R   & R \\
Y \cdot \sigma      & W^{\text{\textasteriskcentered}}                      & -\frac{L}{K} \cdot W^{\text{\textasteriskcentered}} \\
\,Y \cdot \eta^{Y}_{P} & W^{\text{\textasteriskcentered}}                      & R
\end{vmatrix}} \\[0.5cm]
\label{eq:A29}
 \,&=\, \ddfrac{\frac{-Y^{2}  \cdot P \cdot \sigma \cdot ( R \cdot K \cdot \sigma + W^{\text{\textasteriskcentered}} \cdot L \cdot \eta^{Y}_{P})}{ K \cdot \frac{W^{\text{\textasteriskcentered}}}{ \phi_2 \, (\delta + r) \, c \, W^{\phi_{1}} \, \theta^{\phi_{2}-1}}  }  }{  \frac{P^{2} \cdot \sigma \cdot Y^{3} }{L \cdot K}  } \\[0.5cm]
\label{eq:A30}
\,&=\, L \cdot \frac{\phi_2 \cdot (\delta + r) \cdot c \cdot W^{\phi_{1}} \cdot \theta^{\phi_{2}-1}}{W^{\text{\textasteriskcentered}}} \cdot \Big( \,-(1-s_{L}) \cdot \sigma \,-\,  s_{L} \cdot \eta^{Y}_{P} \, \Big)
\end{align}
In a last step, we multiply Equation (\ref{eq:A30}) by $\frac{\theta}{L}$ to arrive at the elasticity of labor demand with respect to labor market tightness:
\begin{equation}
\label{eq:A31}
\eta^{L}_{\theta} \,=\,\, \underbrace{ \phi_{2} \cdot \frac{ (\delta + r) \cdot c \cdot W^{\phi_{1}} \cdot \theta^{\phi_{2}}}{W^{\text{\textasteriskcentered}}} }_{\substack{\text{elasticity of unit labor cost\vphantom{/}} \\ \text{with respect to\vphantom{/}} \\ \text{labor market tightness\vphantom{/}}}} \,\cdot\, \underbrace{\vphantom{\frac{1}{W^{\text{\textasteriskcentered}}}}\Big( \,-(1-s_{L}) \cdot \sigma \,-\, s_{L} \cdot \eta^{Y}_{P}  \, \Big)}_{\substack{\text{elasticity of labor demand\vphantom{/}} \\ \text{with respect to\vphantom{/}} \\ \text{unit labor cost\vphantom{/}}}}
\end{equation}
Analogously to (\ref{eq:A20}), the elasticity equals the elasticity of labor demand with respect to unit labor cost multiplied by the elasticity of unit labor cost to labor market tightness. The latter elasticity is the relative effect of labor market tightness on pre-match hiring cost, $\phi_{2}$, weighted by the share of pre-match hiring costs in unit labor cost. Overall, unconditional labor demand reacts more elastically to changes in labor market tightness, ...
\begin{itemize}[leftmargin=2cm]
\item[1.] ... the higher the elasticity of pre-match hiring cost with respect to labor market tightness, $\phi_{2}$, and
\item[2.] ... the higher the share of amortized pre-match hiring costs in unit labor cost, $\frac{ (\delta + r) \, c  \, W^{\phi_{1}} \, \theta^{\phi_{2}}}{W^{\text{\textasteriskcentered}}}$, and
\item[3.] ... the higher the elasticity of labor demand with respect to unit labor cost.
\end{itemize}
In the absence of pre-match hiring cost, $c=0$, or if labor market tightness had no effect on pre-match hiring cost, $\phi_{2}=0$, the elasticity would equal zero, as in standard models of labor demand.

So far, we have postulated that higher labor market tightness purely raises hiring cost. To counteract congestion in the hiring process, the so-called wage curve propagates that increased labor market tightness also makes firms pay higher wages. In the following, we suppose that the wage rate is a function of labor market tightness. Specifically,
\begin{equation}
\label{eq:A32}
 W  \,\,=\,\, \,w \cdot \theta^{\,\gamma}
\end{equation}
where $w>0$ and $\gamma \geq0$. Given this relationship, we differentiate Equations (\ref{eq:A1}), (\ref{eq:A2}), and (\ref{eq:A3}) with respect to $\theta$ and reformulate the equation system as follows:
\begin{equation}
\label{eq:A33}
\underbrace{
\begin{pmatrix}
Y \cdot \sigma      & -\frac{K}{L} \cdot R   & R \\
Y \cdot \sigma      & W^{\text{\textasteriskcentered}}                      & -\frac{L}{K} \cdot W^{\text{\textasteriskcentered}} \\
\,Y \cdot \eta^{Y}_{P} & W^{\text{\textasteriskcentered}}                      & R
 \end{pmatrix}}_{\textbf{C}}
\cdot
\underbrace{
\begin{pmatrix}
P_{\theta} \\
L_{\theta} \\
K_{\theta}
 \end{pmatrix}}_{\textbf{x}}
=
\underbrace{
\begin{pmatrix}
\,\,Y \cdot P \cdot \sigma \cdot \frac{\gamma \, w \, \theta^{\gamma-1} \,+\, ( \gamma \, \phi_1 \,+\, \phi_2 ) \, (\delta + r) \, c \, w^{\phi_{1}} \,\theta^{\gamma\,\phi_{1} + \phi_{2}-1} }{W^{\text{\textasteriskcentered}}} \,\, \\
0 \\
0
 \end{pmatrix}}_{\textbf{c}}
\end{equation}
We solve the equation system using Cramer's rule and derive the elasticity of labor demand with respect to tightness, taking into account the wage-curve relationship:
\begin{equation}
\label{eq:A34}
\eta^{L}_{\theta} \,=\,\,  \underbrace{ \Bigg(\, \gamma \cdot \frac{W}{W^{\text{\textasteriskcentered}}} \,+\, \big( \gamma \cdot \phi_{1} \,+\, \phi_{2}\big) \cdot \frac{ (\delta + r) \cdot c \cdot W^{\phi_{1}} \cdot \theta^{\phi_{2}}}{W^{\text{\textasteriskcentered}}} \, \Bigg) }_{\substack{\text{elasticity of unit labor cost\vphantom{/}} \\ \text{with respect to\vphantom{/}} \\ \text{labor market tightness\vphantom{/}}}} \,\cdot\, \underbrace{\vphantom{\Bigg(}\Big( \,-(1-s_{L}) \cdot \sigma \,-\, s_{L} \cdot \eta^{Y}_{P}  \, \Big)\vphantom{\Bigg(}}_{\substack{\text{elasticity of labor demand\vphantom{/}} \\ \text{with respect to\vphantom{/}} \\ \text{unit labor cost\vphantom{/}}}}
\end{equation}
The difference between (\ref{eq:A31}) and (\ref{eq:A34}) is twofold. On the one hand, a 1 percent increase in labor market tightness raises the wage rate by $\gamma$ percent. On the other hand, this wage increase amplifies ($\phi_{1}>0$) or reduces ($\phi_{1}<0$) pre-match hiring cost by $\gamma \cdot \phi_{1}$ percent, because pre-match hiring costs depend on the wage rate which is now a function of tightness. Taking these relationships into account, the second of the aforementioned laws only holds when $ \gamma \cdot \phi_{1} \,+\, \phi_{2}>0$. Further, labor demand also reacts more elastically to changes in labor market tightness, ...
\begin{itemize}[leftmargin=2cm]
\item[4.]... the higher the elasticity of the wage rate with respect to labor market tightness, $\gamma$, provided that $\phi_{1}>-\,\frac{W}{ (\delta + r) \, c \, W^{\phi_{1}} \, \theta^{\phi_{2}}}$ , and
\item[5.]... the higher the share of the wage rate in unit labor cost, $\frac{W}{\,W^{\text{\textasteriskcentered}}}$, and
\item[6.]... the higher the elasticity of pre-match hiring cost with respect to the wage rate, $\phi_{1}$.
\end{itemize}
Finally, note that, if labor market tightness exerted no effect on wage (i.e., $\gamma=0$), Equation (\ref{eq:A34}) collapses to (\ref{eq:A31}).

\clearpage

\clearpage

\section{Empirical Two-Stage Least Squares Specification}
\label{sec:B}

Exploiting the plausibly exogenous variation of the Bartik-instruments, we estimate the empirical model (\ref{eq:10}) using two-stage least squares (2SLS). Provided that the exclusion restrictions are fulfilled, 2SLS estimation yields consistent estimates for the parameters of interest. As we estimate effects of two endogenous variables, we run a system of the following two first-stage regressions
\begin{equation}
\label{eq:B1}
    \Delta \, ln \, W_{it} \,=\, \pi_{10} \,+\, \pi_{11} \cdot Z_{W_{it}} \,+\, \pi_{12} \cdot Z_{V_{it}} \,+\, \pi_{13} \cdot Z_{U_{it}} \,+\, \zeta_{1t} \,+\, \Delta \, \varepsilon_{1it}
\end{equation}
\begin{equation}
\label{eq:B2}
    \Delta \, ln \, \theta_{it} \,=\, \pi_{20} \,+\, \pi_{21} \cdot Z_{W_{it}} \,+\, \pi_{22} \cdot Z_{V_{it}} \,+\, \pi_{23} \cdot Z_{U_{it}} \,+\, \zeta_{2t} \,+\, \Delta \, \varepsilon_{2it}
\end{equation}
where both first-stage equations include our three instruments as well as year effects for each year $t$. The first-stage regressions are designed to extract exogenous variation in the regressors of interest. In the second stage,
\begin{equation}
\label{eq:B3}
    \Delta \, ln \,  L_{it} \, = \, \eta_{0} \,+\, \eta^{L}_{W} \cdot \widehat{\Delta \,ln\, W_{it}} \,+\, \eta^{L}_{\theta} \cdot \widehat{\Delta \,ln\, \theta_{it}} \,+\, \zeta_{t} \,+\, \Delta \,\varepsilon_{it}
\end{equation}
we run our empirical model (\ref{eq:12}) on the predictions from the first-stage regressions. When the exogeneity assumptions hold, the residual term $\Delta \,\varepsilon_{it}$ is uncorrelated with the remaining variation in the variables of interest. Note that all instrumental variable estimates are obtained from Stata's ivreg2 command. Thereby, the inference is automatically adjusted for the two-step procedure of 2SLS.

\clearpage

\section{Data: Further Details}
\label{sec:C}

\setcounter{table}{0} 
\setcounter{figure}{0} 

\paragraph{Wage Imputation.} We apply a two-step imputation technique to impute right-censored wages above the upper earnings limit on social security contributions \citep{CardEtAl2013}. In a first step, we calculate fitted wages from a Tobit regression to generate average wages per establishment (excluding the observation at hand). In a second step, we re-estimate the Tobit regression with this variable as an additional covariate, thereby arriving at final imputations. Specifically, we regress log daily wages of full-time workers on age, (square of) log firm size, share of low- and high-skilled workers in the establishment, share of censored observations excluding the  observation at hand as well as binary variables for single-person firms, firms with more than ten full-time employees, German nationality, 5-digit KldB occupation, and 3-digit NUTS region. Separate Tobit models are estimated for each combination of year (2012-2019), gender (2 groups), and education (3 groups).

\paragraph{Hours Imputation.} By default, the only information about working time in the IEB is a binary variable on whether employees are full-time or part-time workers. However, for the years 2010-2014, the IEB additionally includes information on the number of individual working hours. Unfortunately, an indicator whether firms report actual hours (hours worked) or contractual hours (hours paid) is not available. We therefore apply the heuristic from \citet{DustmannEtAl2022} and harmonize the hours information to depict contractual hours plus overtime. In a next step, we pool the available information for 2012-2014 to impute the hours information for the years 2015-2019. Specifically, for each combination of contract type (5 groups), gender (2 groups), and education (3 groups), we regress daily contractual hours (plus overtime) on a set of individual- and establishment-level covariates and use the fitted models to impute missing information on hours for the years 2015-2019. Finally, we divide daily wages by (imputed) daily contractual hours (incl. overtime) to arrive at hourly wages.

\paragraph{Classification of Occupations.} In our three datasets, we utilize information on the German Classification of Occupations (KldB) from the year 2010. The four leading digits describe the type of occupation whereas the fifth digit designates the level of skill requirement (helper, professional, specialist, or expert). Helper occupations require no training or only a maximum of one year's training. The group of professionals includes all activities with industrial, commercial or other vocational training (excluding master craftsmen and technicians). Specialist occupations necessitate a bachelor degree or the completion of master craftsman/technician training. Experts hold a master degree or an equivalent diploma.

\paragraph{Delineation of Commuting Zones.} Ideally, the definition of an region should capture the spatial dimension of economic flows as accurately as possible. Hence, a so-called ``functional region'' is a set of adjacent places where a large fraction of economic activity (e.g., commuting, trade) of resident workers and firms takes place within its boundaries. However, administrative regions, such as districts, are based on politically determined borders and, thus, do not adequately capture economic activities. As a consequence, a large portion of economic flows do not occur within but across borders of these regions.

Relying on administrative regions may result in a mismeasurement of labor market tightness. Suppose there are two cities with asymmetric commuting patterns, implying that a large fraction of workers in region A usually work for firms in the neighboring region B (but not vice versa). In this setting, calculating labor market tightness in region A based on job seekers only from region A disregards the relevant workers in region B, thus overestimating the true value labor market tightness. By contrast, calculating labor market tightness in region B based on vacancies only from region B will underestimate labor market tightness.

To address this problem, we construct functional labor market regions based on observed home-to-work commuting flows. These commuting zones are designed such that there are many connections within zones, and only few connections between zones, thereby minimizing the above-mentioned measurement problem. Specifically, we employ the graph-theoretical method from \citet{KroppSchwengler2016} in order to merge 401 administrative districts (3-digit NUTS regions) to more appropriate commuting zones. The method builds on three steps: First, the user calculates a matrix of bi-directional (i.e., aggregated in- and outward) commuting flows among the administrative regions and calculates their shares relative to the resident labor force. For each region, the largest share with another region is labelled ``dominant flow'' when the region at hand is smaller than the other region, thus highlighting a potential merger. The regions are merged when the share of the dominant flow exceeds a certain threshold, resulting in a consolidated flow matrix. These steps are iteratively repeated until no additional mergers occur. Second, the first step is performed for a large set of different threshold values, with higher values slowing down the merging process. By altering the threshold, the method proposes many possibly meaningful delineations. Then, the method selects the delineation with the highest value of modularity Q \citep{NewmanGirvan2004} -- a measure that is commonly used in network science.\footnote{The modularity approach compares the number of connections inside a cluster with the expected number of connections if the network of connections between clusters was random. Modularity Q equals zero when the delineation is not better than a random delineation. Q approaches the maximum of $Q=1$ when the network is strongly modular (i.e., there are many connections within clusters) and was correctly delineated by the procedure. Values of Q typically range between 0.3 and 0.7.} Third, in the final optimization process, the method ensures that each commuting zone forms a coherent region.

\begin{figure}[!ht]
\centering
\caption{Delineation of Commuting Zones}
\label{fig:C1}
\scalebox{1}{

\includegraphics[page=1,width=0.7\textwidth]{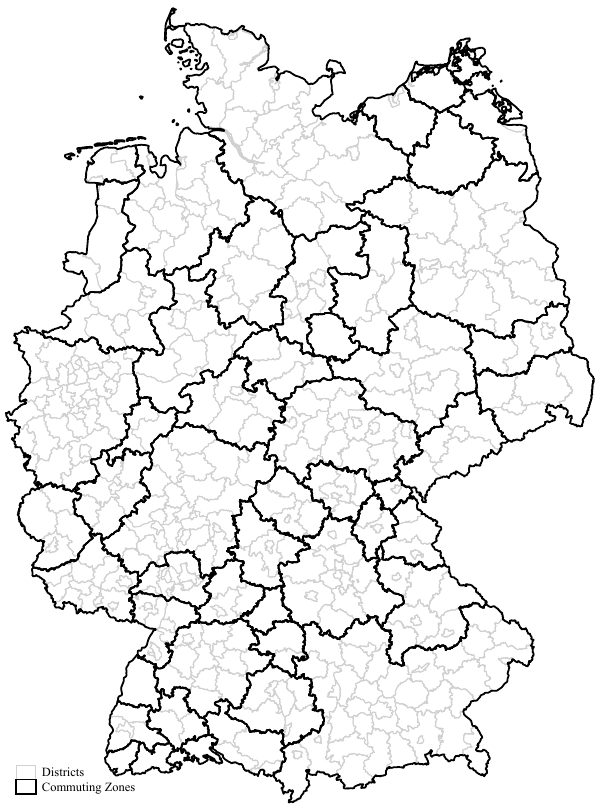}

}
\floatfoot{\footnotesize\textsc{Note. ---} The figure illustrates the delineation of commuting zones based on 401 German districts (NUTS-3 regions). The 401 districts were merged into 51 commuting zones using the graph-theoretical method from \citet{KroppSchwengler2016} and register data on German commuting patterns between 2012 and 2019. NUTS-3 = 3-Digit Statistical Nomenclature of Territorial Units in the European Community. Source: Official Statistics of the German Federal Employment Agency, 2012-2019.}
\end{figure}

For the years 2012-2019, we draw the commuting patterns for all workers in Germany who are subject to social security contributions from the Official Statistics of the Federal Employment Agency. Building on 401 districts (i.e., 3-digit NUTS regions) with initial modularity $Q=0.606$, we apply the graph-theoretical approach and gradually increase the merger threshold by 1 percentage point. For the threshold of 7 percent, modularity turns out to be highest ($Q=0.838$). After five iterations, this threshold yields $R=51$ commuting zones with strong interactions within but few connections between zones (see Figure \ref{fig:C1}). In doing so, we reduce the share of commuters between regions from 38.7 to 10.4 percent.

\paragraph{Registered vs.\ Unregistered Vacancies.} The Official Statistics from the German Federal Employment Agency collect information on only the so-called ``registered'' vacancies (including their 5-digit occupation and district). Registered vacancies are those vacancies that firms passed on to the local employment agency to let them find a suitable job seeker for the open position. For this reason, ``unregistered'' vacancies that were posted only on other channels (e.g., newspaper, private online job boards, etc.) are not part of this statistic.

To determine the total number of registered plus unregistered vacancies for each labor market and year, we make use of additional information from a large-scale business survey in Germany. The IAB Job Vacancy Survey (IAB-JVS) is a representative establishment survey with a focus on labor demand and recruitment behavior \citep{BosslerEtAl2020}. In every fourth quarter of the year, about 15,000 firms agree to disclose the number and structure of their vacancies. In particular, firms are asked to separately state their absolute number of registered and unregistered vacancies and additionally differentiate them by the requirement level of the underlying jobs (i.e., the fifth digit of the KldB occupation variable). Using survey weights, this information allows us to calculate the yearly shares of registered vacancies in all vacancies, separately for helpers, professionals, specialists or experts. However, since the survey differentiates between specialists and experts only since 2015, we pool this information and calculate notification shares for specialists and experts as a whole.

Table \ref{tab:C1} displays the development of notification shares over time. The notification shares vary over time and, on average, decrease with the complexity of the job. While the notification share of helpers is almost 50 percent, the share for professionals tends to be lower but falls in the same order of magnitude. For specialists and experts, around one in three vacancies is registered with the Federal Employment Agency.

\begin{table}[!ht]
\centering
\scalebox{0.90}{
\begin{threeparttable}
\caption{Notification Shares of Registered Vacancies in All Vacancies}
\label{tab:C1}
\begin{tabular}{C{2.5cm}C{4.25cm}C{4.25cm}C{4.25cm}} \hline
\multirow{2.4}{*}{Year} & \multirow{2.4}{*}{Helpers} & \multirow{2.4}{*}{Professionals} & \multirow{2.4}{*}{\shortstack{Specialists \\ and Experts}} \\
& & &   \\[0.2cm] \hline
& & &   \\[-0.2cm]
2012 &    36.0 &    45.0 &  33.6  \\[0.1cm]
2013 &  44.2 &    47.0 &  25.7  \\[0.1cm]
2014 &    48.0 &  41.9 &  29.9  \\[0.1cm]
2015 &  48.1 &  46.5 &  29.3  \\[0.1cm]
2016 &  53.3 &  50.5 &  36.7  \\[0.1cm]
2017 &  52.2 &  46.4 &  31.2  \\[0.1cm]
2018 &  43.9 &  46.2 &  32.8  \\[0.1cm]
2019 &  43.2 &  41.5 &  31.4  \\[0.1cm] \hline
& & &   \\[-0.2cm]
2012-2019 &  46.1 &  45.6 &  31.3  \\[0.2cm] \hline
\end{tabular}
\begin{tablenotes}[para]
\footnotesize\textsc{Note. ---} The table displays the yearly percentage shares of registered vacancies in all vacancies, separately by requirement level of the underlying job (i.e., the 5th digit of the KldB occupation variable). Helper occupations require no training or only a maximum of one year's training. The group of professionals includes all activities with industrial, commercial or other vocational training (excluding master craftsmen and technicians). Specialist occupations necessitate a bachelor degree or the completion of master craftsman/technician training. Experts hold a master degree or an equivalent diploma. Source: IAB Job Vacancy Survey, 2012-2019.
\end{tablenotes}
\end{threeparttable}
}
\end{table}

To quantify the number of registered plus unregistered vacancies in a labor market and year, we proceed in two steps. First, we draw the number of registered vacancies for each combination of 5-digit occupation, commuting zone, and year from the Official Statistics of the Federal Employment Agency. In line with official reporting guidelines, we apply two filters when drawing these data: On the one hand, we disregard vacancies for temporary jobs with less tenure of less than seven days. On the other hand, we exclude subsidized vacancies, vacancies for freelancers, and vacancies from private employment agencies. Second, we divide the resulting number of registered vacancies for each labor market and year by the respective yearly notification share. In doing so, we differentiate notification shares between helpers, professionals, and specialists along with experts based on the requirement level (i.e., the fifth digit) of the underlying KldB occupation.

\paragraph{Descriptive Statistics.}Table \ref{tab:C2} displays descriptive statistics for our model variables (at the firm level). An average establishment employs 7.4 regular full-time and 5.4 regular or marginal part-time workers. On average, average daily gross wages of regular full-time workers per calendar day and firm amount to 84.7 Euro. In terms of average hourly wages, this mean is 15.6 Euro for regular full-time and 12.6 Euro for part-time workers. Our average firm-specific labor market tightness is 0.64, implying that the occupational labor markets of the firm feature two unfilled jobs for every three job seekers. During the period of study, our Bartik instruments deliver an average two-year growth rate in average wages by 5.6-5.7 percent for regular full-time workers and 8.8 percent for (regular and marginal) part-time workers. Our Bartik instruments for the stock of vacancies and job seekers reflect that labor market tightness increased substantially during 2012-2019: the instrument for vacancies features an average two-year growth rate of 17.7 percent whereas the instrument for job seekers implies a rate of shrinkage by 8.8 percent every two years.

\begin{table}[!t]
\centering
\begin{threeparttable}
\caption{Descriptive Statistics}
\label{tab:C2}
\begin{tabular}{L{3.7cm}C{1.3cm}C{1.3cm}C{1.3cm}C{1.3cm}C{1.3cm}C{2cm}} \hline
\multirow{3.4}{*}{}  & \multirow{3.4}{*}{Mean} & \multirow{3.4}{*}{P25} & \multirow{3.4}{*}{P50} & \multirow{3.4}{*}{\shortstack{P75}} & \multirow{3.4}{*}{\shortstack{Stand. \\ Dev.}} & \multirow{3.4}{*}{\shortstack{Obser- \\ vations}} \\
    &   &   &   &   &  & \\
     &   &   &   &   &  & \\ \hline
          &   &   &   &   &  & \\[-0.4cm]
L$^{\text{FT}}$                         & 7.417          & 0                 & 1            & 3                 & 79.60          & 21,689,291             \\
L$^{\text{PT}}$                         & 5.429       & 1              & 2              & 4              & 32.86       & 21,689,291          \\
W$^{\text{FT}}$ (Daily)                 & 84.69     & 57.53     & 77.07     & 102.1     & 43.62     & 12,848,860        \\
W$^{\text{FT}}$ (Hourly)                & 15.56     & 10.46     & 14.11     & 18.85     & 8.100     & 12,848,860        \\
W$^{\text{PT}}$ (Hourly)                & 12.58  & 7.488  & 10.75  & 15.42  & 9.749  & 18,907,646     \\
V/U                                  & 0.636    & 0.127    & 0.329    & 0.740    & 1.201    & 21,689,291       \\ \hline
          &   &   &   &   &  & \\[-0.4cm]
Log L$^{\text{FT}}$                     & 1.246          & 0.000          & 1.099     & 1.946          & 1.260          & 12,848,860             \\
Log L$^{\text{PT}}$                     & 0.965       & 0.000       & 0.693       & 1.609       & 1.023       & 18,907,646          \\
Log W$^{\text{FT}}$ (Daily)             & 4.320     & 4.052     & 4.345     & 4.626     & 0.507     & 12,848,860        \\
Log W$^{\text{FT}}$ (Hourly)            & 2.623     & 2.347     & 2.647     & 2.937     & 0.513     & 12,848,860        \\
Log W$^{\text{PT}}$ (Hourly)            & 2.345  & 2.013  & 2.375  & 2.736  & 0.629  & 18,907,646     \\
Log V/U                     & -1.154\hphantom{-}    & -2.002\hphantom{-}    & -1.078\hphantom{-}    & -0.285\hphantom{-}    & 1.251    & 21,262,679       \\ \hline
          &   &   &   &   &  & \\[-0.4cm]
Z\(_{\text{W}^{\text{FT}}}\) (Daily)    & 0.056          & 0.047          & 0.055          & 0.064          & 0.017          & 16,300,305         \\
Z\(_{\text{W}^{\text{FT}}}\) (Hourly)   & 0.057          & 0.047          & 0.056          & 0.064          & 0.017          & 16,300,305         \\
Z\(_{\text{W}^{\text{PT}}}\) (Hourly)   & 0.088          & 0.060          & 0.074          & 0.115          & 0.039          & 16,300,305         \\
Z\(_{\text{V}}\)                        & 0.177          & 0.061          & 0.194          & 0.302          & 0.228          & 16,300,305          \\
Z\(_{\text{U}}\)                        & -0.088\hphantom{-}          & -0.169\hphantom{-}          & -0.103\hphantom{-}          & -0.014\hphantom{-}          & 0.137          & 16,300,305              \\ \hline
\end{tabular}
\begin{tablenotes}[para]
\footnotesize\textsc{Note. ---} The table shows descriptive statistics for the model variables between 2012 and 2019. All statistics reflect establishment-year observations. The establishment-specific measure of labor market tightness is constructed by weighting the ratio of vacancies to job seekers per labor market by occupational employment in the corresponding establishment. Labor markets refer to combinations of KldB-5 occupations and commuting zones. The instrumental variables refer to shift-share instruments of biennial national changes in employment weighted by past occupational employment in the respective establishment. L = Employment (in Heads). KldB = German Classification of Occupations. PX = Xth Percentile. Stand.\ Dev. = Standard Deviation. U = Job Seekers. V = Vacancies. W = Average Wages (in Euro). Z = Shift-Share Instrument. Sources:  Integrated Employment Biographies $\plus$ Official Statistics of German Federal Employment Agency $\plus$ IAB Job Vacancy Survey, 1999-2019.
\end{tablenotes}
\end{threeparttable}
\end{table}

\clearpage

\section{Labor Market Tightness: Further Evidence}
\label{sec:D}

\setcounter{table}{0} 
\setcounter{figure}{0} 

\vspace*{\fill}

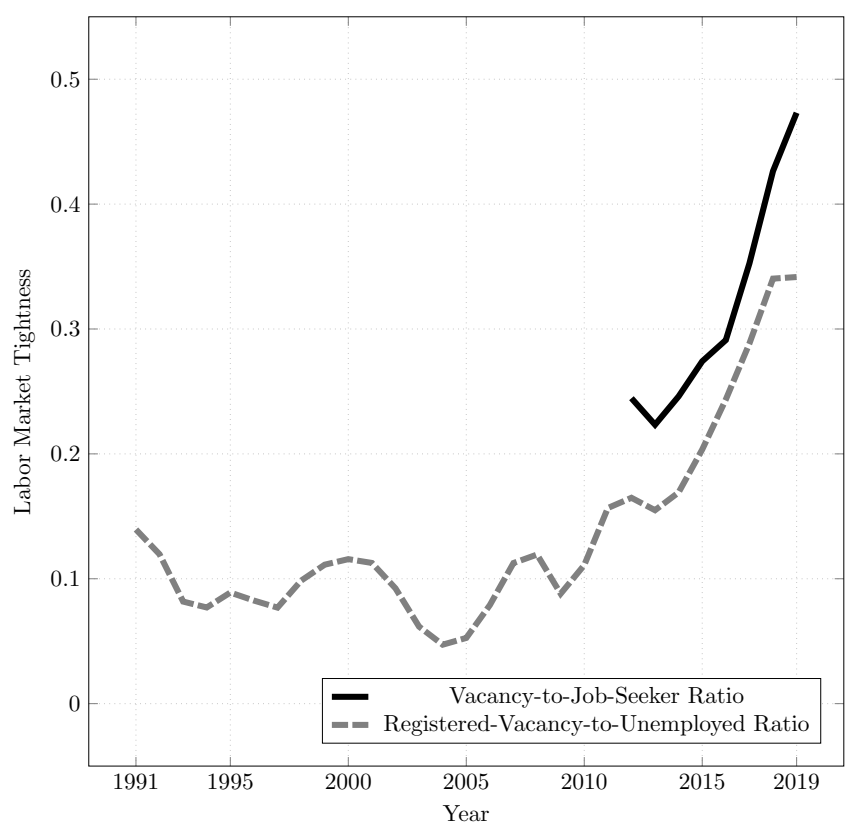
\begin{figure}[!ht]
\centering
\caption{Labor Market Tightness in Germany over Time}
\label{fig:D1}
\scalebox{0.80}{
\begin{tikzpicture}
\begin{axis}[xlabel=Year, ylabel=Labor Market Tightness, xmin=1989, xmax=2021, ymax=0.55, ymin=-0.05, height=14cm, width=14cm, grid=major, legend pos = south east, xtick={1991,1995,2000,2005,2010,2015,2019}, ytick={0,0.1,0.2,0.3,0.4,0.5}, grid style = dotted]
\addplot[mark=none, solid, color=black, line width=1mm] coordinates {  (2012,.24457256) (2013,.22334746) (2014,.24606094) (2015,.27413556) (2016,.29112986) (2017,.35252666) (2018,.42646006) (2019,.47312701) };
\addplot[mark=none, dash pattern = on 7.6pt off 2pt, color=gray, line width=1mm] coordinates {  (1991,.13943379) (1992,.11960001) (1993,.08173164) (1994,.07700071) (1995,.08895709) (1996,.08254041) (1997,.07688753) (1998,.09849158) (1999,.11127768) (2000,.1157065) (2001,.11266185) (2002,.09232483) (2003,.06165151) (2004,.04721222) (2005,.05261526) (2006,.0789534) (2007,.11259947) (2008,.11937818) (2009,.08803564) (2010,.11094563) (2011,.15665711) (2012,.16482818) (2013,.15488903) (2014,.16916645) (2015,.20351033) (2016,.24358828) (2017,.28843191) (2018,.34034148) (2019,.34161475) };
\legend{~Vacancy-to-Job-Seeker Ratio, ~Registered-Vacancy-to-Unemployed Ratio,}
\end{axis}
\end{tikzpicture}
}
\floatfoot{\footnotesize\textsc{Note. ---} The figure illustrates the development of labor market tightness in Germany over time. The solid black line refers to the economy-wide ratio of vacancies to job seekers. To construct this measure, we add up the number of overall vacancies as well as job seekers across all occupation-by-commuting-zone combinations and calculate the ratio of the two sums. The dashed grey line shows the ratio of registered vacancies to unemployed persons and, thus, builds on aggregate information that is available already from 1991 onwards. Source: Official Statistics of the German Federal Employment Agency $\plus$ IAB Job Vacancy Survey, 1991-2019.}
\end{figure}

\vspace*{\fill}

\begin{landscape}

\vspace*{\fill}

\pgfkeys{/pgf/number format/.cd,1000 sep={,}}
\begin{figure}[!ht]
\centering
\caption{Beveridge Curve by Occupational Areas}
\label{fig:D2}
\scalebox{0.74}{
\begin{tikzpicture}
\begin{axis}[ytick={0,200,400,600}, height=15.5cm, width=25cm, grid=major, grid style = dotted, xtick={0,200,400,600,800,1000,1200}, xlabel={Job Seekers (in Thousands)}, ylabel={Vacancies (in Thousands)}, ymin=-50, ymax=650, xmin=-50, xmax=1250, legend pos = north west, y tick label style={/pgf/number format/.cd,fixed,fixed zerofill, precision=0,/tikz/.cd}]
\addplot[mark=*, solid, color=black, mark options={fill=white, scale=1, solid}, line width=0.25mm] coordinates {  (165.059,16.187) (174.16,16.444) (169.087,17.385) (161.008,18.562) (155.581,19.418) (149.633,23.239) (136.244,28.162) (126.989,29.996) };
\node[mark=none, below, color=black] at (165.059,16.187) {2012};
\node[mark=none, left, color=black] at (126.989,29.996) {2019};
\addplot[mark=*, solid, color=gray, mark options={fill=white, scale=1, solid}, line width=0.25mm] coordinates {  (784.879,393.064) (858.579,333.456) (843.582,380.187) (820.487,397.466) (829.214,407.869) (818.882,500.734) (748.311,575.657) (737.66,590.18) };
\node[mark=none, left, color=gray] at (784.879,393.064) {2012};
\node[mark=none, above, color=gray] at (737.66,590.18) {2019};
\addplot[mark=*, solid, color=black, mark options={fill=black, scale=1, solid}, line width=0.25mm] coordinates {  (368.877,105.873) (382.947,104.1) (373.371,103.249) (353.835,105.801) (339.929,119.653) (325.918,146.073) (289.99,160.804) (267.577,181.029) };
\node[mark=none, above, xshift=0.1cm, color=black] at (368.877,105.873) {2012};
\node[mark=none, above, color=black] at (267.577,181.029) {2019};
\addplot[mark=*, solid, color=gray, mark options={fill=gray, scale=1, solid}, line width=0.25mm] coordinates {  (183.662,79.263) (194.863,83.31) (216.099,82.671) (214.333,86.53) (212.018,85.019) (208.591,109.147) (192.001,114.607) (189.402,125.588) };
\node[mark=none, left, color=gray] at (183.662,79.263) {2012};
\node[mark=none, above, color=gray] at (189.402,125.588) {2019};
\addplot[mark=square*, solid, color=black, mark options={fill=white, scale=1, solid}, line width=0.25mm] coordinates {  (1087.265,171.228) (1166.311,144.449) (1185.234,168.561) (1177.275,202.505) (1171.029,222.462) (1174.165,264.918) (1113.82,314.88) (1104.379,330.007) };
\node[mark=none, left, color=black] at (1087.265,171.228) {2012};
\node[mark=none, above, color=black] at (1104.379,330.007)  {2019};
\addplot[mark=square*, solid, color=gray, mark options={fill=white, scale=1, solid}, line width=0.25mm] coordinates { (747.388,156.83) (783.499,151.117) (781.78,173.126) (749.324,181.591) (726.131,179.52) (701.806,218.167) (644.203,245.365) (613.106,256.101) };
\node[mark=none, left, color=gray] at (747.388,156.83) {2012};
\node[mark=none, left, color=gray] at (613.106,256.101)  {2019};
\addplot[mark=square*, solid, color=black, mark options={fill=black, scale=1, solid}, line width=0.25mm] coordinates {  (501.442,102.817) (518.703,103.055) (515.871,103.724) (491.465,110.428) (472.223,128.885) (456.188,132.295) (422.203,146.169) (406.17,169.385) };
\node[mark=none, below, xshift=-0.2cm, color=black] at (501.442,102.817) {2012};
\node[mark=none, above left, color=black] at (406.17,169.385) {2019};
\addplot[mark=square*, solid, color=gray, mark options={fill=gray, scale=1, solid}, line width=0.25mm] coordinates {  (462.191,176.982) (488.233,186.471) (493.271,195.799) (487.492,223.269) (492.664,227.763) (506.878,261.534) (489.098,275.099) (479.896,317.395) };
\node[mark=none, below, color=gray] at (462.191,176.982) {2012};
\node[mark=none, above, color=gray] at (479.896,317.395) {2019};
\legend{Agriculture and Forestry, Mining and Manufacturing, Construction, ~Natural Sciences and Humanities, Traffic and Security, Commercial Services, Law and Administration, Health and Education}
\end{axis}
\end{tikzpicture}
}
\floatfoot{\footnotesize\textsc{Note. ---} The figure shows Beveridge curves by occupational area for Germany between 2012 and 2019. The occupational areas refer to 1-digit KldB occupations. For ease of presentation, we group together natural sciences and humanities. KldB = German Classification of Occupations. Sources: Official Statistics of the German Federal Employment Agency $\plus$ IAB Job Vacancy Survey, 2012-2019.}
\end{figure}
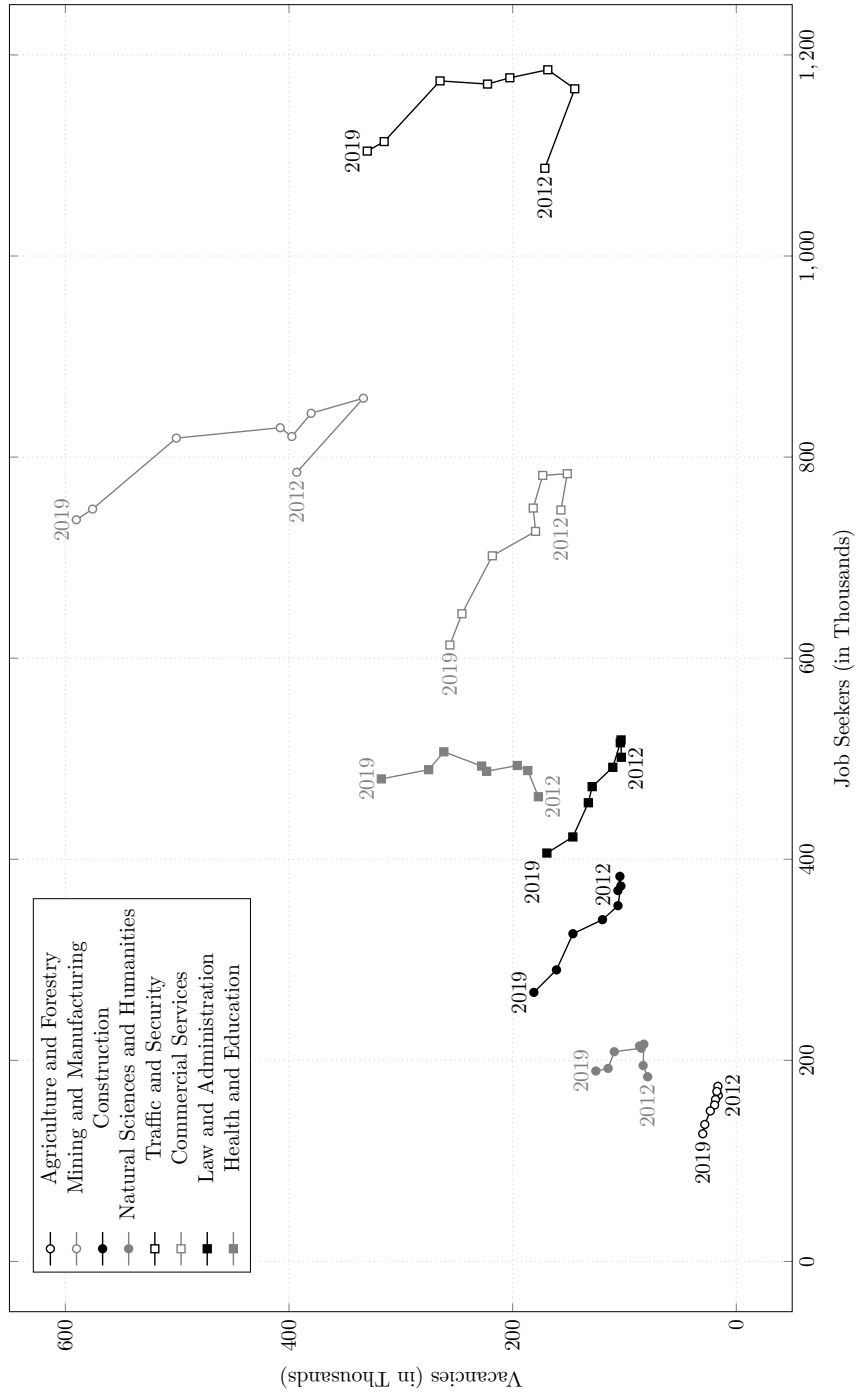
\pgfkeys{/pgf/number format/.cd,1000 sep={}}

\vspace*{\fill}

\clearpage

\begin{figure}[!ht]
\centering
\caption{Labor Market Tightness by Commuting Zones}
\label{fig:D3}
\vspace*{0.5cm}
\begin{subfigure}{0.475\textwidth}
\caption{\normalsize{2012}}
\begin{center}
\includegraphics[page=1,width=0.75\textwidth]{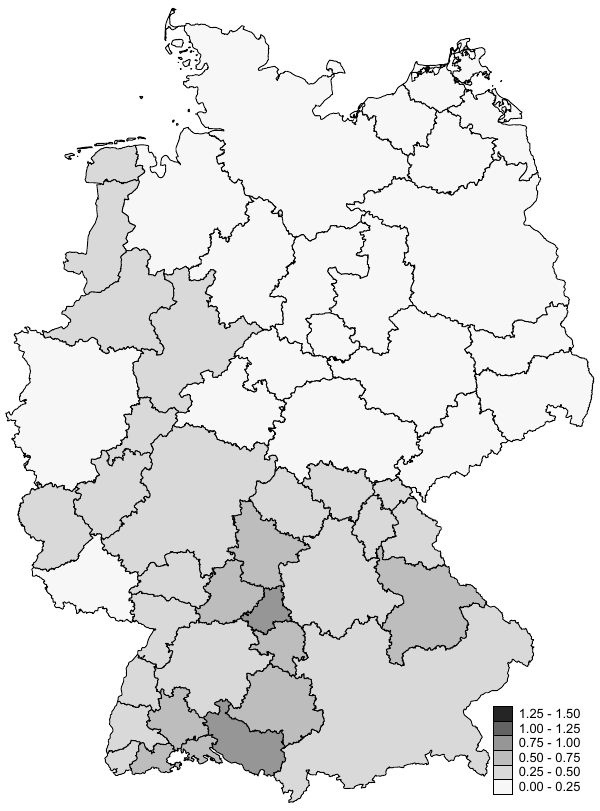}
\end{center}
\end{subfigure}
\hfill
\begin{subfigure}{0.475\textwidth}
\caption{\normalsize{2019}}
\begin{center}
\includegraphics[page=1,width=0.75\textwidth]{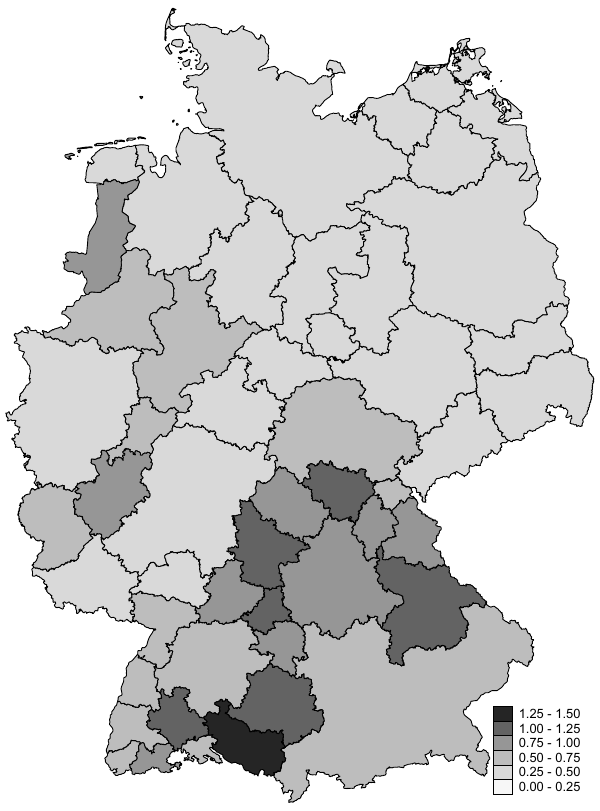}
\end{center}
\end{subfigure}
\floatfoot{\footnotesize\textsc{Note. ---} The figures visualize labor market tightness by commuting zones, separately for the years 2012 and 2019. Sources: Integrated Employment Biographies $\plus$ Official Statistics of the German Federal Employment Agency $\plus$ IAB Job Vacancy Survey, 2012 \& 2019.}
\end{figure}

\end{landscape}

\clearpage

\vspace*{\fill}

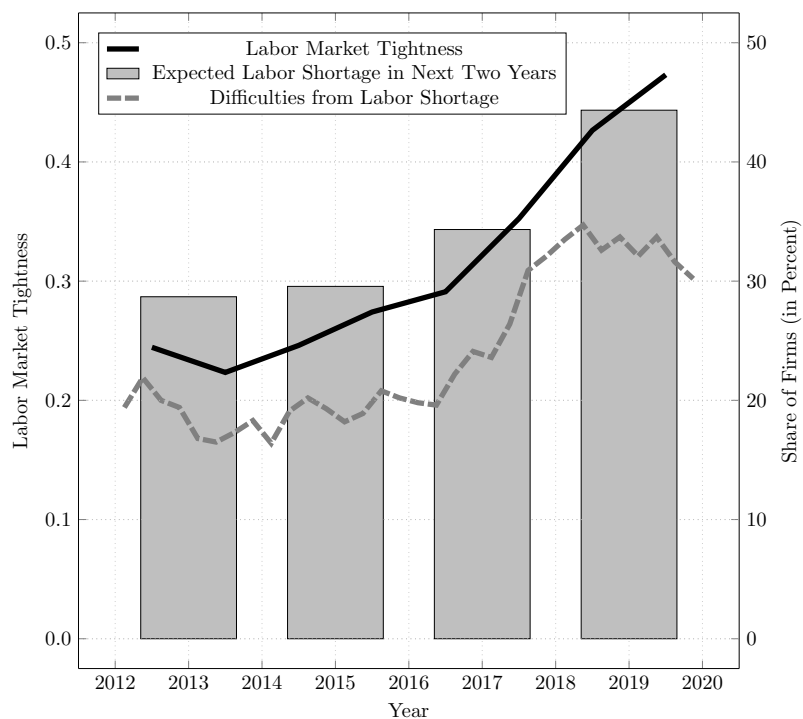
\begin{figure}[!ht]
\centering
\caption{Labor Market Tightness vs.\ Survey Information on Labor Shortage}
\label{fig:D4}

\scalebox{0.70}{
\begin{tikzpicture}

\begin{axis}[xlabel=Year, ylabel={Share of Firms (in Percent)}, xmin=2011.5,xmax=2020.5, ymin=-2.5, ymax=52.5, height=14cm, width=14cm, grid=major, grid style=dotted, ytick={0,10,20,30,40,50}, xtick={2012,2013,2014,2015,2016,2017,2018,2019,2020}, legend pos = north west, y tick label style={/pgf/number format/.cd,fixed,fixed zerofill, precision=0,/tikz/.cd}, x tick label style={/pgf/number format/.cd,fixed,fixed zerofill, precision=0,/tikz/.cd}, axis y line*=right]

\addplot[ybar, area legend,mark=none, solid, color=black, fill=lightgray, bar width=1.8cm,enlarge x limits={abs=1.75cm},] coordinates { (2013,28.69) (2015,29.56) (2017,34.33) (2019,44.34) }; \label{plot 1}

\addplot[mark=none, dash pattern = on 7.6pt off 2pt, color=gray, line width=1mm] coordinates {(2012.125,19.4) (2012.375,21.9) (2012.625,20.0) (2012.875,19.4) (2013.125,16.8) (2013.375,16.5) (2013.625,17.3) (2013.875,18.3) (2014.125,16.4) (2014.375,19.1) (2014.625,20.2) (2014.875,19.3) (2015.125,18.2) (2015.375,18.9) (2015.625,20.8) (2015.875,20.2) (2016.125,19.8) (2016.375,19.6) (2016.625,22.2) (2016.875,24.1) (2017.125,23.6) (2017.375,26.4) (2017.625,30.9) (2017.875,32.1) (2018.125,33.5) (2018.375,34.7) (2018.625,32.6) (2018.875,33.7) (2019.125,32.1) (2019.375,33.7) (2019.625,31.6) (2019.875,30.2) }; \label{plot 2}

\end{axis}

\begin{axis}[xlabel=Year, ylabel={Labor Market Tightness}, xmin=2011.5,xmax=2020.5, ymin=-0.025, ymax=0.525, height=14cm, width=14cm, grid=major, grid style=dotted, ytick={0,0.10,0.20,0.30,0.40,0.50}, xtick={2012,2013,2014,2015,2016,2017,2018,2019,2020}, legend pos = north west, y tick label style={/pgf/number format/.cd,fixed,fixed zerofill, precision=1,/tikz/.cd}, x tick label style={/pgf/number format/.cd,fixed,fixed zerofill, precision=0,/tikz/.cd}, axis y line*=left, axis x line=none]

\addplot[mark=none, solid, color=black, line width=1mm] coordinates { (2012.5,0.2445) (2013.5,0.2234) (2014.5,0.2461) (2015.5,0.2741) (2016.5,0.2911) (2017.5,0.3525) (2018.5,0.4265) (2019.5,0.4731) }; \addlegendentry{~Labor Market Tightness}

\addlegendimage{/pgfplots/refstyle=plot 1}\addlegendentry{~Expected Labor Shortage in Next Two Years}
\addlegendimage{/pgfplots/refstyle=plot 2}\addlegendentry{~Difficulties from Labor Shortage}

\end{axis}

\end{tikzpicture}
}
\vspace{-0.25cm}
\floatfoot{ \footnotesize\textsc{Note. ---} The figure compares economy-wide labor market tightness (left axis) with survey information on the percentage share of firms that face (difficulties from) a shortage of skilled labor (right axis). The IAB Establishment Panel asks firms whether they expect labor shortage in the next two years (grey columns). The KfW-ifo Skilled Labor Barometer surveys whether companies are experiencing adverse impacts on business operations from a shortage of skilled workers (gray dashed line). Source: Official Statistics of the German Federal Employment Agency $\plus$ IAB Establishment Panel $\plus$ KfW-ifo Skilled Labor Barometer, 2012-2019.}
\end{figure}

\vspace*{\fill}


\pgfkeys{/pgf/number format/.cd,1000 sep={,}}
\begin{figure}[!ht]
\caption{Hourly Wage Rate and Pre-Match Hiring Costs}
\label{fig:D5}
\begin{subfigure}{1\textwidth}
\caption{Direct Pre-Match Hiring Costs}
\centering
\scalebox{0.725}{
\begin{tikzpicture}
\begin{axis}[xlabel=Log Hourly Wage Rate, ylabel=Log Hiring Costs, xmin=2, xmax=3.5, ymin=4.75, ymax=8.25, height=8cm, width=16cm, grid=major, legend pos = south east, xtick={2.25,2.75,3.25}, ytick={5,6,7,8}, grid style = dotted, y tick label style={/pgf/number format/.cd,fixed,fixed zerofill, precision=1,/tikz/.cd}]
\addplot[only marks, mark=o, mark options={solid,scale=1}, color=black] coordinates {  (2.1450009,4.9335852) (2.1583502,5.3397946) (2.166039,5.369586) (2.1749473,5.5904369) (2.190758,5.5097089) (2.2025068,4.9615817) (2.2112794,5.2676725) (2.2186279,5.5485096) (2.2349141,5.1101179) (2.2438104,4.8009796) (2.25038,5.9272461) (2.2610133,5.6157622) (2.2767091,5.6662798) (2.2871864,5.2309237) (2.3010819,5.4040136) (2.3091261,5.8336725) (2.3217461,5.7242613) (2.3390055,6.0129333) (2.3486111,5.2991738) (2.3586283,5.3072333) (2.3685737,6.0143132) (2.3816774,5.2985821) (2.3972836,5.8272109) (2.4058418,5.8931804) (2.4201362,5.8828144) (2.4313407,6.3387933) (2.440316,6.4301419) (2.4468908,5.4428344) (2.4568737,5.9192095) (2.4696984,5.7197657) (2.4791567,5.9773989) (2.4841251,5.933002) (2.4910364,6.1055779) (2.5023484,6.1183786) (2.51141,5.736022) (2.524966,6.0963898) (2.5358174,6.0231137) (2.5464787,5.908865) (2.556406,6.0997834) (2.5647154,6.1410952) (2.5698791,4.4551072) (2.5769749,6.0773764) (2.5875065,6.2768703) (2.5968907,6.7807083) (2.6024852,6.3116097) (2.6125407,6.4016118) (2.6214292,6.2020688) (2.6292593,6.2415924) (2.6381826,6.3591981) (2.6518321,6.2251201) (2.6607053,6.5370374) (2.6654198,6.555089) (2.6702456,6.1425247) (2.6769371,6.6431499) (2.6872497,6.101542) (2.6933343,6.7137671) (2.7030027,6.6404252) (2.7100539,6.1990161) (2.7178731,6.6187758) (2.7250142,6.5010347) (2.733845,6.6181731) (2.7422419,6.3896079) (2.7497816,6.2263465) (2.7572932,6.6481428) (2.7693293,6.2150164) (2.77737,6.7652893) (2.7847047,6.5474005) (2.7957551,6.6087708) (2.8039923,6.4621286) (2.8092756,6.7419872) (2.8163793,6.578939) (2.8264971,6.5671172) (2.8356254,6.7658157) (2.8400989,6.8325672) (2.8470845,6.7034173) (2.8558011,6.7578425) (2.8677442,6.7249703) (2.8815589,7.212389) (2.894701,6.7822905) (2.902832,7.1385183) (2.918819,6.9827056) (2.9324031,6.9687285) (2.9478462,7.0130577) (2.959146,7.1810207) (2.9756033,6.8980556) (2.9927287,6.6389232) (3.0057695,7.0003109) (3.0243945,6.9185224) (3.0416327,6.8243551) (3.0611334,7.0902171) (3.0832577,6.8631773) (3.1010587,7.2437978) (3.1275675,7.2889605) (3.1449885,7.1511431) (3.1643305,7.4475565) (3.183182,7.2546206) (3.2029283,7.3383832) (3.2292709,6.9502635) (3.2488146,6.733561) (3.2692649,7.2828593) };
\addplot[domain=2.125:3.375, color=black, solid, line width=0.75mm] {1.024652+1.975215*x};
\legend{~Observations,~Linear Fit}
\end{axis}
\end{tikzpicture}
}
\end{subfigure}
\vskip \baselineskip
\begin{subfigure}{1\textwidth}
\caption{Indirect Pre-Match Hiring Costs}
\centering
\scalebox{0.725}{
\begin{tikzpicture}
\begin{axis}[xlabel=Log Hourly Wage Rate, ylabel=Log Hiring Costs, xmin=2, xmax=3.5, ymin=3.75, ymax=7.25, height=8cm, width=16cm, grid=major, legend pos = south east, xtick={2.25,2.75,3.25}, ytick={4,5,6,7}, grid style = dotted, y tick label style={/pgf/number format/.cd,fixed,fixed zerofill, precision=1,/tikz/.cd}]
\addplot[only marks, mark=o, mark options={solid,scale=1}, color=black] coordinates {  (2.1450772,4.3322043) (2.1557181,4.5284882) (2.1641648,4.670537) (2.1724441,4.3444438) (2.1773367,4.670197) (2.1826396,4.5188537) (2.1986761,4.4712548) (2.2026558,4.022789) (2.2054787,4.5648866) (2.2129099,4.5751047) (2.2233982,4.900476) (2.234422,4.6176338) (2.2434525,4.4518657) (2.2507398,4.5305009) (2.2587099,5.148365) (2.2660153,4.4656377) (2.279433,4.689352) (2.2904117,4.4719658) (2.3003387,4.9406495) (2.3069644,4.67342) (2.3132036,4.850976) (2.3246295,4.5801196) (2.3393016,5.0779099) (2.3475068,4.6920743) (2.3579054,4.6899071) (2.3669534,4.8484836) (2.378298,4.8257933) (2.3916602,5.166574) (2.4019885,4.7404451) (2.4072187,4.7419066) (2.4188137,4.7037287) (2.43012,4.802856) (2.434891,3.5784583) (2.4434097,4.9032664) (2.4486337,4.827929) (2.456517,5.0001974) (2.4673355,5.0478082) (2.4773779,5.2263994) (2.4839973,4.9549618) (2.4895985,4.9718304) (2.4987853,4.9065294) (2.5068076,5.4023533) (2.5171435,5.2579589) (2.5265472,4.8949747) (2.5338004,5.0106344) (2.5441403,4.9991074) (2.5521624,5.1996093) (2.5627408,5.1153116) (2.5688088,4.8611994) (2.5781357,5.3641729) (2.5867028,5.0178843) (2.5943775,5.3981614) (2.602917,5.0633531) (2.6132843,5.3927689) (2.6222134,5.2606487) (2.6306939,5.4151039) (2.6407766,4.9149518) (2.6496773,5.2733006) (2.6573172,5.1766677) (2.6643085,5.5290937) (2.6697519,5.0104241) (2.6750059,5.4206219) (2.6850388,5.2877593) (2.6925263,5.5067148) (2.7022431,5.5488744) (2.7096443,5.0979328) (2.7178454,5.4414048) (2.7252221,5.41115) (2.7361228,5.5364919) (2.7440119,5.6715646) (2.7521813,5.4514251) (2.762347,5.4780111) (2.7724993,5.5403385) (2.7800236,5.299037) (2.7915053,5.4314942) (2.8035076,5.544373) (2.8109074,5.624114) (2.8237822,5.7328544) (2.8359797,5.5570655) (2.8446009,5.7564354) (2.854121,5.7999568) (2.8666153,5.6552973) (2.8803611,5.5648608) (2.8959613,5.7725401) (2.9096978,5.7096415) (2.9277451,5.97189) (2.9464719,5.6902995) (2.9617293,6.0789514) (2.9818673,6.0001235) (3.000165,6.0094938) (3.0183277,5.7039499) (3.0381498,6.087585) (3.0616648,6.1302094) (3.0896294,5.9856234) (3.1204438,6.3860655) (3.1453564,5.6863456) (3.1689532,5.9247937) (3.1964743,6.0954413) (3.2326524,6.1769509) (3.2666936,6.0853062) };
\addplot[domain=2.125:3.375, color=black, solid, line width=0.75mm] {0.7932273+1.693129*x};
\legend{~Observations,~Linear Fit}
\end{axis}
\end{tikzpicture}
}
\end{subfigure}
\vskip \baselineskip
\begin{subfigure}{1\textwidth}
\centering
\caption{Overall Pre-Match Hiring Costs}
\scalebox{0.725}{
\begin{tikzpicture}
\begin{axis}[xlabel=Log Hourly Wage Rate, ylabel= Log Hiring Costs, xmin=2, xmax=3.5, ymin=4.25, ymax=7.75, height=8cm, width=16cm, grid=major, legend pos = south east, xtick={2.25,2.75,3.25}, ytick={4.5,5.5,6.5,7.5}, grid style = dotted, y tick label style={/pgf/number format/.cd,fixed,fixed zerofill, precision=1,/tikz/.cd}]
\addplot[only marks, mark=o, mark options={solid,scale=1}, color=black] coordinates {  (2.1450138,4.8961501) (2.1545148,4.8588634) (2.1640365,5.4764557) (2.1726084,4.6712756) (2.1772904,5.3567538) (2.1848621,5.0941167) (2.2002125,4.7549524) (2.2026563,4.9309282) (2.2106431,5.1055746) (2.2191961,5.3096619) (2.2337787,5.2832909) (2.2410975,5.161973) (2.2491007,5.1128731) (2.2538769,5.1979647) (2.2636459,5.129262) (2.2780058,5.2144084) (2.2881255,5.1886034) (2.2989705,5.1007028) (2.3070426,5.261446) (2.3122039,5.2522268) (2.3211758,5.2564464) (2.3370008,5.6122117) (2.34553,5.3942437) (2.354538,5.2003994) (2.3633027,5.6322598) (2.3752291,5.2813163) (2.3889143,5.4221811) (2.4009895,5.0833921) (2.4051204,5.4982305) (2.4155464,5.0953088) (2.4267592,5.5173635) (2.4340312,5.9484882) (2.4440763,5.3116155) (2.4503314,5.36551) (2.4597251,5.5714192) (2.4689827,5.5857434) (2.4787419,5.9653449) (2.4851596,5.6084976) (2.4917893,5.4981771) (2.501977,5.373014) (2.5102911,5.9186921) (2.5213637,5.6070719) (2.5286868,5.4076405) (2.5379472,5.4047832) (2.5462122,5.5961785) (2.5544674,5.7535105) (2.5637732,5.794158) (2.569623,5.5058498) (2.5787442,5.7759848) (2.586715,5.5534101) (2.594507,6.2070594) (2.6028764,5.8182421) (2.6128659,5.9719605) (2.6231513,6.0746541) (2.6321065,6.1982999) (2.6411593,5.3496494) (2.6504469,5.7699003) (2.6578932,6.2537866) (2.6643422,6.2270546) (2.6697662,5.5639973) (2.6746943,5.9160061) (2.683594,6.0253339) (2.6917455,6.4374118) (2.700223,5.8425031) (2.706984,6.1319957) (2.7134545,5.7252445) (2.7215478,6.2372727) (2.731164,6.235888) (2.7416661,6.2747135) (2.7491181,6.4089065) (2.7563121,6.1778088) (2.7680919,6.1044664) (2.7763498,6.3303342) (2.7850144,6.1152086) (2.7980049,6.4192696) (2.8076131,6.4479036) (2.8164117,6.5530548) (2.8290582,6.4762731) (2.838928,6.7023988) (2.847707,6.3663716) (2.8567221,6.7132697) (2.8694165,6.1146083) (2.8846111,6.6255808) (2.8984847,6.5835948) (2.9148026,6.6481099) (2.9306984,6.7488484) (2.9491885,6.6877651) (2.9655659,6.9252872) (2.9868922,6.8100615) (3.0042529,6.8967795) (3.0252106,6.6912937) (3.0471275,6.8052869) (3.0735261,6.8974209) (3.0972126,7.2010169) (3.1290469,7.1735406) (3.1487882,6.676445) (3.1748061,6.7866859) (3.200346,7.307054) (3.231869,7.0019422) (3.2641602,7.3179245) };
\addplot[domain=2.125:3.375, color=black, solid, line width=0.75mm] {0.2462468+2.162173*x};
\legend{~Observations,~Linear Fit}
\end{axis}
\end{tikzpicture}
}
\end{subfigure}
\linespread{1} \floatfoot{\footnotesize\textsc{Note. ---} The figures show binned scatterplots with 100 markers to depict cross-sectional correlations between the log hourly wage rate upon hiring and the log of direct, indirect, and overall pre-match hiring costs. Whereas direct pre-match hiring costs (in Euro) are asked separately in the IAB Job Vacancy Survey, we calculate indirect pre-match hiring costs (in Euro) by multiplying the search effort (in working hours) by the firm's average hourly wage rate of i) workers in human resource management (KldB-2010 Code: 715) or, if not available, ii) managers or, if not available, iii) all workers. Pre-match hiring costs and the hourly wage rate (upon hiring) were deflated with base year 2015. We trim hourly wages (in Euro) at the 5th and 95th percentile. The numbers of observed successful hires are: 13,085 for direct, 31,933 for indirect, and 26,886 for overall pre-match hiring costs. Sources: Integrated Employment Biographies $\plus$ Official Statistics of Federal Employment Agency $\plus$ IAB Job Vacancy Survey, 2014-2015, 2017-2019.} \linespread{1.5}
\end{figure}
\pgfkeys{/pgf/number format/.cd,1000 sep={}}

\pgfkeys{/pgf/number format/.cd,1000 sep={,}}
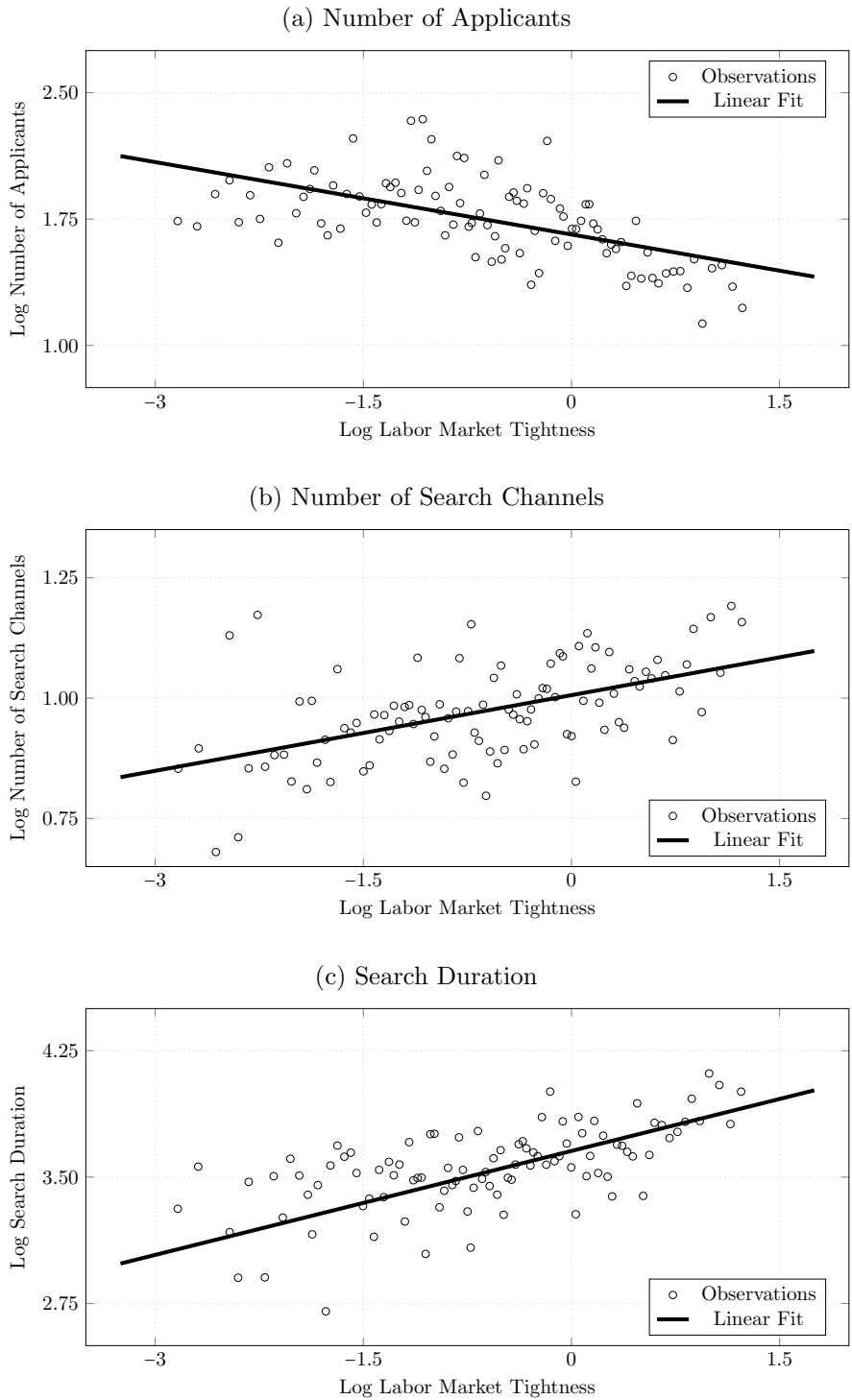
\begin{figure}[!ht]
\caption{Labor Market Tightness and Hiring Indicators}
\label{fig:D6}
\begin{subfigure}{1\textwidth}
\caption{Number of Applicants}
\centering
\scalebox{0.725}{
\begin{tikzpicture}
\begin{axis}[xlabel=Log Labor Market Tightness, ylabel=Log Number of Applicants, xmin=-3.5, xmax=2, ymin=0.75, ymax=2.75, height=8cm, width=16cm, grid=major, legend pos = north east, xtick={-3,-1.5,0,1.5}, ytick={1,1.75,2.5}, grid style = dotted, y tick label style={/pgf/number format/.cd,fixed,fixed zerofill, precision=2,/tikz/.cd}]
\addplot[only marks, mark=o, mark options={solid,scale=1}, color=black] coordinates {  (-2.8378847,1.7382734) (-2.6979482,1.7068665) (-2.5679553,1.8983521) (-2.4644327,1.9803431) (-2.3976953,1.7317077) (-2.3158114,1.8915302) (-2.2446969,1.7508273) (-2.1801856,2.0572348) (-2.1121843,1.6096567) (-2.0493448,2.0811357) (-1.9836812,1.7848115) (-1.931246,1.8821892) (-1.8851697,1.9287024) (-1.8528734,2.039238) (-1.8039215,1.7245911) (-1.7573669,1.653746) (-1.7174549,1.9497387) (-1.6654344,1.6933922) (-1.6193899,1.8991255) (-1.5734884,2.2288179) (-1.5281477,1.8844672) (-1.4808148,1.7878896) (-1.4402014,1.8367008) (-1.4037092,1.7298141) (-1.3706925,1.8384025) (-1.3370115,1.9624194) (-1.3055259,1.940282) (-1.2686962,1.9663141) (-1.2266573,1.9049751) (-1.1896653,1.7405123) (-1.156454,2.3336594) (-1.1283609,1.7309581) (-1.101819,1.9234807) (-1.0715044,2.3426824) (-1.0410084,2.0366914) (-1.0092479,2.2237811) (-.97968286,1.8879246) (-.9423756,1.7987204) (-.90983582,1.6536) (-.88186926,1.9406435) (-.85102963,1.7176238) (-.82514542,2.1247237) (-.80287886,1.8447419) (-.77297151,2.1126826) (-.74075335,1.7048383) (-.71855122,1.7271832) (-.69118619,1.5239415) (-.6597839,1.7833235) (-.62737864,2.01278) (-.60529399,1.7143162) (-.57486701,1.4969147) (-.55049819,1.6483843) (-.52599514,2.0987585) (-.50426215,1.5104831) (-.47744724,1.5770636) (-.4491249,1.88243) (-.41734987,1.9080148) (-.39394107,1.8589344) (-.37242687,1.5478034) (-.34372512,1.8419405) (-.31854635,1.932946) (-.29016763,1.3603423) (-.26435593,1.680528) (-.23371542,1.4292749) (-.20373693,1.9036736) (-.1748533,2.2139688) (-.1481114,1.8688853) (-.11694986,1.6212298) (-.08275539,1.8125561) (-.05755932,1.7651436) (-.02684441,1.591217) (.0036607,1.6922882) (.03348368,1.6909913) (.06925622,1.7397021) (.10215301,1.8377939) (.12799186,1.8384436) (.15728235,1.7235022) (.18863663,1.6891056) (.22285613,1.6301485) (.2541931,1.5477461) (.28638735,1.598574) (.3202655,1.5727091) (.35784632,1.6134578) (.39550763,1.3535292) (.43126929,1.4137262) (.46480602,1.7396252) (.50304401,1.3973734) (.54891378,1.5522505) (.58434689,1.4005144) (.62759483,1.3684421) (.68156642,1.4281399) (.73519105,1.4387933) (.78430295,1.4412884) (.83534628,1.3423429) (.88409209,1.5118016) (.94397539,1.1300102) (1.0131652,1.4580531) (1.0838042,1.4763443) (1.1609485,1.3492237) (1.2311951,1.2227297) };
\addplot[domain=-3.25:1.75, color=black, solid, line width=0.75mm] {1.658716-0.1431904*x};
\legend{~Observations,~Linear Fit}
\end{axis}
\end{tikzpicture}
}
\end{subfigure}
\vskip \baselineskip
\begin{subfigure}{1\textwidth}
\caption{Number of Search Channels}
\centering
\scalebox{0.725}{
\begin{tikzpicture}
\begin{axis}[xlabel=Log Labor Market Tightness, ylabel=Log Number of Search Channels, xmin=-3.5, xmax=2, ymin=0.65, ymax=1.35, height=8cm, width=16cm, grid=major, legend pos = south east, xtick={-3,-1.5,0,1.5}, ytick={0.75,1,1.25}, grid style = dotted, y tick label style={/pgf/number format/.cd,fixed,fixed zerofill, precision=2,/tikz/.cd}]
\addplot[only marks, mark=o, mark options={solid,scale=1}, color=black] coordinates {  (-2.8343606,.85325944) (-2.6867933,.89551562) (-2.5624969,.68035007) (-2.4643426,1.1302097) (-2.4014111,.71097326) (-2.3264916,.8540588) (-2.2617514,1.1727761) (-2.208133,.85747206) (-2.1402676,.88168955) (-2.0718944,.88249773) (-2.0179212,.82678568) (-1.9600762,.99314392) (-1.9072475,.81068605) (-1.871628,.99437362) (-1.833535,.86597699) (-1.7750643,.91350532) (-1.736964,.82574618) (-1.6868213,1.0600806) (-1.6366162,.93750024) (-1.5911191,.92839026) (-1.5484908,.94838655) (-1.4991896,.84778416) (-1.4546494,.86034554) (-1.4209105,.96593434) (-1.3829554,.91429579) (-1.3488301,.96473789) (-1.3125339,.93193567) (-1.2797033,.98414177) (-1.2420855,.95127243) (-1.2018939,.98149872) (-1.1698362,.98550516) (-1.1389565,.94611216) (-1.1091994,1.0836647) (-1.0793686,.97548562) (-1.051694,.96060777) (-1.017524,.86778766) (-.98808074,.92034131) (-.94954884,.98704135) (-.91629899,.85320109) (-.88847095,.95806324) (-.85613346,.88295037) (-.83072948,.97200811) (-.80666023,1.0828604) (-.77779102,.82439202) (-.7454266,.972956) (-.72251296,1.1535482) (-.69803774,.92829674) (-.66708153,.91120625) (-.63698494,.98634213) (-.61565471,.7971468) (-.58660489,.88908339) (-.55904412,1.0420948) (-.53245562,.86466718) (-.50765651,1.067537) (-.4820697,.89232045) (-.45222196,.97612619) (-.41956902,.96565962) (-.39398497,1.0078353) (-.37302822,.95619154) (-.3446857,.89370584) (-.31822267,.9520281) (-.29118553,.97639734) (-.26643601,.9037329) (-.23613068,.99935126) (-.20805943,1.0208778) (-.17758821,1.0192231) (-.14932831,1.071558) (-.11720977,1.0019411) (-.08357117,1.0931627) (-.06186111,1.0866063) (-.03231926,.92515975) (.00064717,.92100006) (.03167304,.82638389) (.05397142,1.1079237) (.08620813,.99444151) (.11491465,1.1347774) (.14350598,1.0615413) (.17271778,1.1054713) (.20096838,.99043393) (.23780276,.93413395) (.27305856,1.0956944) (.30583581,1.0092416) (.34228629,.94992119) (.37931246,.9383772) (.4165445,1.0597813) (.45554307,1.0351424) (.49139285,1.0239874) (.53696781,1.0547767) (.57727659,1.040958) (.62026125,1.0794125) (.67592561,1.0473866) (.73067516,.91283399) (.77974641,1.0137745) (.8309806,1.0698802) (.8791433,1.1437507) (.93913513,.97098559) (1.0039223,1.1680431) (1.0734113,1.0528153) (1.1523958,1.191324) (1.2282346,1.1579767) };
\addplot[domain=-3.25:1.75, color=black, solid, line width=0.75mm] {1.006111+0.0524008*x};
\legend{~Observations,~Linear Fit}
\end{axis}
\end{tikzpicture}
}
\end{subfigure}
\vskip \baselineskip
\begin{subfigure}{1\textwidth}
\centering
\caption{Search Duration}
\scalebox{0.725}{
\begin{tikzpicture}
\begin{axis}[xlabel=Log Labor Market Tightness, ylabel= Log Search Duration, xmin=-3.5, xmax=2, ymin=2.5, ymax=4.5, height=8cm, width=16cm, grid=major, legend pos = south east, xtick={-3,-1.5,0,1.5}, ytick={2.75,3.5,4.25}, grid style = dotted, y tick label style={/pgf/number format/.cd,fixed,fixed zerofill, precision=2,/tikz/.cd}]
\addplot[only marks, mark=o, mark options={solid,scale=1}, color=black] coordinates {  (-2.8374276,3.3118186) (-2.6899104,3.5618114) (-2.5584483,2.0196669) (-2.4633279,3.1735611) (-2.4023955,2.9026175) (-2.3262157,3.4716659) (-2.2620926,2.0314107) (-2.2100151,2.9047072) (-2.1463211,3.5054069) (-2.0805862,3.260129) (-2.0261538,3.6089962) (-1.9614238,3.5090127) (-1.8997936,3.3951757) (-1.8683676,3.1600378) (-1.8281507,3.4524336) (-1.7705679,2.7035334) (-1.7371924,3.5688498) (-1.6866912,3.6870613) (-1.6371644,3.6210933) (-1.590521,3.6451311) (-1.5490052,3.5243502) (-1.502726,3.3279817) (-1.4569894,3.3721979) (-1.4239204,3.1458318) (-1.386471,3.543138) (-1.3527849,3.3810067) (-1.3151757,3.5905871) (-1.2806253,3.5109534) (-1.2399601,3.5745938) (-1.2005343,3.2362242) (-1.1697563,3.7073767) (-1.1392791,3.4813948) (-1.1093833,3.4947536) (-1.0803162,3.497344) (-1.0509958,3.044234) (-1.0170809,3.7545502) (-.98765481,3.7574334) (-.95076895,3.3210151) (-.91682923,3.4189217) (-.8898589,3.5548499) (-.85880792,3.4531589) (-.83387202,3.4768288) (-.81068873,3.7357254) (-.78226167,3.5426881) (-.74913251,3.2959437) (-.72671884,3.0810218) (-.70524395,3.4367998) (-.67396623,3.7734482) (-.64401549,3.4894745) (-.61773443,3.5301716) (-.58958948,3.4469628) (-.56198531,3.611865) (-.53518504,3.395709) (-.51085073,3.6598577) (-.48860186,3.2763088) (-.4581565,3.4958234) (-.43173733,3.4844029) (-.403341,3.5742774) (-.38056621,3.695039) (-.35132924,3.7128904) (-.32487816,3.6705227) (-.29558712,3.5687249) (-.27390015,3.6465619) (-.24395543,3.6243243) (-.21314646,3.8557818) (-.18162738,3.5733807) (-.15392423,4.0076489) (-.12239753,3.5943935) (-.08705677,3.6240225) (-.06264638,3.8308177) (-.03378947,3.6992688) (-.00159306,3.5576937) (.03129885,3.2795084) (.05029022,3.8566952) (.07742524,3.7603076) (.10875453,3.5054836) (.13495341,3.6257029) (.16459388,3.8334656) (.19248535,3.5243771) (.2283501,3.7457833) (.26081219,3.5021567) (.29327044,3.3853145) (.32840586,3.6922829) (.3649528,3.6859314) (.40190482,3.6497667) (.44207451,3.6227932) (.47376737,3.9378698) (.51677227,3.3884003) (.56158596,3.6316988) (.59905434,3.8224206) (.65165216,3.8088856) (.70696223,3.7316427) (.76421076,3.7687123) (.82060039,3.8281281) (.86652082,3.9652469) (.9251287,3.8329945) (.99250561,4.1147723) (1.0659438,4.0462518) (1.1460317,3.8147905) (1.2227739,4.0074887) };
\addplot[domain=-3.25:1.75, color=black, solid, line width=0.75mm] {3.655113+0.2056402*x};
\legend{~Observations,~Linear Fit}
\end{axis}
\end{tikzpicture}
}
\end{subfigure}
\linespread{1} \floatfoot{\footnotesize\textsc{Note. ---} The figures show binned scatterplots with 100 markers to depict cross-sectional correlations between log labor market tightness and the log of the number of applicants, the number of search channels, and search duration (in days). Labor markets are combinations of 5-digit KldB occupations and commuting zones. We trim labor market tightness at the 5th and 95th percentile. The numbers of observed successful hires are: 44,020 for the number of applicants, 52,063 for the number of search channels, and 42,747 for search duration. Sources: Official Statistics of Federal Employment Agency $\plus$ IAB Job Vacancy Survey, 2012-2019.} \linespread{1.5}
\end{figure}

\pgfkeys{/pgf/number format/.cd,1000 sep={}}

\vspace*{\fill}

\clearpage

\vspace*{\fill}

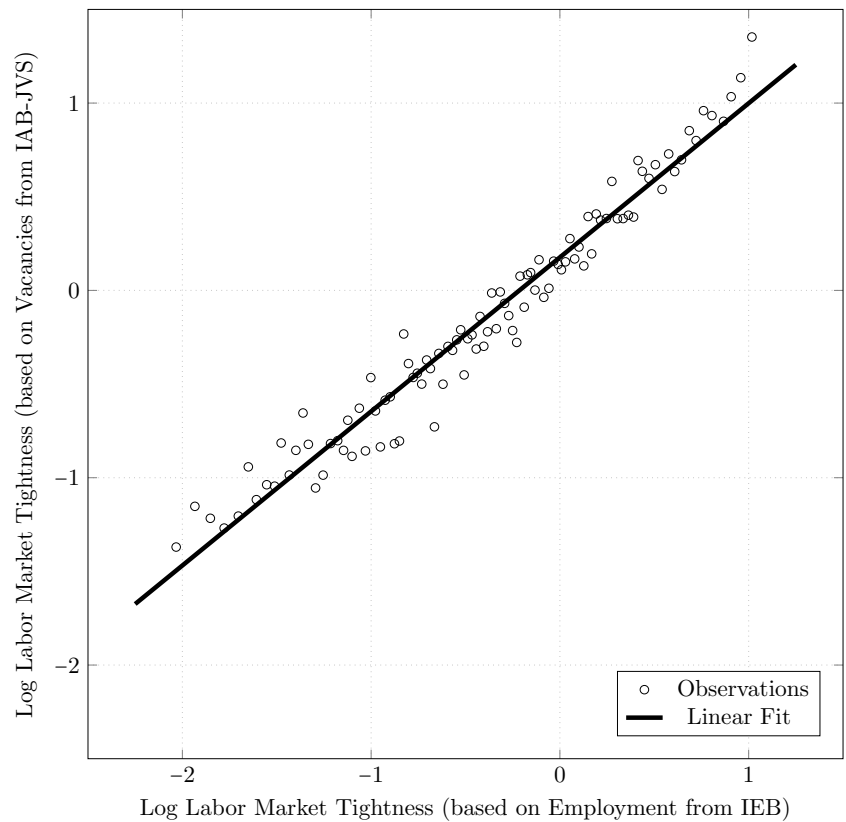
\begin{figure}[!ht]
\caption{Vacancy- vs.\ Employment-Based Firm-Specific Labor Market Tightness}
\label{fig:D7}
\centering
\scalebox{0.80}{
\begin{tikzpicture}
\begin{axis}[xlabel={Log Labor Market Tightness (based on Employment from IEB)},  ylabel={Log Labor Market Tightness (based on Vacancies from IAB-JVS)}, xmin=-2.5, xmax=1.5, ymin=-2.5, ymax=1.5, height=14cm, width=14cm, grid=major, legend pos = south east, xtick={-2,-1,0,1}, ytick={-2,-1,0,1}, grid style = dotted, tick label style={/pgf/number format/.cd,fixed,fixed zerofill, precision=0,/tikz/.cd},  scaled ticks=false]
\addplot[only marks, mark=o, mark options={solid,scale=1}, color=black] coordinates {   (-2.0329082,-1.3701116) (-1.9337039,-1.1527745) (-1.8518358,-1.2166246) (-1.7786123,-1.2687579) (-1.7034744,-1.2047915) (-1.6505995,-.94211406) (-1.6079509,-1.1171309) (-1.5532826,-1.0374767) (-1.511281,-1.0450898) (-1.4766895,-.81458378) (-1.4334438,-.9855929) (-1.3987023,-.85378903) (-1.3609943,-.65440589) (-1.3325661,-.82173836) (-1.2943584,-1.0546398) (-1.2544458,-.98609537) (-1.2150888,-.81756014) (-1.1772823,-.80223376) (-1.1461782,-.85400563) (-1.1236143,-.69312799) (-1.1010923,-.8859511) (-1.0626009,-.62894791) (-1.0300868,-.85682321) (-1.0017204,-.46550846) (-.97723567,-.64346498) (-.95102108,-.83546531) (-.92562348,-.58610934) (-.89910412,-.5681079) (-.87648451,-.81868017) (-.85001153,-.80406708) (-.82742876,-.23267259) (-.80212116,-.3905828) (-.77732366,-.4647342) (-.755472,-.44209149) (-.73191673,-.50010526) (-.70594925,-.3716515) (-.68616331,-.41743916) (-.66463888,-.72837472) (-.6412316,-.33640298) (-.61939144,-.5007779) (-.59292346,-.29891142) (-.56871051,-.32006061) (-.54641569,-.2640813) (-.52559358,-.21026748) (-.50736314,-.45124304) (-.48819026,-.25825769) (-.46562114,-.23747538) (-.44331077,-.31293353) (-.42305675,-.13913333) (-.40317005,-.29819584) (-.3831856,-.22088158) (-.36121109,-.01387901) (-.33690834,-.20461471) (-.3160252,-.00819584) (-.29292467,-.06922905) (-.27120465,-.13458692) (-.25043187,-.21442337) (-.2287107,-.27783221) (-.2110842,.0761397) (-.18909955,-.08970277) (-.17221926,.0832437) (-.15526181,.09497342) (-.13209428,.00167385) (-.11057023,.16346119) (-.08517107,-.03702335) (-.05824137,.01134181) (-.03290042,.15615059) (-.00950208,.13879026) (.00802607,.11011216) (.0294174,.1523798) (.05314385,.27645922) (.07834558,.16820739) (.10094077,.23171584) (.12666927,.13085647) (.14957222,.3942259) (.16829991,.19498865) (.19236897,.40797421) (.21578343,.37506852) (.24650934,.38377264) (.27532136,.58184469) (.30483675,.38248429) (.33509526,.38337353) (.36297512,.40154535) (.39029887,.39143589) (.41405866,.69313842) (.43664008,.63577282) (.47212598,.59792572) (.50581819,.67106187) (.54124868,.53918684) (.57631183,.72855145) (.60819751,.63435388) (.64414603,.69707012) (.68543756,.8525874) (.72160959,.79975587) (.76077205,.95942092) (.80550075,.93318939) (.86712229,.90262449) (.90725768,1.0338838) (.95817935,1.1353928) (1.0172114,1.3527246) };
\addplot[domain=-2.25:1.25, color=black, solid, line width=0.75mm] {0.1764801+0.8225332 *x};
\legend{~Observations,Linear Fit}
\end{axis}
\end{tikzpicture}
}
\floatfoot{\footnotesize\textsc{Note. ---} The figure shows a binned scatterplot with 100 hundred markers to contrast our employment-based measure of log firm-specific labor market tightness with an analogous vacancy-based measure for the very same firm. While our employment-based measure builds on administrative employment shares in the IEB, the vacancy-based measure is constructed from cross-sectional information in the IAB Job Vacancy Survey on a subset of firms' top five occupations with the highest number of unfilled vacancies. The number of observations is 24,323. Sources: Integrated Employment Biographies $\plus$ Official Statistics of the German Federal Employment Agency $\plus$ IAB Job Vacancy Survey, 2012-2019.}
\end{figure}

\vspace*{\fill}

\clearpage

\section{Bartik-Instrument Diagnostics}
\label{sec:E}

\setcounter{table}{0} 
\setcounter{figure}{0} 

\vspace*{-0.5cm}

\begin{table}[!ht]

\small

\caption{Summary of Rotemberg Weights}
\label{tab:E1}
\scalebox{0.975}{
\begin{threeparttable}
\begin{subtable}{1\textwidth}

\caption{\normalsize Effect of Wage Rate}
\centering
\begin{tabular}{L{5.2cm}C{1.6cm}C{1.6cm}C{1.6cm}C{1.6cm}C{1.6cm}} \hline

\multicolumn{6}{l}{\textbf{\multirow{2}{*}{Panel A: Negative and Positive Rotemberg Weights}}} \\
&&&&& \\
                & Sum                    & Mean                  & Share               && \\
Positive        & 1.0908                 & 0.0015                & 0.9232              && \\
Negative        & -0.0908\hphantom{-}    & -0.0002\hphantom{-}   & 0.0768 && \\
Overall         & 1.0000                  & 0.0008                & 1.0000             && \\
&&&&& \\[-0.2cm] \hline

\multicolumn{6}{l}{\textbf{\multirow{2}{*}{Panel B: Correlations}}} \\
&&&&& \\[-0.4cm]

 & $\hat{\alpha}_o$  & $g_o$ & $\hat{\eta}_o$ & $\hat{F}_o$ &  $\sqrt{var(z_o)}$ \\
Rotemberg Weight ($\hat{\alpha}_o$)                     & 1                  &                    &               &       & \\
National Growth Rate ($g_o$)                               & 0.1850 & 1                  &               &       & \\
Just-Identified Coefficient ($\hat{\eta}_o$)             & -0.0845\hphantom{-}              & -0.0340\hphantom{-} &       1       &       & \\
First-Stage F Statistic ($\hat{F}_o$)                               & 0.9118 & 0.0969 & -0.0439\hphantom{-}      &       1       &         \\
Variance of Shares  ($\sqrt{var(z_{o})}$) & 0.4282              & -0.0146\hphantom{-} & -0.1582\hphantom{-}      & 0.4411 & 1\\
&&&&& \\[-0.2cm] \hline

\multicolumn{6}{l}{\textbf{\multirow{2}{*}{Panel C: Rotemberg Weights by Years}}} \\
&&&&& \\
                & Sum                    & Mean    &&& \\
2012 - 2014       & 0.1155                 & 0.0001  &&& \\
2013 - 2015       & 0.2691                 & 0.0002  &&& \\
2014 - 2016       & 0.2716                 & 0.0002  &&& \\
2015 - 2017       & 0.0952                 & 0.0001  &&& \\
2016 - 2018       & 0.1072                 & 0.0001  &&& \\
2017 - 2019       & 0.1414                 & 0.0001  &&& \\

&&&&& \\[-0.2cm] \hline

\multicolumn{6}{l}{\textbf{\multirow{2}{*}{Panel D: Top Five Occupations by Rotemberg Weights}}} \\
&&&&& \\[-0.4cm]
                             & $\hat{\alpha}_o$  & $g_o$ & $\hat{\eta}_o$  & $\hat{F}_o$ & \shortstack{Occ.\vphantom{/} \\ Share} \\
Gastronomy Workers (II)       & 0.1396 & 0.0501 & -0.3495\hphantom{-} & 903.94    & 2.1095 \\
Medical Assistants (II)       & 0.1139 & 0.0264 & -1.3368\hphantom{-} & 941.03    & 1.5520 \\
Hairdressers (II)             & 0.0966 & 0.0496 & -2.0212\hphantom{-} & 506.85    & 0.5531 \\
Cooks (II)                    & 0.0965 & 0.0377 & -0.7576\hphantom{-} & 889.64    & 1.9774 \\
Farmers (I)                   & 0.0789 & 0.0455 &  0.0250             & 339.04    & 0.2859 \\

&&&&& \\[-0.2cm] \hline

\multicolumn{6}{l}{\textbf{\multirow{2}{*}{Panel E: Just-Identified Coefficients by Negative and Positive Rotemberg Weights}}} \\
&&&&& \\
                & \shortstack{$\hat{\alpha}$-Weight. \\ Sum}   & Share  & Mean    && \\
Positive        & -0.4564\hphantom{-}    & 0.6268   & -4.4440\hphantom{-} && \\
Negative        & -0.2717\hphantom{-}    & 0.3732   & 1.3175 && \\
Overall         & -0.7280\hphantom{-}    & 1.0000   & -2.1205\hphantom{-} && \\[0.2cm] \hline
\end{tabular}
\end{subtable}

\begin{tablenotes}[para]
\footnotesize\textsc{Note. ---}  The table displays statistics about the Rotemberg weights underlying our estimated wage effect on labor demand. For ease of computation, we derive the statistics by running specification (2) in Table \ref{tab:1} on a random 50 percent sample of firms. In all cases, we report statistics about the aggregated weights with normalized growth rates (i.e., we subtract the per-period average across occupations). Panel A reports the share, mean, and sum by negative and positive Rotemberg weights. For the occupations with the 100 highest absolute Rotemberg weights, Panel B delivers correlations between the Rotemberg weights, the normalized national two-year growth rates, the just-identified coefficient estimates, the first-stage F statistics of the occupational employment share in the base year, and the standard deviations in the occupational employment shares across firms. Panel C displays the sum of Rotemberg weights across years (in terms of two-year intervals). Panel D describes the top five occupations with the largest Rotemberg weights, including the occupational employment share in the overall labor market (multiplied by 100 for legibility). The Roman number (in parentheses) denotes the level of skill requirements: helpers (I), professionals (II), specialists (III), or experts (IV). Panel E shows how the values of the just-identified coefficients vary by positive and negative Rotemberg weights. CI = Confidence Interval. Occ. = Occupation. Weight. = Weighted. Sources: Integrated Employment Biographies $\plus$ Official Statistics of the German Federal Employment Agency $\plus$ IAB Job Vacancy Survey, 1999-2019.
\end{tablenotes}
\end{threeparttable}
}
\end{table}

\clearpage

\begin{table}[!ht]
\addtocounter{table}{-2}
\small
\caption{Summary of Rotemberg Weights (Cont.)}
\scalebox{0.975}{
\begin{threeparttable}
\begin{subtable}{1\textwidth}
\addtocounter{subtable}{1}
\caption{\normalsize Effect of Labor Market Tightness: Vacancy Instrument}
\centering
\begin{tabular}{L{5.2cm}C{1.6cm}C{1.6cm}C{1.6cm}C{1.6cm}C{1.6cm}} \hline

\multicolumn{6}{l}{\textbf{\multirow{2}{*}{Panel A: Negative and Positive Rotemberg Weights}}} \\
&&&&& \\
                & Sum                    & Mean                  & Share               && \\
Positive        & 1.6418                 & 0.0021                & 0.6078              && \\
Negative        & -0.0979\hphantom{-}    & -0.0003\hphantom{-}   & 0.0362 && \\
Overall         & 1.5439                  & 0.0015                & 0.6440            && \\
&&&&& \\[-0.2cm] \hline

\multicolumn{6}{l}{\textbf{\multirow{2}{*}{Panel B: Correlations}}} \\
&&&&& \\[-0.4cm]

 & $\hat{\alpha}_o$  & $g_o$ & $\hat{\eta}_o$ & $\hat{F}_o$ &  $\sqrt{var(z_o)}$ \\
Rotemberg Weight ($\hat{\alpha}_o$)     & 1                  &                    &               &       & \\
National Growth Rate ($g_o$)                & -0.0597\hphantom{-}  & 1                  &               &       & \\
Just-Identified Coefficient ($\hat{\eta}_o$)   & -0.1327\hphantom{-}           & 0.0859 &       1       &       & \\
First-Stage F Statistic ($\hat{F}_o$)       & 0.5562 & -0.0785\hphantom{-}  & -0.0892\hphantom{-}      &       1       &         \\
Variance of Shares  ($\sqrt{var(z_{o})}$) & 0.5520 & -0.0386\hphantom{-} & -0.0776\hphantom{-}      & 0.4288 & 1\\
&&&&& \\[-0.2cm] \hline

\multicolumn{6}{l}{\textbf{\multirow{2}{*}{Panel C: Rotemberg Weights by Years}}} \\
&&&&& \\
                & Sum                    & Mean    &&& \\
2012 - 2014       & 0.3282                 & 0.0003  &&& \\
2013 - 2015       & 0.3080                 & 0.0003  &&& \\
2014 - 2016       & 0.2336                 & 0.0002  &&& \\
2015 - 2017       & 0.2164                 & 0.0002  &&& \\
2016 - 2018       & 0.2739                 & 0.0003  &&& \\
2017 - 2019       & 0.1839                 & 0.0002  &&& \\

&&&&& \\[-0.2cm] \hline

\multicolumn{6}{l}{\textbf{\multirow{2}{*}{Panel D: Top Five Occupations by Rotemberg Weights}}} \\
&&&&& \\[-0.4cm]
                             & $\hat{\alpha}_o$  & $g_o$ & $\hat{\eta}_o$  & $\hat{F}_o$ & \shortstack{Occ.\vphantom{/} \\ Share} \\
Sale Workers (II)            & 0.1971 & 0.3826 & -0.1788\hphantom{-} & 11,318    & 6.2581 \\
Bankers (II)                 & 0.1201 & 0.1336 & -0.0711\hphantom{-} & 10,004     & 1.4968 \\
Farmers (I)                  & 0.1187 & -0.1799\hphantom{-} & -0.0657\hphantom{-} & 2704.6    & 0.2859 \\
Chimney Sweeps (II)          & 0.0745 & -0.6519\hphantom{-} & -0.0035\hphantom{-} & 2210.8     & 0.0320\\
Construction Workers (I)     & 0.0735 & 0.2912 &  0.0967             & 1733.2     & 1.0320 \\

&&&&& \\[-0.2cm] \hline

\multicolumn{6}{l}{\textbf{\multirow{2}{*}{Panel E: Just-Identified Coefficients by Negative and Positive Rotemberg Weights}}} \\
&&&&& \\
                & \shortstack{$\hat{\alpha}$-Weight.\\ Sum}   & Share  & Mean    && \\
Positive        & -0.0234\hphantom{-}    & 0.5714   & 1.9699 && \\
Negative        & -0.0069\hphantom{-}    & 0.1681   & 1.1189 && \\
Overall         & -0.0302\hphantom{-}    & 0.7395   & 1.7323 && \\[0.2cm] \hline
\end{tabular}
\end{subtable}

\begin{tablenotes}[para]
\footnotesize\textsc{Note. ---}  The table displays statistics about the Rotemberg weights underlying our estimated effect of labor market tightness on labor demand. For ease of computation, we derive the statistics by running specification (3) in Table \ref{tab:1} on a random 50 percent sample of firms. In all cases, we report statistics about the aggregated weights with normalized growth rates (i.e., we subtract the per-period average across occupations). Panel A reports the share, mean, and sum by negative and positive Rotemberg weights. For the occupations with the 100 highest absolute Rotemberg weights, Panel B delivers correlations between the Rotemberg weights, the normalized national two-year growth rates, the just-identified coefficient estimates, the first-stage F statistics of the occupational employment share in the base year, and the standard deviations in the occupational employment shares across firms. Panel C displays the sum of Rotemberg weights across years (in terms of two-year intervals). Panel D describes the top five occupations with the largest Rotemberg weights, including the occupational employment share in the overall labor market (multiplied by 100 for legibility). The Roman number (in parentheses) denotes the level of skill requirements: helpers (I), professionals (II), specialists (III), or experts (IV). Panel E shows how the values of the just-identified coefficients vary by positive and negative Rotemberg weights. CI = Confidence Interval. Occ. = Occupation. Weight. = Weighted. Sources: Integrated Employment Biographies $\plus$ Official Statistics of the German Federal Employment Agency $\plus$ IAB Job Vacancy Survey, 1999-2019.
\end{tablenotes}
\end{threeparttable}
}
\end{table}

\clearpage

\begin{table}[!ht]
\addtocounter{table}{-2}
\small
\caption{Summary of Rotemberg Weights (Cont.)}
\scalebox{0.975}{
\begin{threeparttable}
\begin{subtable}{1\textwidth}
\addtocounter{subtable}{2}
\caption{\normalsize Effect of Labor Market Tightness: Job Seeker Instrument}
\centering
\begin{tabular}{L{5.2cm}C{1.6cm}C{1.6cm}C{1.6cm}C{1.6cm}C{1.6cm}} \hline

\multicolumn{6}{l}{\textbf{\multirow{2}{*}{Panel A: Negative and Positive Rotemberg Weights}}} \\
&&&&& \\
                & Sum                    & Mean                  & Share               && \\
Positive        & 0.2088                 & 0.0004                & 0.0773              && \\
Negative        & -0.7527\hphantom{-}    & -0.0011\hphantom{-}   & 0.2787              && \\
Overall         & -0.5439\hphantom{-}    & -0.0005\hphantom{-}  & 0.3560            && \\
&&&&& \\[-0.2cm] \hline

\multicolumn{6}{l}{\textbf{\multirow{2}{*}{Panel B: Correlations}}} \\
&&&&& \\[-0.4cm]

 & $\hat{\alpha}_o$  & $g_o$ & $\hat{\eta}_o$ & $\hat{F}_o$ &  $\sqrt{var(z_o)}$ \\
Rotemberg Weight ($\hat{\alpha}_o$)     & 1                  &                    &               &       & \\
National Growth Rate ($g_o$)            & 0.0242 & 1                  &               &       & \\
Just-Identified Coefficient ($\hat{\eta}_o$)   & 0.0715   & 0.1770 &       1       &       & \\
First-Stage F Statistic ($\hat{F}_o$)   & -0.7440\hphantom{-} & -0.0082\hphantom{-}  & -0.0327\hphantom{-}      &       1       &         \\
Variance of Shares  ($\sqrt{var(z_{o})}$) & -0.3690\hphantom{-}             & -0.0857\hphantom{-} & -0.1714\hphantom{-}      & 0.4148 & 1\\
&&&&& \\[-0.2cm] \hline

\multicolumn{6}{l}{\textbf{\multirow{2}{*}{Panel C: Rotemberg Weights by Years}}} \\
&&&&& \\
                & Sum                    & Mean    &&& \\
2012 - 2014       & -0.0804\hphantom{-}                 & -0.0001\hphantom{-}  &&& \\
2013 - 2015       & -0.0932\hphantom{-}                 & -0.0001\hphantom{-}  &&& \\
2014 - 2016       & -0.0914\hphantom{-}                 & -0.0001\hphantom{-}  &&& \\
2015 - 2017       & -0.1411\hphantom{-}                 & -0.0001\hphantom{-}  &&& \\
2016 - 2018       & -0.0854\hphantom{-}                 & -0.0001\hphantom{-}  &&& \\
2017 - 2019       & -0.0524\hphantom{-}                 & -0.0000\hphantom{-}  &&& \\

&&&&& \\[-0.2cm] \hline

\multicolumn{6}{l}{\textbf{\multirow{2}{*}{Panel D: Top Five Occupations by Rotemberg Weights}}} \\
&&&&& \\[-0.4cm]
                             & $\hat{\alpha}_o$  & $g_o$ & $\hat{\eta}_o$  & $\hat{F}_o$ & \shortstack{Occ.\vphantom{/} \\ Share} \\
Sale Workers (II)       & -0.0996\hphantom{-} & -0.1982\hphantom{-} & -0.2532\hphantom{-} & 11,975    & 6.2581 \\
Masons (II)             & -0.0459\hphantom{-} & -0.4563\hphantom{-} & -0.0613\hphantom{-} & 2166.7     & 0.9448 \\
Truck Drivers (II)      & -0.0423\hphantom{-} & -0.2329\hphantom{-} & 0.0105 & 3957.7   & 3.1550 \\
Bankers (II)            & -0.0366\hphantom{-} & -0.1941\hphantom{-} & -0.1385\hphantom{-} & 8205.3     & 1.4968 \\
Gardeners (II)          & -0.0360\hphantom{-} & -0.4321\hphantom{-} &  0.0268             & 488.95    & 0.7031 \\

&&&&& \\[-0.2cm] \hline

\multicolumn{6}{l}{\textbf{\multirow{2}{*}{Panel E: Just-Identified Coefficients by Negative and Positive Rotemberg Weights}}} \\
&&&&& \\
                & \shortstack{$\hat{\alpha}$-Weight. \\ Sum}   & Share  & Mean    && \\
Positive        & -0.0069\hphantom{-}    & 0.0149   & -3.5825\hphantom{-} && \\
Negative        & -0.1133\hphantom{-}    & 0.2456   & 0.0219 && \\
Overall         & -0.1202\hphantom{-}    & 0.2605   & -1.4997\hphantom{-} && \\[0.2cm] \hline
\end{tabular}
\end{subtable}

\begin{tablenotes}[para]
\footnotesize\textsc{Note. ---}  The table displays statistics about the Rotemberg weights underlying our estimated effect of labor market tightness on labor demand. For ease of computation, we derive the statistics by running specification (3) in Table \ref{tab:1} on a random 50 percent sample of firms. In all cases, we report statistics about the aggregated weights with normalized growth rates (i.e., we subtract the per-period average across occupations). Panel A reports the share, mean, and sum by negative and positive Rotemberg weights. For the occupations with the 100 highest absolute Rotemberg weights, Panel B delivers correlations between the Rotemberg weights, the normalized national two-year growth rates, the just-identified coefficient estimates, the first-stage F statistics of the occupational employment share in the base year, and the standard deviations in the occupational employment shares across firms. Panel C displays the sum of Rotemberg weights across years (in terms of two-year intervals). Panel D describes the top five occupations with the largest Rotemberg weights, including the occupational employment share in the overall labor market (multiplied by 100 for legibility). The Roman number (in parentheses) denotes the level of skill requirements: helpers (I), professionals (II), specialists (III), or experts (IV). Panel E shows how the values of the just-identified coefficients vary by positive and negative Rotemberg weights. CI = Confidence Interval. Occ. = Occupation. Weight. = Weighted. Sources: Integrated Employment Biographies $\plus$ Official Statistics of the German Federal Employment Agency $\plus$ IAB Job Vacancy Survey, 1999-2019.
\end{tablenotes}
\end{threeparttable}
}
\end{table}

\normalsize

\vfill

\clearpage

\begin{landscape}
\pgfkeys{/pgf/number format/.cd,1000 sep={,}}

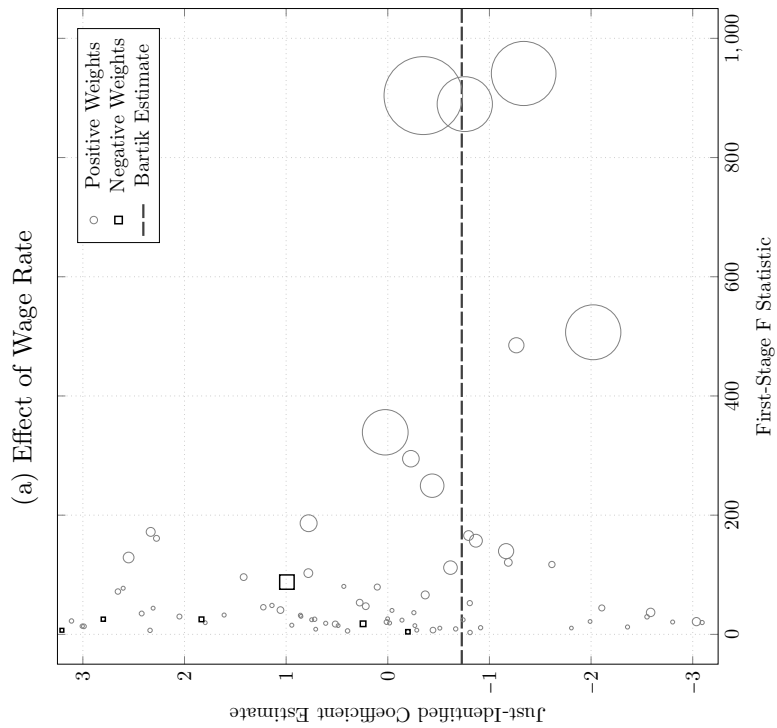
\begin{figure}[!ht]
\centering
\caption{Heterogeneity of Just-Identified Coefficient Estimates}
\label{fig:E1}
\vspace*{0.5cm}
\begin{subfigure}{1\textwidth}
\caption{Effect of Wage Rate}
\centering
\scalebox{0.70}{
\begin{tikzpicture}
\begin{axis}[xlabel=First-Stage F Statistic, ylabel=Just-Identified Coefficient Estimate, xmin=-50, xmax=1050, ymin=-3.25, ymax=3.25,  height=14cm, width=14cm, grid=major, legend pos = north east, xtick={0,200,400,600,800,1000}, ytick={-3,-2,-1,0,1,2,3}, grid style = dotted]
\addplot[scatter, only marks, mark=o, mark options={solid,scale=1}, color=gray, point meta=explicit, scatter/use mapped color={fill=none, fill opacity=0, draw=gray},scatter/@pre marker code/.append style={/tikz/mark size=1+\pgfplotspointmetatransformed/50}] coordinates {   (903.93915,-.34952611) [.13962484] (941.0343,-1.3367988) [.11386675] (506.84952,-2.0212152) [.09655652] (889.63904,-.75760341) [.09648729] (339.03561,.02500887) [.07890134] (52.156094,5.9033208) [.03811422] (249.45236,-.43652955) [.03786326] (98.403465,-5.4515953) [.02629144] (186.46416,.77879024) [.02554664] (294.71289,-.22727583) [.02492748] (485.0426,-1.2649362) [.02240059] (139.64143,-1.1640579) [.02231742] (111.98027,-.61674654) [.01938298] (157.13113,-.8668164) [.01810288] (128.96136,2.5499258) [.01410434] (165.80963,-.79496431) [.01242085] (171.92494,2.3325562) [.01076899] (102.7468,.78167123) [.01044869] (36.86121,-2.5859182) [.00991288] (21.374252,-3.0365245) [.00950112] (66.172798,-.36835426) [.00898322] (120.77717,-1.1857738) [.00853898] (96.063354,1.4179738) [.00679439] (47.24707,.21662888) [.00677468] (27.355598,-3.958173) [.00664015] (40.755215,1.0559934) [.00661133] (53.267826,.2765446) [.00656999] (117.22892,-1.6146108) [.00583015] (79.358917,.10344364) [.00542406] (44.33205,-2.1056597) [.00541275] (17.309742,.51620489) [.00536076] (161.11909,2.2755702) [.00516447] (7.1708899,-.44472295) [.00468626] (71.870338,2.655205) [.00429028] (45.488541,1.2244292) [.00424175] (45.066914,-4.655004) [.00415458] (52.443981,-.80685091) [.00382355] (300.60965,6.0083804) [.00378341] (29.89624,2.0498009) [.00339997] (34.994934,2.4213369) [.00324154] (20.712124,.01190015) [.00318417] (8.9194736,-3.7056482) [.00297738] (13.423342,2.9895182) [.0028963] (29.483305,-2.5506649) [.00250209] (5.8159723,.39575797) [.00243044] (3.1360862,-.80928904) [.00241914] (25.321747,.71992815) [.0024131] (22.475651,3.1118743) [.00239318] (6.7355742,2.3389592) [.00238279] (69.123924,-3.4255476) [.00234259] (9.1469231,-.66805595) [.00220505] (48.529423,1.1396457) [.00218314] (20.791588,3.7315426) [.00213626] (11.1569,-.91310066) [.00209662] (13.754658,3.006228) [.00201824] (19.839788,-3.092798) [.00196554] (36.387554,-.25670895) [.00189753] (12.370636,-2.3586624) [.00183585] (32.05801,.85942692) [.00180612] (18.606661,.61081469) [.00179788] (24.424488,-.74049193) [.00179532] (16.173727,4.2871251) [.0017557] (32.41954,1.6102222) [.00171115] (80.488922,.43259117) [.00169552] (9.2386904,4.9886303) [.00165235] (15.385029,.94438916) [.00164655] (23.885809,-.1395396) [.00160337] (20.716635,-2.80269) [.00157969] (77.428574,2.6033676) [.00157405] (19.011614,-.01783634) [.00154282] (40.188591,-.04121056) [.00153373] (2.0889478,-25.259684) [.00144229] (10.427429,-.51204425) [.00144195] (24.537287,.74869668) [.00144133] (7.0668802,-.2851761) [.00143845] (43.958874,2.3088996) [.00141683] (30.577349,4.8232269) [.00135328] (8.66786,.70900804) [.00128125] (19.562511,1.7960501) [.00127263] (16.315868,-6.0838366) [.00125421] (14.704659,-.26639685) [.00124806] (17.006538,4.5969796) [.00124489] (30.171667,.85429001) [.00124061] (14.715938,.48829216) [.00117865] (21.620405,-1.9900498) [.00117189] (26.251488,.00173011) [.00114689] (10.610194,-1.8064791) [.00107502] };
\addplot[scatter, only marks, thick, mark=square, mark options={solid,scale=1}, color=black, point meta=explicit, scatter/use mapped color={fill=none, fill opacity=0, draw=black},scatter/@pre marker code/.append style={/tikz/mark size=1+\pgfplotspointmetatransformed/50}] coordinates {   (87.733955,.99255598) [.02090371] (17.727972,.24305987) [.00447274] (25.218122,1.8329296) [.00280212] (29.425564,7.6663833) [.00253634] (10.452282,4.2840543) [.00234777] (4.3344221,-.19790791) [.00220144] (19.563606,3.3625171) [.00186492] (25.44553,2.7989867) [.00174006] (4.0388112,-9.5440416) [.00151208] (7.1791267,11.672392) [.00144616] (6.8920736,3.2062538) [.0010956] (21.08564,3.9577937) [.0010945] (.67573011,3.4290776) [.00108275] };
\addplot[loosely dashed, color=darkgray, very thick, dash pattern = on 7.6pt off 2pt] coordinates {(-50,-0.7280) (1050,-0.7280)};
\legend{~Positive Weights,~Negative Weights, ~Bartik Estimate}
\end{axis}
\end{tikzpicture}
}
\end{subfigure}
\floatfoot{\footnotesize\textsc{Note: ---} The figure visualizes the relationship between each instruments’ just-identified coefficient estimate, the first-stage F statistics, and the Rotemberg weights. Each marker refers to a separate instrument’s estimates (occupation share). The figure plots the estimated just-identified coefficents for each instrument (i.e., occupational employment share in the base year) against the respective first-stage F statistic. The size of the markers are proportional to the the absolute value of the Rotemberg weights, with the circles denoting positive weights and the squares denoting negative weights. The horizontal dashed line reflects the overall Bartik estimate. For reasons of parsimony, the figure includes only the 50 instruments with the highest absolute Rotemberg weights, which account for 78.1 percent of the sum of absolute weights. Sources: Integrated Employment Biographies $\plus$ Official Statistics of the German Federal Employment Agency $\plus$ IAB Job Vacancy Survey, 1999-2019. }
\end{figure}

\clearpage

\begin{figure}[!ht]
\addtocounter{figure}{-1}
\centering
\caption{Heterogeneity of Just-Identified Coefficient Estimates (Cont.)}
\vspace*{0.5cm}
\begin{subfigure}{0.475\textwidth}
\addtocounter{subfigure}{1}
\caption{\normalsize{Effect of Labor Market Tightness: Vacancy Instrument}}
\centering
\scalebox{0.70}{
\begin{tikzpicture}
\begin{axis}[xlabel=First-Stage F Statistic, ylabel=Just-Identified Coefficient Estimate, xmin=-1000, xmax=16000, ymin=-0.10, ymax=0.10, height=14cm, width=14cm, grid=major, legend pos = north east, xtick={0,2500,5000,7500,10000,12500,15000}, ytick={-0.09,-0.06,-0.03,0,0.03,0.06,0.09}, grid style = dotted, y tick label style={/pgf/number format/.cd,fixed,fixed zerofill, precision=2,/tikz/.cd}]
\addplot[scatter, only marks, mark=o, mark options={solid,scale=1}, color=gray, point meta=explicit, scatter/use mapped color={fill=none, fill opacity=0, draw=gray},scatter/@pre marker code/.append style={/tikz/mark size=1+\pgfplotspointmetatransformed/50}] coordinates {   (11317.608,-.17877197) [.19705902] (10004.769,-.07113867) [.12007242] (2704.6101,-.06569348) [.1186952] (2210.8147,-.00353116) [.07453567] (1733.1893,.09671432) [.07346752] (3936.0498,-.03328127) [.05162848] (4275.6323,.01506767) [.04953594] (892.66815,-.13379738) [.04841887] (956.03864,.0697493) [.03938916] (3136.1213,-.43862563) [.03406697] (2443.002,.07132643) [.03059648] (4811.9019,-.02270896) [.02932514] (3423.2271,-.04010268) [.02777702] (394.53717,-.00959262) [.02540673] (115.43097,.0268696) [.02283385] (653.20422,-.39465582) [.0203265] (21234.896,.08966081) [.01871969] (966.67047,-.36518505) [.01859727] (101.48582,-.0924529) [.01816803] (1527.2758,-.03757093) [.01518886] (563.83966,-.04426037) [.01493849] (1624.0048,-.02174928) [.01487115] (573.59418,-.03692216) [.01472242] (470.17563,-.21795179) [.01268122] (1019.7386,-.0813105) [.01224861] (393.38055,.14930071) [.01196096] (724.78387,.0085337) [.01171347] (98.766457,-.02220671) [.01120998] (757.58533,.01153447) [.01087433] (956.71869,-.00420869) [.01047225] (222.16689,.09062849) [.01030963] (628.91681,-.00570504) [.00998021] (472.73672,.35862204) [.00907239] (50.415592,-.02913959) [.00865736] (1240.9509,.02513575) [.00801594] (352.61951,.09432561) [.00786503] (595.29486,.03117277) [.00746056] (2612.666,-.02973921) [.00713255] (257.43863,.01093251) [.00669831] (238.68938,-.1712833) [.00636441] (13.866181,.2940135) [.00631209] (1510.6305,.00411763) [.00625273] (644.60974,-.00992541) [.00622742] (198.93024,.42263117) [.0060884] (310.51801,-.19963783) [.00606231] (532.25,-.0235124) [.00602869] (231.79962,-.06079898) [.00578204] (331.94299,.0333625) [.00575694] (93.77916,.01764399) [.00551932] (28.076319,-.17176928) [.00535903] (51.072876,-.76581234) [.00535124] (455.45627,.01683097) [.00526613] (342.064,-.02790767) [.00519168] (80.519257,-.66383356) [.00514792] (15.848269,-.1473892) [.00506346] (147.78334,.02283254) [.00500808] (97.998878,.49699706) [.00470221] (354.83035,.11637682) [.00468494] (38.740601,-.16923408) [.00466912] (705.37085,-.24537851) [.00454323] (274.91531,.0361544) [.00453077] (141.73802,.25129968) [.00450153] (129.49149,.49343204) [.00443555] (248.80806,.79626334) [.00434659] (89.498383,.26903537) [.0037218] (267.05582,-.67381251) [.00372032] (247.85754,-.26421931) [.00362279] (109.71172,-.08625391) [.00356143] (649.47601,.01319072) [.00342194] (123.38459,.48050749) [.00339835] (62.123348,.38443124) [.0033021] (88.034836,-.00093416) [.00324886] (46.683876,-.0439051) [.00309844] (72.534355,.52362192) [.00303146] (11.474039,1.9816022) [.00302707] (173.94252,.06355982) [.00300079] (60.355618,-.06076398) [.00296195] (50.95401,-.02937797) [.00294673] (230.95421,-.00087641) [.00293702] (88.591553,-.09366328) [.00288967] (153.61955,.08267993) [.00288538] (642.07477,.03334107) [.00278364] (91.962502,-.13046528) [.0027432] (189.84798,.021114) [.00270509] (.6024434,-.26004788) [.00262271] (207.87248,.5220328) [.00237018] (504.62265,.02201749) [.00236203] (210.00151,-.05760963) [.00231536]  };
\addplot[scatter, only marks, thick, mark=square, mark options={solid,scale=1}, color=black, point meta=explicit, scatter/use mapped color={fill=none, fill opacity=0, draw=black},scatter/@pre marker code/.append style={/tikz/mark size=1+\pgfplotspointmetatransformed/50}] coordinates {   (306.65994,-.01220352) [.01266441] (154.15729,-.08672002) [.01179171] (124.59718,.39339748) [.00765311] (30.327106,.50722414) [.00734867] (23.950981,.26578203) [.00531905] (2422.6099,-.79443592) [.0044057] (171.81891,.46054751) [.00375624] (7.1336532,-.1387707) [.00359372] (1262.0474,.26833054) [.00304989] (329.38419,.6954698) [.00268875] (28.298464,.38309693) [.00262042] (41.942471,-.30545104) [.00242843]   };
\addplot[loosely dashed, color=darkgray, very thick, dash pattern = on 7.6pt off 2pt] coordinates {(-1000,-0.0302) (16000,-0.0302)};
\legend{~Positive Weights,~Negative Weights, ~Bartik Estimate}
\end{axis}
\end{tikzpicture}
}
\end{subfigure}
\hfill
\begin{subfigure}{0.475\textwidth}
\addtocounter{subfigure}{0}
\caption{\normalsize{Effect of Labor Market Tightness: Job Seeker Instrument}}
\centering
\scalebox{0.70}{
\begin{tikzpicture}
\begin{axis}[xlabel=First-Stage F Statistic, ylabel=Just-Identified Coefficient Estimate, xmin=-1000, xmax=16000, ymin=-0.32, ymax=0.32, height=14cm, width=14cm, grid=major, legend pos = north east, xtick={0,2500,5000,7500,10000,12500,15000}, ytick={-0.30,-0.20,-0.10,0,0.10,0.20,0.30}, grid style = dotted, y tick label style={/pgf/number format/.cd,fixed,fixed zerofill, precision=2,/tikz/.cd}]
\addplot[scatter, only marks, mark=o, mark options={solid,scale=1}, color=gray, point meta=explicit, scatter/use mapped color={fill=none, fill opacity=0, draw=gray},scatter/@pre marker code/.append style={/tikz/mark size=1+\pgfplotspointmetatransformed/50}] coordinates {  (194.58293,.29144371) [.02677071] (2178.1421,.15972687) [.02165673] (613.95935,.39364558) [.01510691] (617.09332,-.05953957) [.01388422] (1513.765,.11440197) [.01004884] (.29712194,.16674596) [.0088714] (792.60162,-.00960251) [.00880295] (775.54462,-.50462735) [.00631193] (68.017471,.8340106) [.00569337] (448.40494,.41457552) [.00566303] (62.168846,.23542261) [.0046569] (298.73898,.07050539) [.00394668] (3.154563,-.10465669) [.00359984] (155.43657,.29770175) [.00330284] (125.06205,-.07743032) [.00322756] (302.59998,-.61979353) [.00301029] (2.3922298,.14352208) [.00264065] (77.200409,-.11242393) [.00263936] (133.03125,-.13112438) [.00238362] (14.066929,.14313439) [.00235721] (113.93295,.05901307) [.00197261] (114.37465,-.32173461) [.00192752] (3.6531892,-.78582364) [.00172703] (225.10051,-.86964124) [.00168634] (.44729796,.30538341) [.00165002] (249.97163,.15768784) [.00164244] (61.411976,.47203863) [.00156842] (104.82086,.58671421) [.00150622]  };
\addplot[scatter, only marks, thick, mark=square, mark options={solid,scale=1}, color=black, point meta=explicit, scatter/use mapped color={fill=none, fill opacity=0, draw=black},scatter/@pre marker code/.append style={/tikz/mark size=1+\pgfplotspointmetatransformed/50}] coordinates {    (11974.907,-.25321865) [.09956982] (2166.6807,-.06132345) [.04591338] (3957.6863,.01050446) [.042302] (8205.3252,-.1385392) [.03659981] (488.95068,.02678912) [.03603784] (440.10211,.28724954) [.0321343] (2431.2402,-.0483581) [.02926697] (1004.3961,-.25300848) [.02160311] (1136.8958,-.0063829) [.01701335] (99.624016,-.51694143) [.01323273] (3110.8972,.06925761) [.01183874] (665.01349,-.01589019) [.0115644] (717.28339,.00708187) [.01110197] (1400.3431,-1.0385022) [.01029374] (507.36038,.21793149) [.00953588] (549.56549,-.42217532) [.00908248] (2300.9307,-.23621607) [.00873499] (482.63168,-.17122528) [.00848883] (122.661,.14204988) [.00811486] (47.635574,.13114513) [.00793592] (270.22137,.01404595) [.00746919] (101.80017,.05148499) [.00739396] (474.11224,-.070043) [.00724249] (1728.8354,.0754518) [.00720783] (72.873978,-1.123626) [.00706197] (39.410439,-.49871805) [.00689943] (668.28735,.01220039) [.0067716] (1598.8555,-.04516887) [.00643032] (468.62839,-.01214287) [.00640088] (533.0401,.40703982) [.00596592] (1050.8436,.0715206) [.00560631] (373.17264,.1758372) [.00528885] (229.66902,.13903004) [.00520779] (220.06514,.03428693) [.00512871] (625.35602,-.01360285) [.00506439] (398.37094,-.00300793) [.00491577] (25.673283,.15308191) [.00452805] (19.97094,.01045735) [.00451233] (365.2926,-.013838) [.00450485] (241.95816,-.06009748) [.00425991] (743.03632,-.31670472) [.00417779] (169.7213,-.29897904) [.00404444] (734.53546,-.02475826) [.00402977] (470.39233,-.04839379) [.0039978] (543.36462,-.08791415) [.00364867] (149.1035,-.92045987) [.0035425] (125.31107,.49476832) [.00347735] (24.725449,.01567353) [.00326409] (176.52933,.42437676) [.00326074] (168.59131,.05681978) [.00309363] (117.83919,.14272758) [.00285869] (78.194183,.09697127) [.00282806] (5.7072406,-.68229294) [.00271018] (35.525082,.7147662) [.00265683] (9.6915178,-3.5714474) [.00251804] (87.314812,-.01392325) [.00251011] (142.37289,-1.3507242) [.00250213] (65.925964,-.44733807) [.00235572] (474.71533,-.71312755) [.0023351] (25.576023,-.51173246) [.00217953] (441.50964,-.51839358) [.00212789] (2.8037226,.36956379) [.00210506] (210.79221,-.19403833) [.0021] (.7852627,-1.109664) [.00199949] (137.49355,.69118595) [.00198435] (54.496693,2.6234782) [.00188569] (84.005348,.11929098) [.00186545] (45.575577,-.1003896) [.00177295] (.00765402,1.1867107) [.0016174] (37.017262,-.22285317) [.00154384] (158.30275,.01085233) [.00153932] (44.606499,.22406067) [.00150119]  };
\addplot[loosely dashed, color=darkgray, very thick, dash pattern = on 7.6pt off 2pt] coordinates {(-1000,-0.1202) (16000,-0.1202)};
\legend{~Positive Weights,~Negative Weights, ~Bartik Estimate}
\end{axis}
\end{tikzpicture}
}
\end{subfigure}
\floatfoot{\footnotesize\textsc{Note: ---} The figure visualizes the relationship between each instruments’ just-identified coefficient estimate, the first-stage F statistics, and the Rotemberg weights. Each marker refers to a separate instrument’s estimates (occupation share). The figure plots the estimated just-identified coefficents for each instrument (i.e., occupational employment share in the base year) against the respective first-stage F statistic. The size of the markers are proportional to the the absolute value of the Rotemberg weights, with the circles denoting positive weights and the squares denoting negative weights. The horizontal dashed line reflects the overall Bartik estimate. For reasons of parsimony, the figure includes only the 100 occupations with the highest absolute Rotemberg weights, which account for 78.9 percent (vacancies) and 52.4 percent (job seekers) of the sum of absolute weights. Sources: Integrated Employment Biographies $\plus$ Official Statistics of the German Federal Employment Agency $\plus$ IAB Job Vacancy Survey, 1999-2019. }
\end{figure}

\pgfkeys{/pgf/number format/.cd,1000 sep={}}
\end{landscape}

\clearpage

\begin{landscape}

\begin{table}[!ht]
\small
\addtocounter{table}{-1}
\caption{Shifters of Demand and Supply for Employment Shares of Top 5 Rotemberg Occupations}
\label{tab:E2}

\begin{subtable}{1\textwidth}

\caption{\normalsize Effect of Wage Rate}
\centering

\begin{tabular}{L{4cm}C{2.2cm}C{2.2cm}C{2.2cm}C{2.2cm}C{2.2cm}} \hline
\multirow{3}{*}{\diagbox[height=3.4\line, innerwidth=4cm, linecolor=white]{}{\shortstack{ \\ \hspace{0.1cm}Employment \\ \hspace{0.1cm}Share of ... }}} & \multirow{3.4}{*}{\shortstack{(1)  \\ Gastronomy \\ Workers  }} & \multirow{3.4}{*}{\shortstack{(2)  \\ Medical \vphantom{/} \\ Assistants \vphantom{/}  }} & \multirow{3.4}{*}{\shortstack{(3)  \\ Hairdressers \vphantom{/}  }} & \multirow{3.4}{*}{\shortstack{(4)  \\ Cooks \vphantom{/}  }} & \multirow{3.4}{*}{\shortstack{(5)  \\ Farmers \vphantom{/}  }}  \\
&&&&&  \\
&&&&&  \\[0.2cm] \hline
&&&&& \\[-0.3cm]
\multirow{2.4}{*}{Log Productivity} &  \multirow{2.4}{*}{\shortstack{\hphantom{***}-0.855***\hphantom{-} \\ (0.250)}} & \multirow{2.4}{*}{\shortstack{-0.051\hphantom{-}\\ (0.387)}} & \multirow{2.4}{*}{\shortstack{\hphantom{**}-0.581**\hphantom{-}\\ (0.233)}} &   \multirow{2.4}{*}{\shortstack{\hphantom{***}-0.988***\hphantom{-}\\ (0.289)}} & \multirow{2.4}{*}{\shortstack{\hphantom{*}-1.077*\hphantom{-}\\(0.560)}}  \\
&&&&&   \\
\multirow{2.4}{*}{Log Investments} & \multirow{2.4}{*}{\shortstack{0.010\\(0.102)}} & \multirow{2.4}{*}{\shortstack{\hphantom{*}-0.538*\hphantom{-}\\(0.313)}} & \multirow{2.4}{*}{\shortstack{0.097\\(0.166)}} &  \multirow{2.4}{*}{\shortstack{0.188\\(0.217)}} & \multirow{2.4}{*}{\shortstack{\hphantom{***}0.819***\\(0.228)}}  \\
&&&&&   \\
\multirow{2.4}{*}{Business Expectations} &  \multirow{2.4}{*}{\shortstack{0.002\\(0.003)}} & \multirow{2.4}{*}{\shortstack{-0.001\hphantom{-}\\(0.014)}} & \multirow{2.4}{*}{\shortstack{0.00006\\(0.00390)}} &   \multirow{2.4}{*}{\shortstack{0.002\\(0.008)}} & \multirow{2.4}{*}{\shortstack{\hphantom{***}-0.024***\hphantom{-}\\(0.009)}} \\
&&&&&   \\
\multirow{2.4}{*}{Female Share} &  \multirow{2.4}{*}{\shortstack{\hphantom{***}1.675***\\(0.428)}} & \multirow{2.4}{*}{\shortstack{\hphantom{***}14.20***\\(1.851)}} & \multirow{2.4}{*}{\shortstack{\hphantom{***}2.456***\\(0.880)}} &   \multirow{2.4}{*}{\shortstack{\hphantom{***}2.952***\\(0.846)}} & \multirow{2.4}{*}{\shortstack{\hphantom{***}-3.322***\hphantom{-}\\(1.077)}}  \\
&&&&&   \\
\multirow{2.4}{*}{Foreign Share} &  \multirow{2.4}{*}{\shortstack{\hphantom{***}6.842***\\(2.321)}} & \multirow{2.4}{*}{\shortstack{2.262\\(2.467)}} & \multirow{2.4}{*}{\shortstack{2.032\\(1.823)}} &   \multirow{2.4}{*}{\shortstack{\hphantom{***}7.613***\\(0.846)}} & \multirow{2.4}{*}{\shortstack{\hphantom{-}-1.661\hphantom{-}\\\hphantom{-}(1.420)}} \\
&&&&&  \\[0.2cm]
Firm Size  & Yes &  Yes & Yes & Yes & \hphantom{-}Yes   \\[0.2cm] \hline
&&&&&  \\[-0.2cm]
F: Productivity & \hphantom{***}11.7*** & 0.02 & \hphantom{**}6.23** & \hphantom{***}11.7*** & \hphantom{*}3.70*   \\
&&&&&  \\
F: Demand Shifters & 0.15 & 1.50 & 0.25 & 0.38 & \hphantom{***}7.32***  \\
&&&&&  \\
F: Supply Shifters & \hphantom{***}11.7*** & \hphantom{***}29.4*** & \hphantom{**}4.17** & \hphantom{***}8.55*** & \hphantom{-***}4.84***  \\[0.2cm]
Observations & 4,728 & 4,728 & 4,728 & 4,728 & \hphantom{-}4,728 \\[0.2cm] \hline
&&&&&  \\[-0.2cm]
Identifying Variation & 0.118 & 0.096 & 0.082 & 0.082 & \hphantom{-}0.067  \\[0.2cm]
2015 Minimum Wage Bite & 0.620 & 0.143 & 0.546 & 0.313 & 0.460  \\[0.2cm] \hline
\end{tabular}
\end{subtable} \vspace*{-0.3cm}
\floatfoot{\footnotesize\textsc{Note: ---}  The table displays weighted least squares regressions of the top five occupational employment shares (in the firms' base year), as defined by column titles, on explanatory variables in the year after the predetermined share was fixed. Apart from productivity, the set of covariates includes variables that are likely to shift labor demand (investments and business expectations) or labor supply (share of female or foreign workers). In all specifications, we control for ten firm size categories. The F Statistics refer to tests of (joint) significance of the productivity variable, the set of labor demand variables, or the set of labor supply variables.  The last row delivers the occupations' relative weight in the Bartik estimator. * = p$<$0.10. ** = p$<$0.05. *** = p$<$0.01. Sources: Integrated Employment Biographies $\plus$ IAB Establishment Panel, 1999-2019.}
\end{table}

\clearpage

\begin{table}[!ht]
\small
\addtocounter{table}{-1}
\caption{Shifters of Demand and Supply for Employment Shares of Top 5 Rotemberg Occupations (Cont.)}

\begin{subtable}{1\textwidth}
\addtocounter{subtable}{1}
\caption{\normalsize Effect of Labor Market Tightness: Vacancy Instrument}

\centering

\begin{tabular}{L{4cm}C{2.2cm}C{2.2cm}C{2.2cm}C{2.2cm}C{2.2cm}} \hline
\multirow{3}{*}{\diagbox[height=3.4\line, innerwidth=4cm, linecolor=white]{}{\shortstack{ \\ \hspace{0.1cm}Employment \\ \hspace{0.1cm}Share of ... }}} & \multirow{3.4}{*}{\shortstack{(1)  \\ Sales \\ Workers  }} & \multirow{3.4}{*}{\shortstack{(2)  \\ Bankers  \vphantom{/}  }} & \multirow{3.4}{*}{\shortstack{(3)  \\ Farmers \vphantom{/}  }} & \multirow{3.4}{*}{\shortstack{(4)  \\ Chimney \vphantom{/} \\ Sweeps \vphantom{/}}} & \multirow{3.4}{*}{\shortstack{(5)  \\ Construction \vphantom{/} \\ Workers  \vphantom{/} }}  \\
&&&&&  \\
&&&&&  \\[0.2cm] \hline
&&&&& \\[-0.3cm]
\multirow{2.4}{*}{Log Productivity} &  \multirow{2.4}{*}{\shortstack{-0.567\hphantom{-}\\(0.469)}} & \multirow{2.4}{*}{\shortstack{0.047\\(0.034)}} & \multirow{2.4}{*}{\shortstack{\hphantom{*}-1.077*\hphantom{-}\\(0.560)}} &   \multirow{2.4}{*}{\shortstack{\hphantom{**}-0.443**\hphantom{-}\\(0.212)}} & \multirow{2.4}{*}{\shortstack{-0.009\hphantom{-}\\(0.140)}}  \\
&&&&&   \\
\multirow{2.4}{*}{Log Investments} & \multirow{2.4}{*}{\shortstack{-0.247\hphantom{-}\\(0.306)}} & \multirow{2.4}{*}{\shortstack{-0.036\hphantom{-}\\(0.026)}} & \multirow{2.4}{*}{\shortstack{\hphantom{**}0.819***\\(0.228)}} &  \multirow{2.4}{*}{\shortstack{-0.041\hphantom{-}\\(0.192)}} & \multirow{2.4}{*}{\shortstack{0.064\\(0.148)}}  \\
&&&&&   \\
\multirow{2.4}{*}{Business Expectations} &  \multirow{2.4}{*}{\shortstack{-0.007\hphantom{-}\\(0.011)}} & \multirow{2.4}{*}{\shortstack{-0.0005\hphantom{-}\\(0.0003)}} & \multirow{2.4}{*}{\shortstack{\hphantom{***}-0.024***\\(0.009)}} &   \multirow{2.4}{*}{\shortstack{-0.006*\\(0.004)}} & \multirow{2.4}{*}{\shortstack{-0.0003\hphantom{-}\\(0.004)}} \\
&&&&&   \\
\multirow{2.4}{*}{Female Share} &  \multirow{2.4}{*}{\shortstack{\hphantom{***}9.686***\\(1.544)}} & \multirow{2.4}{*}{\shortstack{0.146\\(0.132)}} & \multirow{2.4}{*}{\shortstack{\hphantom{***}-3.322***\hphantom{-}\\(1.077)}} &   \multirow{2.4}{*}{\shortstack{\hphantom{**}-1.443**\hphantom{-}\\(0.690)}} & \multirow{2.4}{*}{\shortstack{\hphantom{***}-2.876***\hphantom{-}\\(0.715)}}  \\
&&&&&   \\
\multirow{2.4}{*}{Foreign Share} &  \multirow{2.4}{*}{\shortstack{\hphantom{*}-3.585*\hphantom{-}\\(1.869)}} & \multirow{2.4}{*}{\shortstack{-0.035\hphantom{-}\\(0.046)}} & \multirow{2.4}{*}{\shortstack{-1.661\hphantom{-}\\(1.420)}} &   \multirow{2.4}{*}{\shortstack{\hphantom{**}-1.801**\hphantom{-}\\(0.711)}} & \multirow{2.4}{*}{\shortstack{2.506\\(2.183)}} \\
&&&&&  \\[0.2cm]
Firm Size  & Yes &  Yes & Yes & Yes & \hphantom{-}Yes   \\[0.2cm] \hline
&&&&&  \\[-0.2cm]
F: Productivity & 1.46 & 1.90 & \hphantom{*}3.70* & \hphantom{*}4.35* & 0.00   \\
&&&&&  \\
F: Demand Shifters & 0.63 & 1.33 & \hphantom{***}7.32*** & 1.51 & 0.13  \\
&&&&&  \\
F: Supply Shifters & \hphantom{***}21.9*** & 0.74 & \hphantom{***}4.84*** & \hphantom{**}3.21** & \hphantom{***}8.09***  \\[0.2cm]

Observations & 4,728 & 4,728 & 4,728 & 4,728 & \hphantom{-}4,728 \\[0.2cm]
\hline
&&&&&  \\[-0.2cm]
Identifying Variation & 0.113 & 0.069 & 0.068 & 0.043 & 0.042  \\[0.2cm]
\hline
\end{tabular}
\end{subtable} \vspace*{-0.3cm}
\floatfoot{\footnotesize\textsc{Note: ---} The table displays weighted least squares regressions of the top five occupational employment shares (in the firms' base year), as defined by column titles, on explanatory variables in the year after the predetermined share was fixed. Apart from productivity, the set of covariates includes variables that are likely to shift labor demand (investments and business expectations) or labor supply (share of female or foreign workers). In all specifications, we control for ten firm size categories. The F Statistics refer to tests of (joint) significance of the productivity variable, the set of labor demand variables, or the set of labor supply variables.  The last row delivers the occupations' relative weight in the Bartik estimator. * = p$<$0.10. ** = p$<$0.05. *** = p$<$0.01. Sources: Integrated Employment Biographies $\plus$ IAB Establishment Panel, 1999-2019.}
\end{table}

\clearpage

\begin{table}[!ht]
\small
\addtocounter{table}{-1}
\caption{Shifters of Demand and Supply for Employment Shares of Top 5 Rotemberg Occupations (Cont.)}

\begin{subtable}{1\textwidth}
\addtocounter{subtable}{2}
\caption{\normalsize Effect of Labor Market Tightness: Job Seeker Instrument}

\centering

\begin{tabular}{L{4cm}C{2.2cm}C{2.2cm}C{2.2cm}C{2.2cm}C{2.2cm}} \hline
\multirow{3}{*}{\diagbox[height=3.4\line, innerwidth=4cm, linecolor=white]{}{\shortstack{ \\ \hspace{0.1cm}Employment \\ \hspace{0.1cm}Share of ... }}} & \multirow{3.4}{*}{\shortstack{(1)  \\ Sales  \vphantom{/} \\ Workers  \vphantom{/} }} & \multirow{3.4}{*}{\shortstack{(2)  \\ Masons  \vphantom{/}  }} & \multirow{3.4}{*}{\shortstack{(3)  \\ Truck \vphantom{/} \\ Drivers \vphantom{/}  }} & \multirow{3.4}{*}{\shortstack{(4)  \\ Bankers \\ \vphantom{/}  }} & \multirow{3.4}{*}{\shortstack{(5)  \\ Gardeners  \vphantom{/}  }}  \\
&&&&&  \\
&&&&&  \\[0.2cm] \hline
&&&&& \\[-0.3cm]
\multirow{2.4}{*}{Log Productivity} &  \multirow{2.4}{*}{\shortstack{-0.567\hphantom{-}\\(0.469)}} & \multirow{2.4}{*}{\shortstack{-0.309\hphantom{-}\\(0.281)}} & \multirow{2.4}{*}{\shortstack{\hphantom{*}-0.769*\hphantom{-}\\ (0.421)}} &   \multirow{2.4}{*}{\shortstack{0.047\\ (0.034)}} & \multirow{2.4}{*}{\shortstack{-0.767\hphantom{-}\\(0.526)}}  \\
&&&&&   \\
\multirow{2.4}{*}{Log Investments} & \multirow{2.4}{*}{\shortstack{-0.247\hphantom{-}\\(0.306)}} & \multirow{2.4}{*}{\shortstack{-0.454\hphantom{-}\\(0.340)}} & \multirow{2.4}{*}{\shortstack{\hphantom{***}1.783***\\(0.319)}} &  \multirow{2.4}{*}{\shortstack{-0.036\hphantom{-}\\(0.026)}} & \multirow{2.4}{*}{\shortstack{0.356\\(0.251)}}  \\
&&&&&   \\
\multirow{2.4}{*}{Business Expectations} &  \multirow{2.4}{*}{\shortstack{-0.007\hphantom{-}\\(0.011)}} & \multirow{2.4}{*}{\shortstack{0.001\\(0.005)}} & \multirow{2.4}{*}{\shortstack{\hphantom{**}-0.032**\hphantom{-}\\(0.012)}} &   \multirow{2.4}{*}{\shortstack{-0.0005\hphantom{-}\\(0.0003)}} & \multirow{2.4}{*}{\shortstack{\hphantom{**}-0.019**\hphantom{-}\\(0.009)}} \\
&&&&&   \\
\multirow{2.4}{*}{Female Share} &  \multirow{2.4}{*}{\shortstack{\hphantom{***}9.686***\\(1.544)}} & \multirow{2.4}{*}{\shortstack{\hphantom{***}-4.025***\hphantom{-}\\(1.150)}} & \multirow{2.4}{*}{\shortstack{\hphantom{***}-6.231***\hphantom{-}\\(0.948)}} &  \multirow{2.4}{*}{\shortstack{0.146\\(0.132)}} & \multirow{2.4}{*}{\shortstack{-1.516\hphantom{-}\\(1.131)}}  \\
&&&&&   \\
\multirow{2.4}{*}{Foreign Share} &  \multirow{2.4}{*}{\shortstack{\hphantom{*}-3.585*\hphantom{-}\\(1.869)}} & \multirow{2.4}{*}{\shortstack{-0.668\hphantom{-}\\(1.858)}} & \multirow{2.4}{*}{\shortstack{-0.572\hphantom{-}\\(2.060)}} &   \multirow{2.4}{*}{\shortstack{-0.035\hphantom{-}\\(0.046)}} & \multirow{2.4}{*}{\shortstack{0.324\\(1.948)}} \\
&&&&&  \\[0.2cm]
Firm Size  & Yes &  Yes & Yes & Yes & Yes   \\[0.2cm] \hline
&&&&&  \\[-0.2cm]
F: Productivity & 1.46 & 1.21 & \hphantom{*}3.34* & 1.90 & 2.13   \\
&&&&&  \\
F: Demand Shifters & 0.63 & 1.11 & \hphantom{***}15.6*** & 1.33 & \hphantom{*}2.49*  \\
&&&&&  \\
F: Supply Shifters & \hphantom{***}21.9*** & \hphantom{***}5.25*** & \hphantom{***}22.8*** & 0.74 & 0.91  \\[0.2cm]

Observations & 4,728 & 4,728 & 4,728 & 4,728 & \hphantom{-}4,728 \\[0.2cm]
\hline
&&&&&  \\[-0.2cm]
Identifying Variation & 0.104 & 0.048 & 0.044 & 0.038 & 0.037  \\[0.2cm]
\hline
\end{tabular}
\end{subtable} \vspace*{-0.3cm}
\floatfoot{\footnotesize\textsc{Note: ---} The table displays weighted least squares regressions of the top five occupational employment shares (in the firms' base year), as defined by column titles, on explanatory variables in the year after the predetermined share was fixed. Apart from productivity, the set of covariates includes variables that are likely to shift labor demand (investments and business expectations) or labor supply (share of female or foreign workers). In all specifications, we control for ten firm size categories. The F Statistics refer to tests of (joint) significance of the productivity variable, the set of labor demand variables, or the set of labor supply variables.  The last row delivers the occupations' relative weight in the Bartik estimator. * = p$<$0.10. ** = p$<$0.05. *** = p$<$0.01. Sources: Integrated Employment Biographies $\plus$ IAB Establishment Panel, 1999-2019.}
\end{table}

\end{landscape}

\clearpage

\section{Regression Results: Further Evidence}
\label{sec:F}

\setcounter{table}{0} 
\setcounter{figure}{0} 

\begin{figure}[!ht]
\caption{First-Stage Regressions}
\label{fig:F1}
\begin{subfigure}{1\textwidth}
\caption{Wages}
\centering
\scalebox{0.725}{
\begin{tikzpicture}
\begin{axis}[xlabel=Z\(_{\text{W}^{\text{FT}}}\), ylabel= $\Delta$ Log W$^{\text{FT}}$, xmin=-0.01, xmax=0.11, ymin=0.01, ymax=0.11, height=8cm, width=16cm, grid=major, legend pos = south east, xtick={0,0.02,0.04,0.06,0.08,0.1}, ytick={0.02,0.04,0.06,0.08,0.1}, grid style = dotted, y tick label style={/pgf/number format/.cd,fixed,fixed zerofill, precision=2,/tikz/.cd}, x tick label style={/pgf/number format/.cd,fixed,fixed zerofill, precision=2,/tikz/.cd}]
\addplot[only marks, mark=o, mark options={solid,scale=1}, color=black] coordinates {  (.00260047,.03955764) (.01896946,.04321265) (.02576875,.04491893) (.029859,.04691682) (.03198348,.04869709) (.0335369,.04537552) (.03494535,.04932824) (.03623819,.04724056) (.03708618,.04755289) (.037906,.04661266) (.03869947,.04520552) (.03945285,.04806257) (.04002408,.04707823) (.04051278,.05909671) (.04100432,.04939587) (.04144624,.05855127) (.04194588,.04747409) (.04248095,.04696743) (.04293461,.04516155) (.04330869,.04921323) (.043772,.04651774) (.04421189,.04908404) (.04465603,.04925198) (.04516365,.04569469) (.04554528,.04923409) (.04590538,.05062488) (.04631885,.04844899) (.04670011,.0470191) (.04693207,.05803629) (.04714152,.04903143) (.04752512,.05246475) (.0478912,.0483453) (.0482813,.05032408) (.04862854,.05188914) (.04896211,.05039406) (.04924951,.05573582) (.04936766,.05318482) (.04971158,.04898714) (.05004288,.05378664) (.05040525,.05185663) (.05064599,.04884624) (.05095236,.04977132) (.0513423,.05135081) (.05168017,.05448398) (.05198941,.0518883) (.05237074,.0522209) (.05272909,.05324077) (.05310721,.05568633) (.05339977,.05308086) (.05368846,.05212328) (.05396961,.0605032) (.05411927,.05167529) (.05439181,.06401043) (.05459678,.05291988) (.05492584,.05357133) (.05516604,.05794906) (.05538003,.05248186) (.05572933,.05301483) (.05599843,.05850678) (.05618779,.05460279) (.05653535,.05311382) (.05679814,.05249556) (.05711437,.05535342) (.05747335,.05749966) (.0578399,.05587289) (.05830469,.05601666) (.05883446,.06074901) (.05902802,.05609705) (.05932132,.05684325) (.05960616,.06279204) (.05987476,.05631567) (.06001503,.05553642) (.0603528,.05780276) (.06090156,.06002058) (.06126823,.06823112) (.06172375,.05686377) (.06229988,.06035921) (.06287331,.06247286) (.06338932,.06051536) (.06402919,.06167061) (.06467256,.06195412) (.06515514,.06242876) (.06591851,.06243887) (.0667166,.06324908) (.06740864,.06464334) (.06816762,.06045289) (.06896103,.06608118) (.06969283,.06630684) (.07051658,.06852157) (.07140145,.06966441) (.07268769,.06985673) (.07383539,.07236043) (.07477499,.07130309) (.07594182,.06856027) (.07725171,.08266397) (.07956473,.07680876) (.08303511,.08770352) (.0878882,.08347594) (.09297314,.08725844) (.10365934,.09363359) };
\addplot[domain=-0.005:0.105, color=black, solid, line width=0.75mm] {.6073528813708711*x+.0237086095353697};
\legend{~Observations,~Linear Fit}
\end{axis}
\end{tikzpicture}
}
\end{subfigure}
\vskip \baselineskip
\begin{subfigure}{1\textwidth}
\caption{Vacancies}
\centering
\scalebox{0.725}{
\begin{tikzpicture}
\begin{axis}[xlabel=Z\(_{\text{V}}\), ylabel= $\Delta$ Log V/U, xmin=-0.85, xmax=0.85, ymin=-0.6, ymax=1.1, height=8cm, width=16cm, grid=major, legend pos = south east, xtick={-0.75,-0.5,-0.25,0.25,0,0.50,0.75}, ytick={-0.5,-0.25,0.25,0,0.50,0.75,1.00}, grid style = dotted, y tick label style={/pgf/number format/.cd,fixed,fixed zerofill, precision=2,/tikz/.cd}]
\addplot[only marks, mark=o, mark options={solid,scale=1}, color=black] coordinates {  (-.68505645,-.15062116) (-.38506129,-.05248878) (-.26849604,-.0004595) (-.24326806,-.08074746) (-.182979,-.00438296) (-.13642733,.18307465) (-.11175872,-.00968479) (-.08691371,.01983872) (-.06510865,.03075109) (-.04468112,.03530545) (-.02788155,-.01331329) (-.02069791,.02291392) (-.01054338,.23028567) (-.00277563,.1091103) (.00495974,.08600092) (.01384233,.14537148) (.02405061,.06331794) (.0315979,.14154141) (.03981137,.0582229) (.0470814,.12953389) (.05500978,.12455348) (.05952774,.20433666) (.06745851,.13017607) (.07563015,.09510578) (.08195788,.18059047) (.08549189,.19360007) (.09013066,.18342298) (.09705148,.19579189) (.10349649,.1734838) (.10945906,.21029796) (.11527221,.21686175) (.12233252,.14673916) (.1276394,.23425993) (.1341759,.19043598) (.14022675,.32664165) (.14485531,.23899306) (.14905687,.22657117) (.15669543,.23742302) (.15915734,.19408141) (.16166569,.20108886) (.16627669,.26077801) (.1718749,.26270571) (.17726545,.22756597) (.18249752,.25028381) (.18812975,.30713597) (.1911197,.25542632) (.19551381,.26457486) (.20037133,.26047364) (.20508026,.28541312) (.2105293,.29205734) (.21500687,.35981065) (.21867666,.29482085) (.22453074,.28382364) (.22719337,.33068827) (.2283131,.34921634) (.23104101,.35150042) (.23471648,.32134047) (.24001934,.36095491) (.24208155,.3212443) (.24652423,.31593975) (.24922536,.29249382) (.2514922,.12907773) (.2541436,.34413227) (.25819725,.31839895) (.26351944,.31793064) (.26598036,.39486197) (.27048114,.301539) (.27613556,.39944839) (.28204024,.36483169) (.28840238,.337107) (.29378781,.35228923) (.29958215,.38444239) (.30538127,.36984062) (.31281796,.39386177) (.31337619,.40248683) (.31457353,.40298951) (.31830278,.40197042) (.32486448,.36234653) (.33243647,.36878204) (.34109354,.37647924) (.35098156,.37517542) (.36257374,.42133486) (.37204647,.38200113) (.38221017,.36643997) (.39059502,.38656357) (.39911929,.39965227) (.41409317,.35964835) (.42510682,.37522095) (.42759418,.40942454) (.44665864,.32480961) (.46182773,.51388478) (.47368619,.46962658) (.49171579,.50897968) (.51475495,.38364336) (.54048926,.4967528) (.61058289,.5210135) (.94474345,.48567978) (.,.) (.,.) (.,.) };
\addplot[domain=-0.80:0.80, color=black, solid, line width=0.75mm] {.5873321959647914*x+.1462748710956576};
\legend{~Observations,~Linear Fit}
\end{axis}
\end{tikzpicture}
}
\end{subfigure}
\vskip \baselineskip
\begin{subfigure}{1\textwidth}
\centering
\caption{Job Seekers}
\scalebox{0.725}{
\begin{tikzpicture}
\begin{axis}[xlabel=Z\(_{\text{U}}\), ylabel=$\Delta$ Log V/U, xmin=-0.85, xmax=0.85, ymin=-0.6, ymax=1.1, height=8cm, width=16cm, grid=major, legend pos = north east, xtick={-0.75,-0.5,-0.25,0.25,0,0.50,0.75}, ytick={-0.5,-0.25,0,0.25,0.50,0.75,1.00}, grid style = dotted, y tick label style={/pgf/number format/.cd,fixed,fixed zerofill, precision=2,/tikz/.cd}]
\addplot[only marks, mark=o, mark options={solid,scale=1}, color=black] coordinates {  (-.46726367,.33767751) (-.36667424,.36437741) (-.32597303,.45986956) (-.29382035,.46906012) (-.27040321,.42553645) (-.25671673,.33703244) (-.24388549,.37834987) (-.23558284,.37536395) (-.22857708,.34321231) (-.22060458,.38997358) (-.21690111,.48060066) (-.2158782,.32009029) (-.21223286,.41957784) (-.20609745,.37150437) (-.19957441,.32866415) (-.1990241,.35624668) (-.19592302,.34873539) (-.1897535,.32944655) (-.1856515,.4054597) (-.18487768,.27655256) (-.18216082,.31640109) (-.17874153,.25044793) (-.17681417,.29377621) (-.17294502,.32528573) (-.1695786,.32570148) (-.16839021,.3656354) (-.16787139,.50669652) (-.16588961,.34623545) (-.16392608,.30412397) (-.16066249,.36764613) (-.15886571,.34840482) (-.1572804,.29783291) (-.15442055,.30677342) (-.15101819,.3131426) (-.1476219,.29808027) (-.14397348,.31476182) (-.14111277,.27604547) (-.13881931,.26450342) (-.13649824,.31901214) (-.13374744,.26948413) (-.13073365,.30944583) (-.12752229,.31120902) (-.12399375,.27150148) (-.1213432,.29453099) (-.11980984,.23538253) (-.11926943,.240365) (-.11692574,.3014546) (-.11369383,.25990525) (-.11060521,.23080608) (-.10725578,.27319792) (-.10285254,.26715219) (-.09901135,.28704742) (-.0950537,.25864276) (-.09215622,.23514767) (-.08865534,.28578606) (-.08554321,.3414059) (-.08311594,.24506335) (-.07862296,.22406287) (-.07429297,.20506515) (-.06924188,.17753474) (-.06643995,.3444199) (-.06215555,.22880203) (-.05678137,.27133486) (-.05197003,.2498171) (-.04733708,.22639333) (-.04370864,.22597446) (-.04084788,.16119583) (-.03535068,.28680143) (-.03287202,.34293753) (-.03015492,.17937112) (-.02704165,.17338866) (-.02324237,.24131086) (-.01966148,.01240313) (-.017278,.11577448) (-.01353159,.0733878) (-.00804864,.14325804) (-.00344295,.20147948) (.00089706,.19251287) (.00461583,.08762682) (.0105264,.09080749) (.01732011,.23927905) (.02036689,.08066852) (.0252016,.24601027) (.03140175,.17052111) (.03565357,.14946038) (.04229749,.10866338) (.04897047,.21451654) (.0566621,.10800866) (.06141675,.1059133) (.06861831,.12279314) (.07653053,.22478952) (.08673407,.11577687) (.09566255,.03082907) (.1108173,.14856204) (.12604888,-.00169908) (.13696589,-.1696296) (.19372365,.01128253) (.49803883,-.00146725) (.,.) (.,.) };
\addplot[domain=-0.80:0.80, color=black, solid, line width=0.75mm] {-.6865692167717847*x+.1894931640674387};
\legend{~Observations,~Linear Fit}
\end{axis}
\end{tikzpicture}
}
\end{subfigure}
\floatfoot{\footnotesize\textsc{Note. ---} The figures show binned scatterplots with 100 hundred markers to visualize the underlying variation of the first-stage regressions. L = Employment. FT = Full-Time. U = Jobs Seekers. V = Vacancies. W = Average Daily Wages. Z = Shift-Share Instrument. Sources: Integrated Employment Biographies $\plus$ Official Statistics of German Federal Employment Agency $\plus$ IAB Job Vacancy Survey, 1999-2019.}
\end{figure}
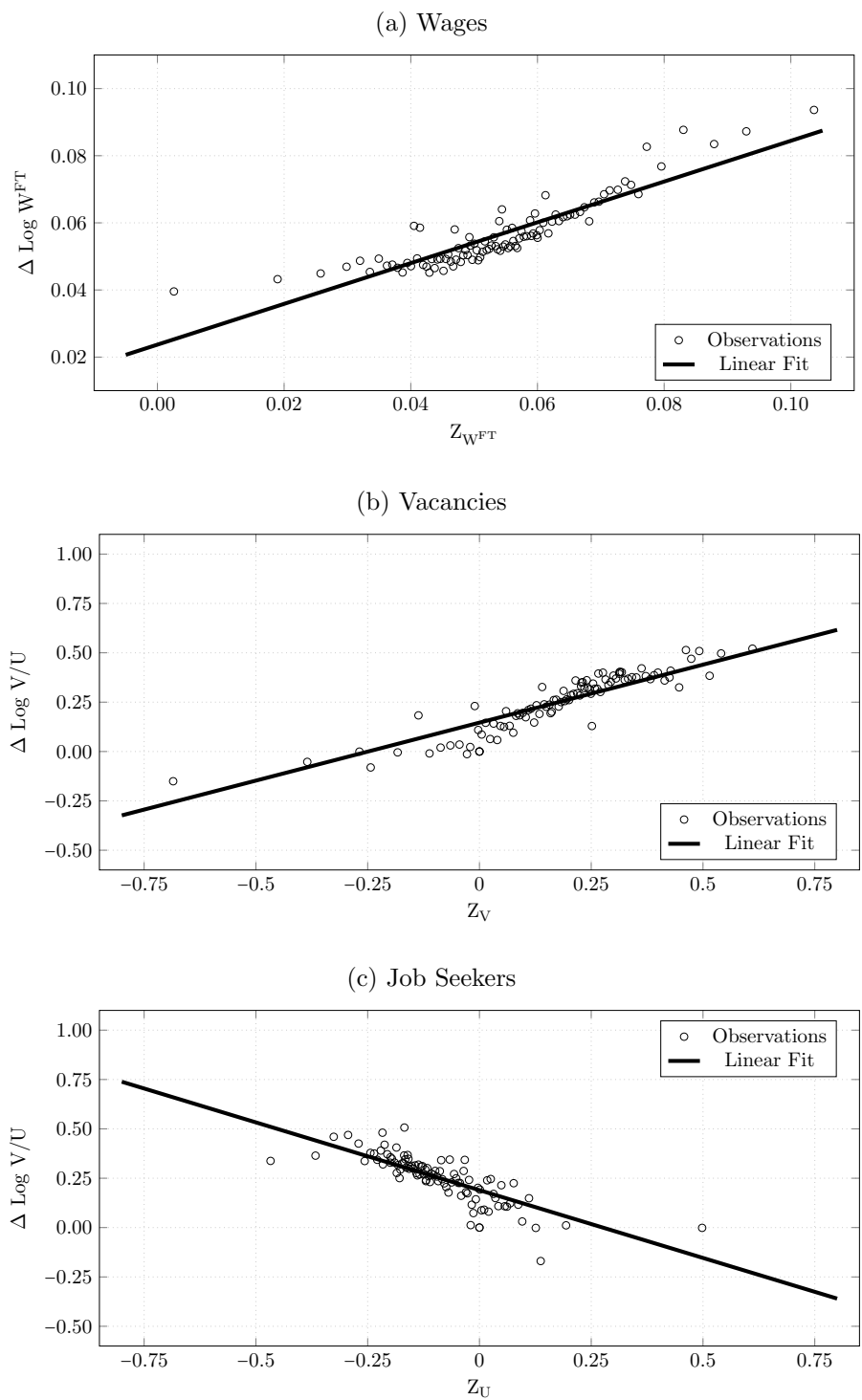

\clearpage

\vspace*{3cm}
\begin{table}[!ht]
\centering
\scalebox{0.90}{
\begin{threeparttable}
\caption{Reduced-Form Regressions}
\label{tab:F1}
\begin{tabular}{L{4cm}C{2.7cm}C{2.7cm}C{2.7cm}C{2.7cm}} \hline
\multirow{3.4}{*}{} & \multirow{3.4}{*}{\shortstack{(1) \\ $\Delta$ Log L$^{\text{FT}}$ \vphantom{/}  }} & \multirow{3.4}{*}{\shortstack{(2) \\ $\Delta$ Log L$^{\text{FT}}$ \vphantom{/}  }} & \multirow{3.4}{*}{\shortstack{(3) \\ $\Delta$ Log L$^{\text{FT}}$ \vphantom{/}  }} & \multirow{3.4}{*}{\shortstack{(4) \\ $\Delta$ Log L$^{\text{FT}}$\vphantom{/}  }} \\
&&&& \\
&&&& \\[0.2cm] \hline
&&&& \\[-0.3cm]
\multirow{2.4}{*}{ Z\(_{\text{W}^{\text{FT}}}\)  } &  \multirow{2.4}{*}{\shortstack{\hphantom{***}-0.466***\hphantom{-} \\ (0.013)}} & &   &   \multirow{2.4}{*}{\shortstack{\hphantom{***}-0.455***\hphantom{-} \\ (0.013)}} \\
&&&&  \\
\multirow{2.4}{*}{ Z\(_{\text{V}}\)  } & & \multirow{2.4}{*}{\shortstack{\hphantom{***}-0.014***\hphantom{-} \\ (0.001)}}  &   & \multirow{2.4}{*}{\shortstack{\hphantom{***}-0.011***\hphantom{-} \\ (0.001)}}    \\
&&&&  \\
\multirow{2.4}{*}{ Z\(_{\text{U}}\)  } & &    &  \multirow{2.4}{*}{\shortstack{\hphantom{***}0.061*** \\ (0.002)}} & \multirow{2.4}{*}{\shortstack{\hphantom{***}0.059*** \\ (0.002)}} \\
&&&&  \\[0.2cm]
Fixed Effects  & Year &  Year & Year &  Year   \\[0.2cm] \hline
&&&& \\[-0.2cm]
Observations &  7,993,993          & 7,993,993        &  7,993,993          & 7,993,993               \\[0.2cm]
Clusters &  1,801,671          & 1,801,671        &  1,801,671          & 1,801,671     \\[0.2cm] \hline
\end{tabular}
\begin{tablenotes}[para]
\footnotesize\textsc{Note. ---} The table displays the underlying reduced-form regressions of the IV estimations in Columns (2), (3), and (4) from Table \ref{tab:2}. The instrumental variables refer to shift-share instruments of biennial national changes in occupations in occupations weighted by past occupational employment in the respective establishment. The lag difference is two years. Labor markets are combinations of 5-digit KldB occupations and commuting zones. Standard errors (in parentheses) are clustered at the establishment level. L = Employment. U = Job Seekers. V = Vacancies. Z = Shift-Share Instrument. * = p$<$0.10. ** = p$<$0.05. *** = p$<$0.01. Sources: Integrated Employment Biographies $\plus$ Official Statistics of the German Federal Employment Agency $\plus$ IAB Job Vacancy Survey, 1999-2019.
\end{tablenotes}
\end{threeparttable}
}
\end{table}

\clearpage

\vspace*{3cm}
\begin{table}[!ht]
\centering
\scalebox{0.90}{
\begin{threeparttable}
\caption{Labor Demand Effects by Lag Difference}
\label{tab:F2}
\begin{tabular}{L{4cm}C{3.75cm}C{3.75cm}C{3.75cm}} \hline
\multirow{3.4}{*}{} & \multirow{3.4}{*}{\shortstack{(1) \\ $\Delta$ Log L$^{\text{FT}}$ }} & \multirow{3.4}{*}{\shortstack{(2) \\ $\Delta$ Log  L$^{\text{FT}}$ }} & \multirow{3.4}{*}{\shortstack{(3) \\ $\Delta$ Log  L$^{\text{FT}}$ }} \\
&&& \\
&&& \\[0.2cm] \hline
&&& \\[-0.3cm]
\multirow{2.4}{*}{$\Delta$ Log W$^{\text{FT}}$} &  \multirow{2.4}{*}{\shortstack{\hphantom{***}-0.617***\hphantom{-} \\ (0.022)}}  & \multirow{2.4}{*}{\shortstack{\hphantom{***}-0.730***\hphantom{-} \\ (0.022)}} &  \multirow{2.4}{*}{\shortstack{\hphantom{***}-0.772***\hphantom{-} \\ (0.023)}}  \\
&&& \\
\multirow{2.4}{*}{$\Delta$ Log V/U} &  \multirow{2.4}{*}{\shortstack{\hphantom{***}-0.023***\hphantom{-}\\ (0.002)}}  & \multirow{2.4}{*}{\shortstack{\hphantom{***}-0.051***\hphantom{-}\\ (0.002)}} &  \multirow{2.4}{*}{\shortstack{\hphantom{***}-0.058***\hphantom{-}\\ (0.002)}}   \\
&&& \\[0.2cm]
Fixed Effects  & Year &  Year  & Year  \\[0.2cm] \hline
&&& \\[-0.2cm]
Instruments  & Z\(_{\text{W}^{\text{FT}}}\), Z\(_{\text{V}}\), Z\(_{\text{U}}\) & Z\(_{\text{W}^{\text{FT}}}\), Z\(_{\text{V}}\), Z\(_{\text{U}}\) & Z\(_{\text{W}^{\text{FT}}}\), Z\(_{\text{V}}\), Z\(_{\text{U}}\)  \\[0.2cm]
Lag Difference  & 1 Year &  2 Years & 3 Years  \\[0.2cm]
Observations &  9,988,682          & 7,993,993        &  6,277,010                  \\[0.2cm]
Clusters &  2,057,324          & 1,801,671        &  1,600,914                  \\[0.2cm]
F: $\Delta$ Log W$^{\text{FT}}$ & 2,790  & 3,322   & 3,344   \\[0.2cm]
F: $\Delta$ Log V/U & 24,539 & 30,380 & 29,733  \\[0.2cm] \hline
\end{tabular}
\begin{tablenotes}[para]
\footnotesize\textsc{Note. ---} The table displays IV regressions of differences in log employment (of regular full-time workers) per establishment on differences in the log of average daily wages and the log of labor market tightness. The instrumental variables refer to shift-share instruments of national changes in occupations weighted by past occupational employment in the respective establishment. The lag of the first-differences estimator (in years) differs across specifications. Labor markets are combinations of 5-digit KldB occupations and commuting zones. Standard errors (in parentheses) are clustered at the establishment level. F = F Statistics of Excluded Instruments. FT = Full-Time. KldB = German Classification of Occupations. L = Employment. U = Job Seekers. V = Vacancies. W = Average Daily Wages. Z = Shift-Share Instrument. * = p$<$0.10. ** = p$<$0.05. *** = p$<$0.01. Sources: Integrated Employment Biographies $\plus$ Official Statistics of the German Federal Employment Agency $\plus$ IAB Job Vacancy Survey, 1999-2019.
\end{tablenotes}
\end{threeparttable}
}
\end{table}


\begin{landscape}

\vspace*{1cm}
\begin{table}[!ht]
\centering
\scalebox{0.90}{
\begin{threeparttable}
\caption{Labor Demand Effects by Specification}
\label{tab:F3}
\begin{tabular}{L{4cm}C{2.7cm}C{2.7cm}C{2.7cm}C{2.7cm}C{2.7cm}C{2.7cm}C{2.7cm}} \hline
\multirow{3.4}{*}{} & \multirow{3.4}{*}{\shortstack{(1) \\ $\Delta$ Log L$^{\text{FT}}$ }} & \multirow{3.4}{*}{\shortstack{(2) \\ $\Delta$ Log  L$^{\text{FT}}$ }} & \multirow{3.4}{*}{\shortstack{(3) \\ $\Delta$ Log  L$^{\text{FT}}$ }} & \multirow{3.4}{*}{\shortstack{(4) \\ $\Delta$ Log  L$^{\text{FT}}$ }} & \multirow{3.4}{*}{\shortstack{(5) \\ $\Delta$ Log  L$^{\text{FT}}$ }} & \multirow{3.4}{*}{\shortstack{(6) \\ $\Delta$ Log  L$^{\text{FT}}$ }} & \multirow{3.4}{*}{\shortstack{(7) \\ $\Delta$ Log  L$^{\text{FT}}$ }} \\
&&&&&&& \\
&&&&&&& \\[0.3cm] \hline
&&&&&&& \\[-0.3cm]
\multirow{2.4}{*}{$\Delta$ Log W$^{\text{FT}}$} &  \multirow{2.4}{*}{\shortstack{\hphantom{***}-0.783***\hphantom{-} \\ (0.038)}} & \multirow{2.4}{*}{\shortstack{\hphantom{***}-0.719***\hphantom{-} \\ (0.022)}} &  \multirow{2.4}{*}{\shortstack{\hphantom{***}-0.717***\hphantom{-} \\ (0.022)}} &  \multirow{2.4}{*}{\shortstack{\hphantom{***}-0.759***\hphantom{-} \\ (0.022)}} &  \multirow{2.4}{*}{\shortstack{\hphantom{***}-0.743***\hphantom{-} \\ (0.022)}} &  \multirow{2.4}{*}{\shortstack{\hphantom{***}-0.751***\hphantom{-} \\ (0.022)}} &  \multirow{2.4}{*}{\shortstack{\hphantom{***}-0.760***\hphantom{-} \\ (0.024)}}  \\
&&&&&&&\\
\multirow{2.4}{*}{$\Delta$ Log V/U} &  \multirow{2.4}{*}{\shortstack{\hphantom{***}-0.067***\hphantom{-} \\ (0.002)}}  &  \multirow{2.4}{*}{\shortstack{\hphantom{***}-0.050***\hphantom{-} \\ (0.002)}} &  \multirow{2.4}{*}{\shortstack{\hphantom{***}-0.052***\hphantom{-} \\ (0.002)}}  &   \multirow{2.4}{*}{\shortstack{\hphantom{***}-0.046***\hphantom{-} \\ (0.002)}}   &   \multirow{2.4}{*}{\shortstack{\hphantom{***}-0.053***\hphantom{-} \\ (0.002)}}  &   \multirow{2.4}{*}{\shortstack{\hphantom{***}-0.050***\hphantom{-} \\ (0.002)}}  &   \multirow{2.4}{*}{\shortstack{\hphantom{***}-0.059***\hphantom{-} \\ (0.002)}}  \\
&&&&&&& \\[0.2cm]
Fixed Effects  & Year $\times$ Industry  &  Year $\times$ CZ  & Year & Year & Year & Year & Year \\[0.2cm] \hline
&&&&&&& \\[-0.2cm]
Instruments  & Z\(_{\text{W}^{\text{FT}}}\), Z\(_{\text{V}}\), Z\(_{\text{U}}\) & Z\(_{\text{W}^{\text{FT}}}\), Z\(_{\text{V}}\), Z\(_{\text{U}}\) & Z\(_{\text{W}^{\text{FT}}}\), Z\(_{\text{V}}\), Z\(_{\text{U}}\) & Z\(_{\text{W}^{\text{FT}}}\), Z\(_{\text{V}}\), Z\(_{\text{U}}\) & Z\(_{\text{W}^{\text{FT}}}\), Z\(_{\text{V}}\), Z\(_{\text{U}}\) & Z\(_{\text{W}^{\text{FT}}}\), Z\(_{\text{V}}\), Z\(_{\text{U}}\) & Z\(_{\text{W}^{\text{FT}}}\), Z\(_{\text{V}}\), Z\(_{\text{U}}\) \\[0.2cm]
Wage Measure  & Mean &  Mean  & P50 & Mean & Mean & Mean & Mean  \\[0.2cm]
Vacancy Measure  & Overall &  Overall  & Overall & Registered & Overall & Overall & Overall  \\[0.2cm]
\multirow{2.4}{*}{\shortstack{Labor Market Definition}}  & \multirow{2.4}{*}{\shortstack{KldB-5 \\ $\times$ CZ}} &  \multirow{2.4}{*}{\shortstack{KldB-5 \\ $\times$ CZ}}  & \multirow{2.4}{*}{\shortstack{KldB-5 \\ $\times$ CZ}} & \multirow{2.4}{*}{\shortstack{KldB-5 \\ $\times$ CZ}} & \multirow{2.4}{*}{\shortstack{KldB-3/5 \\ $\times$ CZ}} & \multirow{2.4}{*}{\shortstack{Adj.\ KldB-5 \\ $\times$ CZ}} & \multirow{2.4}{*}{\shortstack{KldB-5 \\ $\times$ NUTS-3}} \\
&&&&&&& \\[0.2cm]
Observations &  7,993,993          & 7,993,993        &  7,993,993    &  7,993,993    & 8,068,327        &  8,130,231      &  7,499,651                \\[0.2cm]
Clusters &  1,801,671          & 1,801,671        &  1,801,671     &  1,801,671     & 1,812,829        &  1,821,306     &  1,723,597               \\[0.2cm]
F: $\Delta$ Log W$^{\text{FT}}$ & 1,168  & 3,431   & 3,209  & 3,321  & 3,263   & 3,249  &  3,073   \\[0.2cm]
F: $\Delta$ Log V/U & 17,381 & 30,048 & 30,380 & 27,826 & 40,883 & 66,365 &  12,069   \\[0.2cm] \hline
\end{tabular}
\begin{tablenotes}[para]
\footnotesize\textsc{Note. ---} The table displays IV regressions of differences in log employment (of regular full-time workers) per establishment on differences in the log of average daily wages and the log of labor market tightness. The instrumental variables refer to shift-share instruments of biennial national changes in occupations weighted by past occupational employment in the respective establishment. Industry refers to the 2-digit Classification of Economic Activities in the European Community (NACE). Standard errors (in parentheses) are clustered at the establishment level. Adj. = Flow-Adjusted. CZ = Commuting Zone. F = F Statistics of Excluded Instruments. FT = Full-Time. KldB-X = X-Digit German Classification of Occupations. KldB-3/5 = 3-Digit KldB Occupation by Level of Skill Requirement. L = Employment. P50 = Median. NUTS-X = X-Digit Statistical Nomenclature of Territorial Units. U = Job Seekers. V = Vacancies. W = Average Daily Wages. Z = Shift-Share Instrument. * = p$<$0.10. ** = p$<$0.05. *** = p$<$0.01. Sources: Integrated Employment Biographies $\plus$ Official Statistics of the German Federal Employment Agency $\plus$ IAB Job Vacancy Survey, 1999-2019.
\end{tablenotes}
\end{threeparttable}
}
\end{table}

\end{landscape}


\vspace*{3cm}
\begin{table}[!ht]
\centering
\scalebox{0.90}{
\begin{threeparttable}
\caption{Labor Demand Effects by Establishment Size}
\label{tab:F4}
\begin{tabular}{L{4cm}C{2.7cm}C{2.7cm}C{2.7cm}C{2.7cm}} \hline
\multirow{3.4}{*}{} & \multirow{3.4}{*}{\shortstack{(1) \\ $\Delta$ Log L$^{\text{FT}}$ }} & \multirow{3.4}{*}{\shortstack{(2) \\ $\Delta$ Log  L$^{\text{FT}}$ }} & \multirow{3.4}{*}{\shortstack{(3) \\ $\Delta$ Log  L$^{\text{FT}}$ }} & \multirow{3.4}{*}{\shortstack{(4) \\ $\Delta$ Log  L$^{\text{FT}}$ }}  \\
&&&& \\
&&&& \\[0.2cm] \hline
&&&& \\[-0.3cm]
\multirow{2.4}{*}{$\Delta$ Log W$^{\text{FT}}$} &  \multirow{2.4}{*}{\shortstack{\hphantom{***}-0.830***\hphantom{-} \\ (0.115)}}  & \multirow{2.4}{*}{\shortstack{\hphantom{***}-0.518***\hphantom{-} \\ (0.027)}} &  \multirow{2.4}{*}{\shortstack{\hphantom{***}-0.906***\hphantom{-} \\ (0.040)}} &  \multirow{2.4}{*}{\shortstack{\hphantom{***}-0.760***\hphantom{-} \\ (0.196)}}  \\
&&&& \\
\multirow{2.4}{*}{$\Delta$ Log V/U} &  \multirow{2.4}{*}{\shortstack{\hphantom{***}-0.054***\hphantom{-}\\ (0.011)}}  & \multirow{2.4}{*}{\shortstack{\hphantom{***}-0.045***\hphantom{-}\\ (0.002)}} &  \multirow{2.4}{*}{\shortstack{\hphantom{***}-0.082***\hphantom{-}\\ (0.003)}}  &  \multirow{2.4}{*}{\shortstack{\hphantom{***}-0.059***\hphantom{-}\\ (0.011)}} \\
&&&& \\[0.2cm]
Fixed Effects  & Year &  Year  & Year & Year \\[0.2cm]  \hline
&&&& \\[-0.2cm]
Instruments  & Z\(_{\text{W}^{\text{FT}}}\), Z\(_{\text{V}}\), Z\(_{\text{U}}\) & Z\(_{\text{W}^{\text{FT}}}\), Z\(_{\text{V}}\), Z\(_{\text{U}}\) & Z\(_{\text{W}^{\text{FT}}}\), Z\(_{\text{V}}\), Z\(_{\text{U}}\) & Z\(_{\text{W}^{\text{FT}}}\), Z\(_{\text{V}}\), Z\(_{\text{U}}\)\\[0.2cm]
Weighted Regression  & Yes &  No  & No & No \\
\multirow{2.4}{*}{Sample} & \multirow{2.4}{*}{\shortstack{All \vphantom{/} \\[-0.1cm] Establishments}} &  \multirow{2.4}{*}{\shortstack{Small \vphantom{/} \\[-0.1cm] Establishments}} & \multirow{2.4}{*}{\shortstack{Medium-Sized \vphantom{/} \\[-0.1cm] Establishments}} & \multirow{2.4}{*}{\shortstack{Large \vphantom{/} \\[-0.1cm] Establishments}} \\
&&&& \\[0.2cm]
Observations &  7,993,993          & 4,976,471        &  2,735,624          & 281,898               \\[0.2cm]
Clusters &  1,801,671          & 1,215,334        &  535,843            & 50,494                \\[0.2cm]
F: $\Delta$ Log W$^{\text{FT}}$ & 628    & 1,631   & 2,121  & 178    \\[0.2cm]
F: $\Delta$ Log V/U & 1,814  & 18,406 & 11,645 & 1,304  \\[0.2cm] \hline
\end{tabular}
\begin{tablenotes}[para]
\footnotesize\textsc{Note. ---} The table displays IV regressions of differences in log employment (of regular full-time workers) per establishment on differences in the log of average daily wages and the log of labor market tightness. The instrumental variables refer to shift-share instruments of biennial national changes in occupations weighted by past occupational employment in the respective establishment. The lag difference is two years. Labor markets are combinations of 5-digit KldB occupations and commuting zones. Regression weights reflect the number of workers of an establishment. We calculate time-constant establishment size categories from the unit-specific median of employees across available years: small (1-9 workers), medium (10-99 workers), and large (at least 100 workers). Standard errors (in parentheses) are clustered at the establishment level.  F = F Statistics of Excluded Instruments. FT = Full-Time. KldB = German Classification of Occupations. L = Employment. U = Job Seekers. V = Vacancies. W = Average Daily Wages. Z = Shift-Share Instrument. * = p$<$0.10. ** = p$<$0.05. *** = p$<$0.01. Sources: Integrated Employment Biographies $\plus$ Official Statistics of the German Federal Employment Agency $\plus$ IAB Job Vacancy Survey, 1999-2019.
\end{tablenotes}
\end{threeparttable}
}
\end{table}

\clearpage

\vspace*{3cm}
\begin{table}[!ht]
\centering
\scalebox{0.90}{
\begin{threeparttable}
\caption{Labor Demand Effects by Territory and AKM Effects}
\label{tab:F5}
\begin{tabular}{L{4cm}C{2.7cm}C{2.7cm}C{2.7cm}C{2.7cm}} \hline
\multirow{3.4}{*}{} & \multirow{3.4}{*}{\shortstack{(1) \\ $\Delta$ Log L$^{\text{FT}}$ }} & \multirow{3.4}{*}{\shortstack{(2) \\ $\Delta$ Log  L$^{\text{FT}}$ }} & \multirow{3.4}{*}{\shortstack{(3) \\ $\Delta$ Log  L$^{\text{FT}}$ }} & \multirow{3.4}{*}{\shortstack{(4) \\ $\Delta$ Log  L$^{\text{FT}}$ }}  \\
&&&& \\
&&&& \\[0.2cm] \hline
&&&& \\[-0.3cm]
\multirow{2.4}{*}{$\Delta$ Log W$^{\text{FT}}$} &  \multirow{2.4}{*}{\shortstack{\hphantom{***}-0.657***\hphantom{-} \\ (0.027)}}  & \multirow{2.4}{*}{\shortstack{\hphantom{***}-0.842***\hphantom{-} \\ (0.033)}} &  \multirow{2.4}{*}{\shortstack{\hphantom{***}-0.641***\hphantom{-} \\ (0.034)}} &  \multirow{2.4}{*}{\shortstack{\hphantom{***}-0.351***\hphantom{-} \\ (0.041)}}  \\
&&&& \\
\multirow{2.4}{*}{$\Delta$ Log V/U} &  \multirow{2.4}{*}{\shortstack{\hphantom{***}-0.049***\hphantom{-}\\ (0.002)}}  & \multirow{2.4}{*}{\shortstack{\hphantom{***}-0.058***\hphantom{-}\\ (0.004)}} &  \multirow{2.4}{*}{\shortstack{\hphantom{***}-0.028***\hphantom{-}\\ (0.003)}}  &  \multirow{2.4}{*}{\shortstack{\hphantom{***}-0.070***\hphantom{-}\\ (0.002)}} \\
&&&& \\[0.2cm]
Fixed Effects  & Year &  Year  & Year & Year \\[0.2cm]  \hline
&&&& \\[-0.2cm]
Instruments  & Z\(_{\text{W}^{\text{FT}}}\), Z\(_{\text{V}}\), Z\(_{\text{U}}\) & Z\(_{\text{W}^{\text{FT}}}\), Z\(_{\text{V}}\), Z\(_{\text{U}}\) & Z\(_{\text{W}^{\text{FT}}}\), Z\(_{\text{V}}\), Z\(_{\text{U}}\) & Z\(_{\text{W}^{\text{FT}}}\), Z\(_{\text{V}}\), Z\(_{\text{U}}\)\\[0.1cm]

\multirow{2.4}{*}{Sample} & \multirow{2.4}{*}{\shortstack{West \\Germany}} &   \multirow{2.4}{*}{\shortstack{East \\Germany}} & \multirow{2.4}{*}{\shortstack{Low AKM \\ Effects}} & \multirow{2.4}{*}{\shortstack{High AKM \\ Effects}} \\
&&&& \\[0.2cm]
Observations &  6,566,797          & 1,427,196        &  3,817,245          & 4,176,748             \\[0.2cm]
Clusters &  1,486,423          & 321,747          &  911,127            & 890,545               \\[0.2cm]
F: $\Delta$ Log W$^{\text{FT}}$ & 2,146  & 1,732   & 1,030  & 1,418  \\[0.2cm]
F: $\Delta$ Log V/U & 26,326 & 4,687  & 12,906 & 17,812 \\[0.2cm] \hline
\end{tabular}
\begin{tablenotes}[para]
\footnotesize\textsc{Note. ---} The table displays IV regressions of differences in log employment (of regular full-time workers) per establishment on differences in the log of average daily wages and the log of labor market tightness. The instrumental variables refer to shift-share instruments of biennial national changes in occupations weighted by past occupational employment in the respective establishment. The lag difference is two years. Labor markets are combinations of 5-digit KldB occupations and commuting zones. We separate employers into low- and high-productivity firms depending on whether their respective AKM wage effect lies below or above the median. Standard errors (in parentheses) are clustered at the establishment level. F = F Statistics of Excluded Instruments. FT = Full-Time. KldB = German Classification of Occupations. L = Employment. U = Job Seekers. V = Vacancies. W = Average Daily Wages. Z = Shift-Share Instrument. * = p$<$0.10. ** = p$<$0.05. *** = p$<$0.01. Sources: Integrated Employment Biographies $\plus$ Official Statistics of the German Federal Employment Agency $\plus$ IAB Job Vacancy Survey, 1999-2019.
\end{tablenotes}
\end{threeparttable}
}
\end{table}

\clearpage

\clearpage

\section{Flow-Adjusted Labor Market Tightness}
\label{sec:G}
\setcounter{table}{0} 
\setcounter{figure}{0} 

The calculation of meaningful measures of labor market tightness requires a definition of the relevant labor market. When studying labor market tightness, researchers normally use regions, and in rare cases, also occupations to delineate markets. In doing so, however, labor markets are implicitly divided into mutually exhaustive segments. This rules out that vacancies and job seekers in related markets can serve as valuable outside options for workers and firms. Consequently, if the labor market is defined too narrow (broad), the absolute number of vacancies and job seekers in the market will be underestimated (overestimated).

Regarding the spatial division, administrative regions (such as federal states or districts) do not necessarily overlap with the true regional scope of the underlying labor market. To address this problem, we follow previous research and construct functional labor market regions to capture commuting flows more adequately than administrative regions. In terms of the occupational division of the labor market, researchers normally rely on available classifications of occupations. However, in contrast to regions, it is not standard practice to group together related sub-occupations to improve the delineation of the occupational classifications.

To overcome this shortcoming, we implement a data-driven approach and consider vacancies and job seekers in differently classified occupations as additional vacancies and job seekers outside the focal occupation. Specifically, we follow \citet{Arnold2021}, who accounts for jobs in neighboring markets when calculating indices of labor market concentration, and transfer his method to the setting of labor market tightness. Specifically, we calculate a flow-adjusted version of labor market tightness that builds on mobility patterns across labor markets to attribute weights to vacancies and job seekers in neighboring occupations. The underlying idea is that the relative value of vacancies and job seekers in different occupations can be inferred from labor market flows within and between these occupations. Let $P(h|o)$ denote the probability that a worker in occupation $o$ in year $t$ is employed in occupation $h$ in year $t+1$. When working in (employing) occupation $o$, the worker's (firm's) relative value of a vacancy (job seeker) in occupation $h$ (compared to occupation $o$) then is:
\begin{equation}
\label{eq:G1}
\omega_{oh} = \frac{P(h|o)}{P(o|o)} \cdot \frac{L_{o}}{L_{h}}
\end{equation}
To infer weights from these flows, it is necessary to take into account that inflows from one market to another depend on the relative size of the markets. Therefore, we normalize relative transition probabilities by employment in the respective occupations. Note that, by construction, the occupation under study always receives unit weight, i.e., $\omega_{oo}=1$. When determining the transition probabilities with the employee-level administrative IEB data, we pool mobility patterns over commuting zones and the years 2012-2019 to arrive at a meaningful and stable weighting matrix.

Given the weighting matrix, the flow-adjusted number of vacancies in occupation $o$ and region $r$
\begin{equation}
\label{eq:G2}
\tilde{V}_{ort} = \sum_{h=1}^{H} \omega_{oh} V_{ort}
\end{equation}
is calculated as the weighted sum of vacancies in the same occupation and all other occupations in region $r$. By construction, the number of flow-adjusted vacancies always exceeds the number of actual vacancies in a labor market because the flow adjustment takes into account that workers can fill vacancies not only in the same occupation but also in neighboring occupations.

In a similar fashion, the flow-adjusted number of job seekers in occupation $o$ and region $r$
\begin{equation}
\label{eq:G3}
\tilde{U}_{ort} = \sum_{h=1}^{H} \omega_{oh} U_{ort}
\end{equation}
is calculated as the weighted sum of job seekers in the same occupation and all other occupations in region $r$. Thus, the number of flow-adjusted job seekers always exceeds the number of actual job seekers in a labor market because the flow adjustment takes into account that firms can recruit job seekers not only from the same occupation but also from neighboring occupations.

Unlike for occupations, our flow adjustment disregards vacancies and job seekers from neighboring regions since we already account for spatial transitions by delineating labor market regions based on observed commuting flows. Thus, units in the same occupation and commuting zone receive unit weight, units in all other occupations but the same commuting zone receive the weight $\omega_{oh}$, and occupations in all other commuting zones receive zero weight. When there are no flows between occupations, the neighboring occupations receive zero weight and the flow-based number of vacancies and job seekers collapses to the factual number of vacancies and job seekers in the labor market. When there are random flows between occupations, all occupations receive the same weight and the flow-based number of vacancies and job seekers collapses to the factual number of vacancies and job seekers in the entire commuting zone.

Mirroring Equation (\ref{eq:13}), we obtain our firm-specific measure of flow-adjusted labor market tightness
\begin{equation}
\label{eq:G4}
\tilde{\theta}_{it} = \sum_{o=1}^{O} \frac{L_{oit}}{L_{it}}  \cdot  \frac{\tilde{V}_{ort}}{\tilde{U}_{ort}}
\end{equation}
by dividing the flow-adjusted number of vacancies by the flow-adjusted number of job seekers for every occupation and weighting these flow-adjusted measures of labor market tightness with the occupational employment shares in the respective firm.

\clearpage

\section{Discussion: Further Evidence}
\label{sec:H}

\setcounter{table}{0} 
\setcounter{figure}{0} 

In this section, we discuss the implications of our findings in three further analyses. In the first analysis, we examine the wage and employment effects of the 2015 introduction of a statutory minimum wage in Germany in light of our regression results. In the second analysis, we calculate the employment effect from the doubling in labor market tightness between 2012 and 2019. In the third analysis, we explore whether firms made concessions in terms of lower wages and skills to reduce hiring frictions that originate from higher labor market tightness.

\paragraph{Minimum Wage Introduction in 2015.} For the first time, Germany introduced a national minimum wage on January 1, 2015. The minimum wage was set at 8.50 Euro per hour and strongly bit into the wage distribution, raising wages of about 17.8 percent of the workforce. In fact, the minimum wage was introduced mid-way during our period of analysis. Hence, variation from the minimum wage introduction is part of the variation leading to our elasticity estimates. In line with our Bartik-style identification strategy, the minimum wage caused an effective wage shift that strongly differed by occupation \citep{Friedrich2020}.

We use our estimated labor demand elasticities to predict employment effects from minimum wage induced wage changes in simulation exercises. Such a simulation of employment effects is particularly helpful for an assessment of policy effects before a minimum wage is introduced or raised. Prior to the minimum wage introduction in Germany, the simulation by \citet{KnabeEtAl2014} has had the most controversial impact in the scientific and public debate, especially since their results predicted substantial disemployment effects ranging between 425,000 and 910,000 jobs depending on the postulated market structure.\footnote{In an alternative simulation, \citet{MuellerSteiner2013} arrive at a negative employment effect of 490,000 workers in their preferred scenario. \citet{ArniEtAl2014} predict a loss of 570,000 jobs. Moreover, \citet{HenzelEngelhardt2014} gauge the disemployment effect to range between 470,000 and 1.4 million workers, depending on the underlying own-wage elasticity of labor demand (-0.1 vs.\ -0.8).} In their influential study, they provide a careful description of the wage distribution before the wage floor came into effect. They take the relative wage gaps for bins of workers which were paid below the minimum wage. To calculate employment effects for these group of workers, the authors interact the wage gaps with a uniform own-wage elasticity of labor demand of -0.75, which is retrieved from early reviews on elasticity estimates for Germany \citep{SinnEtAl2006,RagnitzThum2007}. In fact, the debate on the ex-ante simulation by \citet{KnabeEtAl2014} still continues since ex-post evaluations do not detect disemployment effects of the predicted size \citep{AhlfeldtEtAl2018,CaliendoEtAl2018,BosslerGerner2020,DustmannEtAl2022}.

While the assumed labor demand elasticity (-0.75) in the study of \citet{KnabeEtAl2014} is strikingly similar to our baseline own-wage elasticity at the firm level (-0.71), we can expand their simulation approach in various aspects: First, we can account for observed wage effects after the minimum wage introduction. In doing so, we account for non-compliance and spillovers which, of course, were not available when ex-ante simulations were debated. Second, we can apply separate own-wage elasticities of labor demand for full-time and part-time employment, which turns out to be an important distinction given the heterogeneities in Table \ref{tab:4}. Third, as the national statutory minimum wage applies to all employers in Germany, it will affect aggregate labor demand. As suggested by our results, such aggregated changes will alter labor market tightness via search externalities. Hence, building on our aggregate own-wage elasticity of labor demand, we can incorporate the feedback mechanism that limits employment effects of the minimum wage.

To simulate the aggregated employment effect, we need to estimate the causal wage effect of the minimum wage introduction, which we will then multiply by the own-wage elasticity of labor demand. A naive difference-in-difference estimation would likely overestimate the wage effect as low-wage workers may feature more positive earnings growth than high-wage workers. Hence, we estimate the wage effect from the following worker-level difference-in-difference-in-difference (DiDiD) specification:
\begin{equation}
\label{eq:H1}
\Delta^{t+2} \,ln\,W_{jt} \,=\, \beta_{0} \,+\, \beta_{1} \cdot \text{bite}_{jt} \,+\, \beta_{2} \cdot \text{cohort}_{jt} \,+\, \beta_{DiDiD} \cdot \text{bite}_{jt} \cdot \text{cohort}_{jt} \,+\, \varepsilon_{jt}
\end{equation}
The dependent variable is the change in log hourly wages over the upcoming two years of individual $j$ in cohort $t$, which is either the year 2012 or 2014.  
The bite variable measures the percentage difference between the wage of affected workers (in either 2012 or 2014) and the 2015 minimum wage level. The bite is zero for wages of unaffected workers above the minimum wage level. The coefficient $\beta_{1}$ captures general wage growth of affected workers, irrespective of the minimum wage introduction. The cohort dummy takes the value 1 for workers in 2014, when the minimum wage was upcoming, and 0 for workers in 2012. The coefficient $\beta_{2}$ captures wage growth of the 2014 cohort relative to the 2012 cohort independent of the minimum wage bite. We are interested in $\beta_{DiDiD}$ which is the wage growth of workers affected by the minimum wage i) relative to the wage growth of unaffected workers and ii) relative to wage growth from before the minimum wage introduction. Hence, $\beta_{DiDiD}$ yields the causal wage effect of the minimum wage on the treated workers.

We estimate Equation (\ref{eq:H1}) by OLS on the universe of administrative employment records in Germany (see Section \ref{sec:4}).\footnote{The only difference in terms of data to our analysis in Section \ref{sec:6} is a restriction to workers with a single job. This restriction is required to calculate worker-level wage growth.} The estimated DiDiD wage effect is 0.396 log points (standard error: 0.002) for affected full-time workers and 0.341 log points (standard error: 0.002) for affected part-time workers. These effects closely match the wage effects of \citet{BosslerSchank2022}, who identify wage effects of the minimum wage introduction from regional variation with a size of 0.4 log points. While the estimated wage effects are treatment effects on the treated (i.e., effects for workers who received an hourly wages below the threshold prior to the minimum wage introduction), we are interested in the aggregate wage effect. We calculate the aggregate wage effect of the minimum wage by multiplying the DiDiD-based wage effects with the average bite of the respective group of workers (1.7 percent for full-time and 9.6 percent for part-time workers): $\hat{\beta}_{DiDiD} \cdot \overline{\text{bite}}_{t=2014}$. We arrive at an aggregate wage growth of 0.7 and 3.3 percent for full-time and part-time workers, respectively.

Given the aggregate minimum wage effects on wages along with our estimated own-wage elasticities of labor demand, we simulate aggregated employment effects of the minimum wage introduction from the following equation:
\begin{equation}\label{eq:H2}
\underbrace{\vphantom{\big(}\,\,\,\,\,\,\,\Delta\, L\,\,\,\,\,\,\,}_{\substack{ \text{aggregate} \\ \text{minimum wage effect} \\ \text{on employment} }}  \,=\, \underbrace{\,\,\,\,\,\,\,\,\,\,\,\tilde{\eta}^{L}_{W}\,\,\,\,\,\,\,\,\,\,\,}_{\substack{\text{aggregate} \\ \text{own-wage elasticity} \\ \text{of labor demand} }} \,\,\,\, \cdot \,\,\,\, \underbrace{\,\,\big(\,\beta_{DiDiD} \cdot \overline{\text{bite}}_{t=2014}\,\big)\,\,}_{\substack{\text{aggregate} \\ \text{minimum wage effect} \\ \text{on wages} }} \,\,\,\, \cdot \,\,\,\,  \underbrace{\vphantom{\big(}\,\,\,L_{t=2014}\,\,\,}_{\text{workforce}}
\end{equation}
Table \ref{tab:H1} delivers the simulation results. Whereas Columns (1) and (2) display simulated minimum wage effects based on own-wage elasticities of labor demand at the firm level (i.e., without applying the feedback cycle), Columns (3) and (4) use the estimated aggregate own-wage elasticities of labor demand to account for search externalities. In both cases, we estimate separate effects for full- and part-time workers. As the own-wage elasticities of labor demand and the aggregate wage effects are estimated statistics, we draw the parameters from the underlying effect distributions. Based on 10,000 draws, we simulate standard errors of the predicted employment effects.

\begin{table}[!ht]
\centering
\scalebox{0.90}{
\begin{threeparttable}
\caption{Employment and Wage Effects of the German Statutory Minimum Wage}
\label{tab:H1}
\begin{tabular}{L{4.8cm}C{2.5cm}C{2.5cm}C{2.5cm}C{2.5cm}} \hline
\multirow{3.4}{*}{} & \multirow{3.4}{*}{\shortstack{(1) \\ L$^{\text{FT}}$ }} & \multirow{3.4}{*}{\shortstack{(2) \\  L$^{\text{PT}}$ }} & \multirow{3.4}{*}{\shortstack{(3) \\  L$^{\text{FT}}$ }} & \multirow{3.4}{*}{\shortstack{(4) \\ L$^{\text{PT}}$ }}  \\
&&&& \\
&&&& \\[0.2cm] \hline
&&&& \\[-0.3cm]
\multirow{2.4}{*}{\shortstack{Individual-Firm WELD}} &  \multirow{2.4}{*}{\shortstack{\hphantom{***}-0.713***\hphantom{-} \\ (0.021)}} & \multirow{2.4}{*}{\shortstack{-0.067\hphantom{-} \\ (0.060)}} &  &  \\
&&&&\\[0.2cm]
\multirow{2.4}{*}{\shortstack{Aggregate WELD}} &  &  & \multirow{2.4}{*}{\shortstack{\hphantom{***}-0.494***\hphantom{-} \\ (0.022)}} & \multirow{2.4}{*}{\shortstack{-0.046\hphantom{-} \\ (0.042)}} \\
&&&& \\[0.2cm] \hline
&&&& \\[-0.4cm]
\multirow{2.4}{*}{\shortstack[l]{Minimum Wage Effects \\ on Wages}} &  \multirow{2.4}{*}{\shortstack{\hphantom{***}0.0069***\hphantom{-} \\ (0.00004)}} & \multirow{2.4}{*}{\shortstack{\hphantom{***}0.0328***\hphantom{-} \\ (0.0003)}} & \multirow{2.4}{*}{\shortstack{\hphantom{***}0.0069***\hphantom{-} \\ (0.00004)}}  & \multirow{2.4}{*}{\shortstack{\hphantom{***}0.0328***\hphantom{-} \\ (0.0003)}}  \\
&&&& \\[0.2cm] \hline
&&&& \\[-0.4cm]
\multirow{2.4}{*}{\shortstack[l]{Minimum Wage Effects \\ on Employment}} &  \multirow{2.4}{*}{\shortstack{\hphantom{***}-96,432***\hphantom{-} \\ (2,916)}} & \multirow{2.4}{*}{\shortstack{-29,867\hphantom{-} \\ (27,054)}}   & \multirow{2.4}{*}{\shortstack{\hphantom{***}-66,757***\hphantom{-} \\ (3,053)}}  &   \multirow{2.4}{*}{\shortstack{-21,087\hphantom{-} \\ (19,008)}}   \\
&&&& \\[0.2cm] 
\multirow{2.4}{*}{\shortstack[l]{Minimum Wage Effects \\ on Total Employment}} &  \multicolumn{2}{c}{\multirow{2.4}{*}{\shortstack{\hphantom{***}-126,299***\hphantom{-} \\ (27,216)}}}
 &  \multicolumn{2}{c}{\multirow{2.4}{*}{\shortstack{\hphantom{***}-87,844***\hphantom{-} \\ (19,218)}}}   \\
&&&& \\[0.2cm]\hline
\end{tabular}
\begin{tablenotes}[para]
\footnotesize\textsc{Note. ---} The table presents simulation results for employment effects of the 2015 introduction of the nation-wide minimum wage in Germany. Following Equation (\ref{eq:H2}), we interact the individual-firm or aggregate own-wage elasticity of labor demand (first and second row) from Table \ref{tab:4} and Equation (\ref{eq:11}) with aggregate minimum wage effects on wages from Equation (\ref{eq:H1}) (third row). Columns (1) and (2) use individual-firm own-wage elasticities of labor demand whereas Columns (3) and (4) incorporate the aggregate own-wage elasticities of labor demand, for full- and part-time workers respectively. We retrieve standard errors of the simulated employment effects by drawing 10,000 realizations from the underlying effects distributions of the estimated own-wage elasticities of labor demand and the estimated minimum wage effect on wages. FT = Full-Time. L = Employment. PT = Part-Time. WELD = Own-Wage Elasticity of Labor Demand. * = p$<$0.10. ** = p$<$0.05. *** = p$<$0.01. Source: Integrated Employment Biographies, 2012-2016.
\end{tablenotes}
\end{threeparttable}
}
\end{table}

The baseline simulation absent search externalities yields a negative effect on full-time employment of $-96,\!432$ workers, which is broadly in line with \citet{KnabeEtAl2014} who report disemployment effects of 160,000 (competitive model) and 40,000 (monopsony model) for full-time workers. In Table \ref{tab:H2}, we compare heir findings with the results from our simulation in more detail. However, we find a much smaller effect on part-time employment that is only $-29,867$ and statistically insignificant, which stems from the small own-wage elasticity of labor demand of this group. In total, our simulation yields a disemployment effect of 126,299 workers. Crucially, however, this effect disregards that an aggregate decline in labor demand reduces labor market tightness.


When incorporating the feedback mechanism, the aggregate decline reduces to 66,757 full-time and 21,087 part-time workers, which adds up to an overall disemployment effect of 87,844 workers. This much smaller disemployment effect mirrors evidence from ex-post evaluations of the minimum wage \citep{Bruttel2019,CaliendoEtAl2019}.
While our effects are somewhat smaller than the estimates in \citet{CaliendoEtAl2018}, they slightly exceed the estimated aggregated employment effect in \citet{BosslerGerner2020}. \citet{DustmannEtAl2022} argue that most of the firm-level employment reduction is offset by job mobility to competing employers. Thus, their finding of reallocation effects closely matches our reasoning about search externalities, namely that a labor demand reduction facilitates employment expansions at other firms.

\begin{table}[!ht]
\centering
\scalebox{0.90}{
\begin{threeparttable}
\caption{Comparison of Minimum Wage Simulations}
\label{tab:H2}
\begin{tabular}{L{7.5cm}C{2.5cm}C{2.5cm}C{2.5cm}} \hline
\multirow{3.4}{*}{} & \multirow{3.4}{*}{\shortstack{Full-Time \\ Employment }} & \multirow{3.4}{*}{\shortstack{Part-Time \\ Employment }} & \multirow{3.4}{*}{\shortstack{Overall \\ Employment}}   \\
&&& \\
&&& \\[0.2cm] \hline
&&& \\[-0.3cm]
\multirow{2.4}{*}{\shortstack[l]{ WELD = -0.75 \\ Ex-Ante Wage Gap (SOEP) }} &  \multirow{2.4}{*}{\shortstack{-160,203}} & \multirow{2.4}{*}{\shortstack{-750,514}} &
\multirow{2.4}{*}{\shortstack{-910,717}}\\
&&&\\[0.2cm] 

\multirow{2.4}{*}{\shortstack[l]{WELD = -0.75 \\ Ex-Post Wage Effect (IEB)  }} & \multirow{2.4}{*}{\shortstack{-102,596}} & \multirow{2.4}{*}{\shortstack{-340,147}} &
\multirow{2.4}{*}{\shortstack{-442,743}} \\
&&&\\[0.2cm] 

\multirow{2.4}{*}{\shortstack[l]{Estimated Invididual-Firm WELDs \\ Ex-Post Wage Effect (IEB) }} & \multirow{2.4}{*}{\shortstack{-96,432}} & \multirow{2.4}{*}{\shortstack{-29,867}} &
\multirow{2.4}{*}{\shortstack{-126,299}}    \\
&&& \\[0.2cm] 

\multirow{2.4}{*}{\shortstack[l]{Estimated Aggregate WELDs  \\ Ex-Post Wage Effect (IEB) }} &\multirow{2.4}{*}{\shortstack{-66,757}} & \multirow{2.4}{*}{\shortstack{-21,087}} &
\multirow{2.4}{*}{\shortstack{-87,843}}    \\
&&& \\[0.2cm] \hline

\end{tabular}
\begin{tablenotes}[para]
\footnotesize\textsc{Note. ---} The table displays the employment effects from different simulations of the 2015 national minimum wage introduction in Germany. The first row presents the results from the simulation in Knabe, Schöb, and Thum (\citeyear{KnabeEtAl2014}, Table 9a). The study is based on a single own-wage elasticity of -0.75 and assumes that wage increases mirror ex-ante wage gaps prior to the minimum wage introduction (based on SOEP data). In the second row, we apply the approach from \citet{KnabeEtAl2014} to observed minimum wage effects on wages (based on IEB data).  In the third row, we instead make use of our estimated individual-firm own-wage elasticities of labor demand for full-time (-0.71) and part-time workers (-0.08). The last row mirrors our preferred simulation from Table \ref{tab:H1} in which we replace the individual-firm elasticities by our estimates for the aggregate own-wage elasticities for full- (-0.50) and part-time workers (-0.06). IEB = Integrated Employment Biographies. SOEP = German Socio-Economic Panel. WELD = Own-Wage Elasticity of Labor Demand. Sources: \citet{KnabeEtAl2014} $\plus$ Integrated Employment Biographies, 2012-2016.
\end{tablenotes}
\end{threeparttable}
}
\end{table}

The literature offers several explanations to rationalize the absence of large negative employment effects of minimum wages with theory \citep{Schmitt2015}. Analyses in terms of headcount employment (extensive margin) may underestimate the overall employment effect when minimum wages spark off reductions in working hours (intensive margin). Product price increases can buffer higher personnel cost \citep{AaronsonEtAl2008}. An increasing labor productivity may enable firms to pay the minimum wage \citep{RileyBondibene2017}. Moreover, a fraction of firms may not comply with the minimum wage legislation \citep{AshenfelterSmith1979}. In monopsonistic labor markets, modest wage floors can even stimulate employment \citep{Stigler1946}. Our analysis provides an additional explanation for this puzzle: aggregate reductions in employment lower labor market tightness through search externalities which in turn facilitates recruitment for firms that want to hire. We demonstrate that this channel reduces the disemployment effect of minimum wages by about thirty percent.

\paragraph{Employment Trends and Labor Market Tightness.} In a second application, we quantify the overall impact of the doubling of labor market tightness between 2012 and 2019 on aggregate employment. Such a quantification is highly relevant as there was an emerging public opinion that the increased tightness was posing a severe problem to the German labor market.
To the best of our knowledge, there is no causal evidence on the absolute magnitude of employment effects of the increased labor market tightness.

In a counterfactual analysis, we compare the observed aggregated employment growth with a hypothetical scenario in which labor market tightness did not change. For the counterfactual, we use our estimated tightness effects on labor demand and simulate the development of aggregate employment conditional on an unchanged labor market tightness during the period of analysis (i.e., we fix the ratio of vacancies to job seekers at the level of 2012).
In this hypothetical scenario, for each year, we multiply the observed relative change in tightness by our estimated elasticities of labor demand with respect to tightness (separately for full- and part-time workers), and subtract the respective sum from the factual stock of employment.

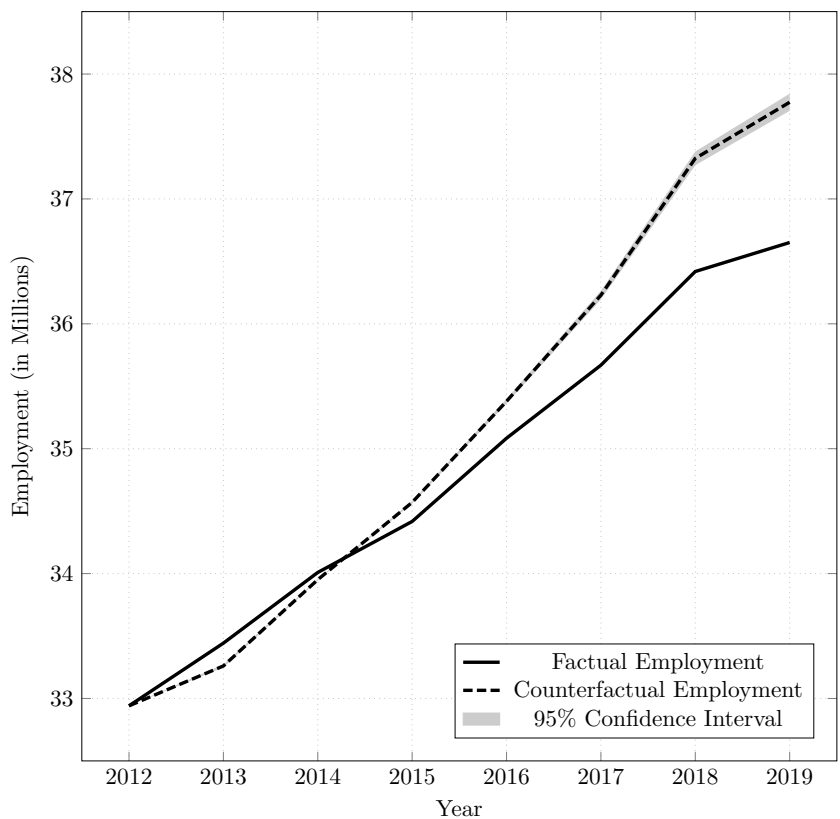
\begin{figure}[!ht]
\centering
\caption{Labor Market Tightness and Employment Trends}
\label{fig:H1}
\scalebox{0.80}{
\begin{tikzpicture}
\begin{axis}[xlabel=Year, ylabel={Employment (in Millions)}, xmin=2011.5,xmax=2019.5, ymin=32.5, ymax=38.5, height=14cm, width=14cm, grid=major, grid style=dotted, ytick={33,34,35,36,37,38}, xtick={2012,2013,2014,2015,2016,2017,2018,2019}, legend pos = south east, y tick label style={/pgf/number format/.cd,fixed,fixed zerofill, precision=0,/tikz/.cd}, x tick label style={/pgf/number format/.cd,fixed,fixed zerofill, precision=0,/tikz/.cd}]
\addplot[mark=none, solid, color=black, line width=1.6pt] coordinates {  (2012,32.941101) (2013,33.44294) (2014,34.008938) (2015,34.41737) (2016,35.083866) (2017,35.668858) (2018,36.41888) (2019,36.651081) };
\addplot[mark=none, dash pattern = on 4.4pt off 2pt, color=black, line width=1.6pt] coordinates {  (2012,32.941101) (2013,33.260265) (2014,33.951328) (2015,34.570015) (2016,35.379581) (2017,36.229523) (2018,37.324829) (2019,37.774082) };
\addplot[draw=none, name path = A] coordinates {  (2012,32.941101) (2013,33.24868) (2014,33.946568) (2015,34.56015) (2016,35.361244) (2017,36.194633) (2018,37.269867) (2019,37.705395) };
\addplot[draw=none, name path = B] coordinates {  (2012,32.941101) (2013,33.271847) (2014,33.956093) (2015,34.579884) (2016,35.397919) (2017,36.264416) (2018,37.379791) (2019,37.84277) };
\addplot[color=gray!40!white] fill between[of=A and B];
\legend{~Factual Employment,~Counterfactual Employment,,, ~95\% Confidence Interval}
\end{axis}
\end{tikzpicture}
}
\floatfoot{\footnotesize\textsc{Note. ---} The figure contrasts the factual trend of total employment in Germany with a hypothetical trend that simulates employment if labor market tightness was fixed at its 2012 level. We simulate full- and part-time employment separately and add them up to predict total employment. We draw 10,000 realizations from the distribution of the estimated labor demand elasticity with respect to labor market tightness to calculate standard errors for the simulated employment effect. The grey shade indicates 95 percent confidence intervals. Employment refers to the number of jobs (as opposed to individual workers) which are subject to social security contributions. This number refers to the total number of jobs minus civil servants, family workers, apprentices, and people in partial retirement schemes. Sources: Integrated Employment Biographies $\plus$ Official Statistics of the German Federal Employment Agency $\plus$ IAB Job Vacancy Survey, 2012-2019.}
\end{figure}

Figure \ref{fig:H1} displays the factual and counterfactual development of total employment, which is the sum of full- and part-time jobs.\footnote{Given the IEB data, employment refers to the number of jobsich  (as opposed to individual workers) that are subject to social security contributions. This number refers to the total number of jobs minus civil servants, family workers, apprentices, and people in partial retirement schemes.}
While observed total employment rose from 32.9 million employees in 2012 to 36.7 million jobs in 2019, it could have risen to 37.8 million jobs if labor market tightness had not changed. Thus, in the absence of increasing labor market tightness, employment could have grown by an additional 1.1 million jobs until 2019. Figure \ref{fig:H1} delivers a separate analysis of full-time and part-time employment, showing that the growth of both kinds of employment was slowed by the increase in labor market tightness between 2012 and 2019. In sum, our results imply that the increase in labor market tightness considerably dampened the positive trend in labor demand, underlining the importance of hiring frictions in tight labor markets.

As labor market tightness is not solely a market outcome, our finding of labor shortage raises the question of potential policy interventions. First, at given labor market tightness, improvements in the matching efficiency would result in more hires (e.g., better public and private employment services). Second, given our negative tightness effect, any measures that lift the number of job seekers (relative to vacancies) would stimulate employment. In the short and medium run, appropriate policies could i) encourage inactive individuals to search for jobs, ii) stimulate female labor supply, iii) allow for immigration of workers, or iv) raise the effective retirement age. In the long run, higher birth rates would have a positive effect on the working population. Apart from policy measures, firms can partly circumvent labor shortage by i) substituting labor by capital (e.g., machines), ii) increasing productivity or working hours of incumbent workers, iii) reducing outflow of workers (e.g., by raising wages or improving non-wage job amenities), or iv) making wage or skill concessions upon hiring.

\afterpage{
\begin{landscape}
\begin{figure}[!t]
\centering
\caption{Labor Market Tightness and Employment Trends by Labor Outcome}
\label{fig:H2}
\vspace{0.1cm}
\begin{subfigure}{0.475\textwidth}
\caption{Full-Time Workers}
\centering
\scalebox{0.70}{
\begin{tikzpicture}
\begin{axis}[xlabel=Year, ylabel={Employment (in Millions)}, xmin=2011.5,xmax=2019.5, ymin=18.75, ymax=22.25, height=14cm, width=14cm, grid=major, grid style=dotted, ytick={19,19.5,20,20.5,21,21.5,22}, xtick={2012,2013,2014,2015,2016,2017,2018,2019}, legend pos = south east, y tick label style={/pgf/number format/.cd,fixed,fixed zerofill, precision=1,/tikz/.cd}, x tick label style={/pgf/number format/.cd,fixed,fixed zerofill, precision=0,/tikz/.cd}]
\addplot[mark=none, solid, color=black, line width=1.6pt] coordinates {  (2012,19.360292) (2013,19.505922) (2014,19.717863) (2015,19.899733) (2016,20.18586) (2017,20.435862) (2018,20.839214) (2019,20.9289) };
\addplot[mark=none, dash pattern = on 4.4pt off 2pt, color=black, line width=1.6pt] coordinates {  (2012,19.360292) (2013,19.393694) (2014,19.682507) (2015,19.993425) (2016,20.367437) (2017,20.780184) (2018,21.395611) (2019,21.618454) };
\addplot[draw=none, name path = A] coordinates {  (2012,19.360292) (2013,19.384344) (2014,19.679014) (2015,19.98554) (2016,20.352428) (2017,20.751644) (2018,21.350519) (2019,21.562147) };
\addplot[draw=none, name path = B] coordinates {  (2012,19.360292) (2013,19.403042) (2014,19.685999) (2015,20.001312) (2016,20.382448) (2017,20.808723) (2018,21.440699) (2019,21.674761) };
\addplot[color=gray!40!white] fill between[of=A and B];
\legend{~Factual Employment,~Counterfactual Employment,,, ~95\% Confidence Interval}
\end{axis}
\end{tikzpicture}
}
\end{subfigure}
\hfill
\begin{subfigure}{0.475\textwidth}
\caption{Part-Time Workers}
\centering
\scalebox{0.70}{
\begin{tikzpicture}
\begin{axis}[xlabel=Year, ylabel={Employment (in Millions)}, xmin=2011.5,xmax=2019.5, ymin=13.25, ymax=16.75, height=14cm, width=14cm, grid=major, grid style=dotted, ytick={13.5,14,14.5,15,15.5,16,16.5}, xtick={2012,2013,2014,2015,2016,2017,2018,2019}, legend pos = south east, y tick label style={/pgf/number format/.cd,fixed,fixed zerofill, precision=1,/tikz/.cd}, x tick label style={/pgf/number format/.cd,fixed,fixed zerofill, precision=0,/tikz/.cd}]
\addplot[mark=none, solid, color=black, line width=1.6pt] coordinates {  (2012,13.580809) (2013,13.937017) (2014,14.291071) (2015,14.517637) (2016,14.898007) (2017,15.232993) (2018,15.579667) (2019,15.722178) };
\addplot[mark=none, dash pattern = on 4.4pt off 2pt, color=black, line width=1.6pt] coordinates {  (2012,13.580809) (2013,13.86657) (2014,14.268823) (2015,14.576591) (2016,15.012143) (2017,15.449339) (2018,15.92922) (2019,16.15563) };
\addplot[draw=none, name path = A] coordinates {  (2012,13.580809) (2013,13.860029) (2014,14.266445) (2015,14.571106) (2016,15.001665) (2017,15.429731) (2018,15.897198) (2019,16.116089) };
\addplot[draw=none, name path = B] coordinates {  (2012,13.580809) (2013,13.873111) (2014,14.2712) (2015,14.582075) (2016,15.022621) (2017,15.468948) (2018,15.961242) (2019,16.195169) };
\addplot[color=gray!40!white] fill between[of=A and B];
\legend{~Factual Employment,~Counterfactual Employment,,, ~95\% Confidence Interval}
\end{axis}
\end{tikzpicture}
}
\end{subfigure}
\floatfoot{\footnotesize\textsc{Note. ---} The figure contrasts factual trends of employment in Germany with hypothetical trends that simulate employment if labor market tightness was fixed at its 2012 level. We simulate trends separately for full-time and part-time employment. Part-time employment encompasses regular part-time and marginal part-time workers. We draw 10,000 realizations from the effect distribution of the respective labor demand elasticity with respect to tightness to calculate standard errors for the simulated employment effects. The grey shade indicates 95\% confidence intervals. Employment refers to the number of jobs (as opposed to individual workers) which are subject to social security contributions. This number refers to the total number of jobs minus civil servants, family workers, apprentices, and people in partial retirement schemes. Sources: Integrated Employment Biographies $\plus$ Official Statistics of the German Federal Employment Agency $\plus$ IAB Job Vacancy Survey, 2012-2019.}
\end{figure}
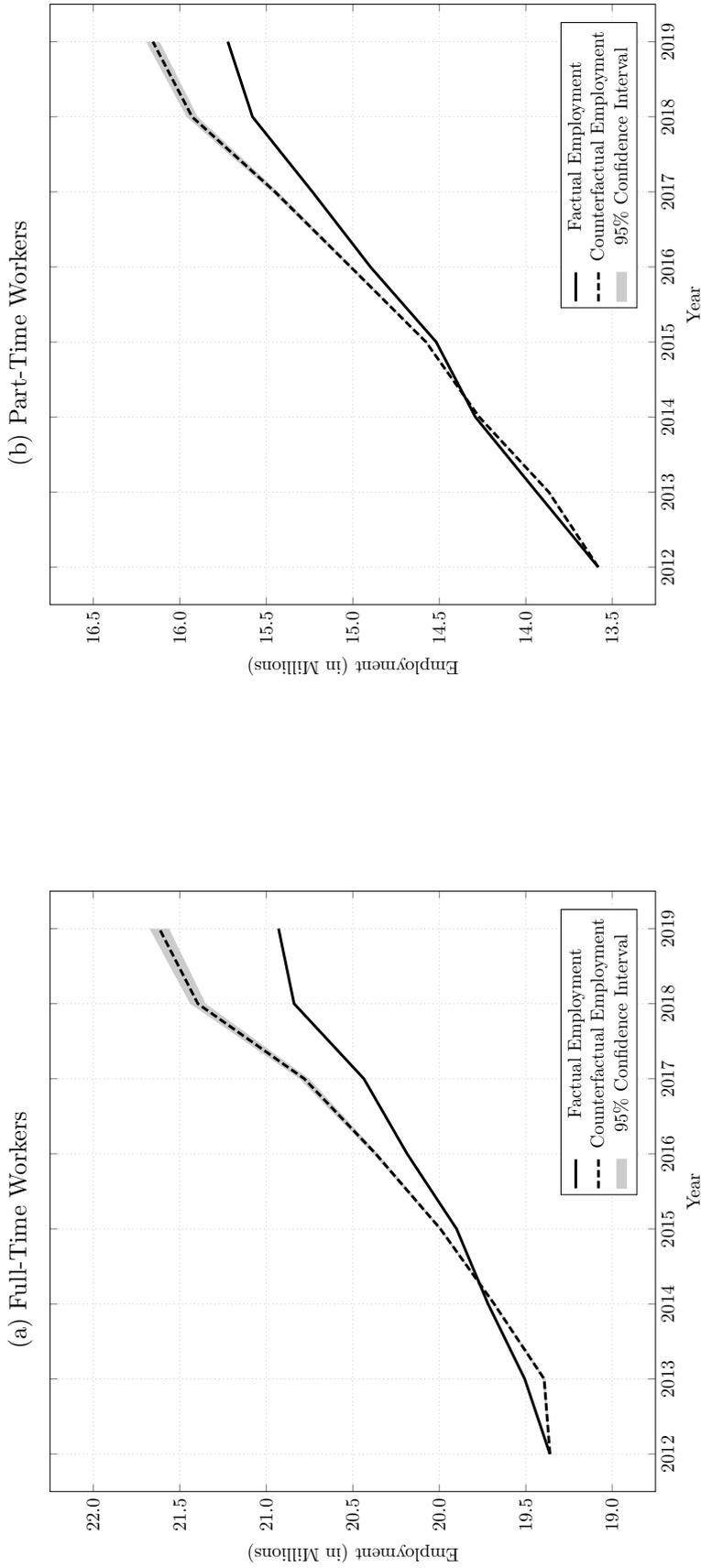

\end{landscape}
}

\paragraph{Wage and Skill Concessions.} In practice, firms facing higher labor market tightness do not necessarily have to settle for lower employment levels. Instead, these firms could still manage to expand or retain the workforce by making concessions, e.g., by raising wages or by recruiting workers with lower skills. The theoretical literature discusses extensively the positive effect of higher labor market tightness on wages, which is commonly referred to as the ``wage curve'' \citep{BlanchflowerOswald1995,Card1995}. First, in search-and-matching models, employers pay higher wages as the value of a filled position increases with labor market tightness \citep{MortensenPissarides1999}. Second, in bargaining models, high labor market tightness enables workers to extract a larger fraction of the overall surplus due to more outside options \citep{Nash1950,NickellAndrews1983}. Third, according to the efficiency wage hypothesis, firms may raise wages above market-clearing levels to retain incumbent workers and attract hires \citep{Stiglitz1974,Yellen1984}.

We empirically address the conjecture that firms need to make concessions to maintain their employment (growth) in tight labor markets. Building on the same instrumental variable approach as in our analysis of labor demand, we regress the average wage level of firms and the fraction of unskilled workers in a firm on our measure of firm-specific labor market tightness. Table \ref{tab:H3} shows the respective results for full-time and part-time employment. In Columns (1) and (2), we observe a positive effect of higher labor market tightness on the wage level in a firm: on average, the doubling in tightness (i.e., a 100 percent increase in the vacancy-to-job-seekers ratio) raises mean wages of full-time (part-time) workers in a firm by 0.9 (0.5) percent. Although the wage response is significantly positive for both groups of workers, the magnitude of these effects is fairly small: wages of full-time workers only increased by about a fifth of the negative employment response and only by a tenth when looking at part-timers. This result mirrors a relatively flat but positively sloped wage curve, which is in line with previous findings from Germany \citep{BaltagiEtAl2009,BellmannBlien2001}. Rather than tightness, these studies relate wages only to the unemployment rate, which is a less comprehensive measure for workers' outside options. Our flat wage curve indicates an only limited role of outside options in the bargaining process, which can be rationalized on two grounds. First, firms may rely on wage posting instead of wage bargaining. Second, both parties may resume rather than terminate unsuccessful negotiations \citep{HallMilgrom2008} in the consensus-based system of industrial relations in Germany \citep{DustmannEtAl2014}.

\begin{table}[!ht]
\centering
\scalebox{0.90}{
\begin{threeparttable}
\caption{Wage and Skill Concessions}
\label{tab:H3}
\begin{tabular}{L{4cm}C{2.7cm}C{2.7cm}C{2.7cm}C{2.7cm}} \hline
\multirow{4.4}{*}{} & \multirow{4.4}{*}{\shortstack{(1) \\ $\Delta$ Log W$^{\text{FT}}$ }} & \multirow{4.4}{*}{\shortstack{(2) \\ $\Delta$ Log  W$^{\text{PT}}$ }} & \multirow{4.4}{*}{\shortstack{(3) \\ Share of \\ Unskilled in \\ FT Workers }} & \multirow{4.4}{*}{\shortstack{(4) \\ Share of \\ Unskilled in \\ PT Workers }} \\
&&&& \\
&&&& \\
&&&& \\[0.2cm] \hline
&&&& \\[-0.3cm]
\multirow{2.4}{*}{$\Delta$ Log W$^{\text{FT}}$} &    &  & \multirow{2.4}{*}{\shortstack{\hphantom{***}-0.031***\hphantom{-} \\ (0.006)}}   &   \\
&&&& \\
\multirow{2.4}{*}{$\Delta$ Log W$^{\text{PT}}$} &    &  &    &  \multirow{2.4}{*}{\shortstack{\hphantom{***}-0.062***\hphantom{-} \\ (0.023)}} \\
&&&& \\
\multirow{2.4}{*}{$\Delta$ Log V/U} &  \multirow{2.4}{*}{\shortstack{\hphantom{***}0.009***\\ (0.001)}}  & \multirow{2.4}{*}{\shortstack{\hphantom{***}0.005***\\ (0.002)}} & \multirow{2.4}{*}{\shortstack{\hphantom{***}0.003***\\ (0.001)}} & \multirow{2.4}{*}{\shortstack{\hphantom{**}0.001**\\ (0.001)}}  \\
&&&& \\[0.2cm]
Fixed Effects  & Year &  Year  & Year & Year \\[0.2cm] \hline
&&&& \\[-0.2cm]
Instruments  & Z\(_{\text{V}}\), Z\(_{\text{U}}\) & Z\(_{\text{V}}\), Z\(_{\text{U}}\)  & Z\(_{\text{W}^{\text{FT}}}\), Z\(_{\text{V}}\), Z\(_{\text{U}}\) & Z\(_{\text{W}^{\text{PT}}}\), Z\(_{\text{V}}\), Z\(_{\text{U}}\) \\[0.2cm]
Observations &  7,993,993             & 11,448,610          &  7,589,549             & 10,107,198               \\[0.2cm]
Clusters &  1,801,671             & 2,693,588           &  1,693,188            & 2,407,445                \\[0.2cm]
F: $\Delta$ Log W$^{\text{FT}}$ &  &    & 2,825  &  \\[0.2cm]
F: $\Delta$ Log W$^{\text{PT}}$ &  &    &  & 61      \\[0.2cm]
F: $\Delta$ Log V/U & 45,522 & 46,424 & 28,327 & 27,168 \\[0.2cm] \hline
\end{tabular}
\begin{tablenotes}[para]
\footnotesize\textsc{Note. ---} The table displays IV regressions of differences in measures of wage and skill concessions per establishment on differences in the log of average hourly wages and the log of labor market tightness. The instrumental variables refer to shift-share instruments of biennial national changes in occupations weighted by past occupational employment in the respective establishment. The lag difference is two years. Labor markets are combinations of 5-digit KldB occupations and commuting zones. Full-time employment includes regular full-time workers whereas part-time employment encompasses regular part-time and marginal part-time workers. Unskilled workers have neither completed vocational education nor have acquired a university degree. Standard errors (in parentheses) are clustered at the establishment level. F = F Statistics of Excluded Instruments. FT = Full-Time. KldB = German Classification of Occupations. L = Employment. PT = Part-Time. U = Job Seekers. V = Vacancies. W = Average Hourly Wages. Z = Shift-Share Instrument. * = p$<$0.10. ** = p$<$0.05. *** = p$<$0.01. Sources: Integrated Employment Biographies $\plus$ Official Statistics of the German Federal Employment Agency $\plus$ IAB Job Vacancy Survey, 1999-2019.
\end{tablenotes}
\end{threeparttable}
}
\end{table}

We cross-validate our finding of relatively small wage increases by examining additional information from the IAB Job Vacancy Survey on whether firms were willing to accept wage concessions upon hiring. Between 2012 and 2019, employers report wage concession in 15.7 percent of all hires. A naive regression of a binary variable for wage concessions on log labor market tightness suggests that the doubling of tightness raised the probability of a wage concession upon hiring by only 3.4 percentage points.

Columns (3) and (4) display the regressions for the fraction of unskilled workers. We define unskilled workers as employees who neither have completed vocational training nor hold a university degree. In 2012, the average share of unskilled workers in full-time and part-time employment across firms was 6.4 and 9.5 percent, respectively. The results imply that wage increases lowered the share of unskilled workers in full-time employment, reflecting positive returns to skills. For the effect of labor market tightness, we arrive at a significantly positive semi-elasticity of 0.003 for full-time workers and 0.001 for part-time workers. Hence, firms were willing to make some skill concessions. The massive increase in labor market tightness (by 100 percent) resulted in an increase in the share of unskilled workers in full-time (part-time) employment by 0.3 (0.1) percentage points.\footnote{Similarly, \citet{Koelling2020a} finds that German establishments which report labor shortages between 2004 and 2014 employ ceteris paribus more low- and medium-skilled but less high-skilled workers.} 
According to the IAB Job Vacancy Survey, firms hired workers with lower skills than originally demanded in 9.9 percent of new matches between 2012 and 2019. In line with little skill concessions, pooled OLS regressions imply that the doubling of labor market tightness raised the probability of hiring a worker with lower skills by only 1.8 percentage points.

Our estimates suggest that the extent of firms' wage and skills concession was fairly small in practice, providing an explanation for the markedly negative effect of labor market tightness on employment. Crucially, however, the results do not shed light on whether profit-maximizing firms were not willing or, alternatively, were not able to make substantial concessions. On the one hand, firms with monopoly or monopsony power dispose of rents but have an incentive to stay small. On the other hand, firms without rents would incur losses when raising wages above the worker's marginal value product.

\clearpage
\addcontentsline{toc}{section}{References} 
\printbibliography[title=References] 

\end{refsection}
\end{appendix}

\end{document}